\PassOptionsToPackage{unicode}{hyperref}
\PassOptionsToPackage{hyphens}{url}
\PassOptionsToPackage{dvipsnames,svgnames,x11names}{xcolor}
\documentclass[twoside,11pt]{article}

\usepackage{jmlr2e}
\usepackage{amsmath,amssymb}
\usepackage{iftex}
\ifPDFTeX
  \usepackage[T1]{fontenc} 
  \usepackage[utf8]{inputenc}
  \usepackage{textcomp} % provide euro and other symbols
\else % if luatex or xetex
  \usepackage{unicode-math}
  \defaultfontfeatures{Scale=MatchLowercase}
  \defaultfontfeatures[\rmfamily]{Ligatures=TeX,Scale=1}
\fi
\usepackage{lmodern}
\ifPDFTeX\else  
\fi
\IfFileExists{upquote.sty}{\usepackage{upquote}}{}
\IfFileExists{microtype.sty}{% use microtype if available
  \usepackage[]{microtype}
  \UseMicrotypeSet[protrusion]{basicmath} % disable protrusion for tt fonts
}{}
\usepackage{xcolor}
\usepackage{colortbl}
\definecolor{highlightColor}{RGB}{232,242,252}
\makeatletter
\ifx\paragraph\undefined\else
  \let\oldparagraph\paragraph
  \renewcommand{\paragraph}{
    \@ifstar
      \xxxParagraphStar
      \xxxParagraphNoStar
  }
  \newcommand{\xxxParagraphStar}[1]{\oldparagraph*{#1}\mbox{}}
  \newcommand{\xxxParagraphNoStar}[1]{\oldparagraph{#1}\mbox{}}
\fi
\ifx\subparagraph\undefined\else
  \let\oldsubparagraph\subparagraph
  \renewcommand{\subparagraph}{
    \@ifstar
      \xxxSubParagraphStar
      \xxxSubParagraphNoStar
  }
  \newcommand{\xxxSubParagraphStar}[1]{\oldsubparagraph*{#1}\mbox{}}
  \newcommand{\xxxSubParagraphNoStar}[1]{\oldsubparagraph{#1}\mbox{}}
\fi
\makeatother

\usepackage{longtable,booktabs,array}
\usepackage{calc} % for calculating minipage widths
\usepackage{etoolbox}
\makeatletter
\patchcmd\longtable{\par}{\if@noskipsec\mbox{}\fi\par}{}{}
\makeatother
\IfFileExists{footnotehyper.sty}{\usepackage{footnotehyper}}{\usepackage{footnote}}
\makesavenoteenv{longtable}
\usepackage{graphicx}
\usepackage{wrapfig}
\usepackage{rotating}
\usepackage{multirow}
\makeatletter
\def\maxwidth{\ifdim\Gin@nat@width>\linewidth\linewidth\else\Gin@nat@width\fi}
\def\maxheight{\ifdim\Gin@nat@height>\textheight\textheight\else\Gin@nat@height\fi}
\makeatother
\setkeys{Gin}{width=\maxwidth,height=\maxheight,keepaspectratio}
\makeatletter
\def\fps@figure{htbp}
\makeatother

\makeatletter
\@ifpackageloaded{caption}{}{\usepackage{caption}}
\AtBeginDocument{%
\ifdefined\contentsname
  \renewcommand*\contentsname{Table of contents}
\else
  \newcommand\contentsname{Table of contents}
\fi
\ifdefined\listfigurename
  \renewcommand*\listfigurename{List of Figures}
\else
  \newcommand\listfigurename{List of Figures}
\fi
\ifdefined\listtablename
  \renewcommand*\listtablename{List of Tables}
\else
  \newcommand\listtablename{List of Tables}
\fi
\ifdefined\figurename
  \renewcommand*\figurename{Figure}
\else
  \newcommand\figurename{Figure}
\fi
\ifdefined\tablename
  \renewcommand*\tablename{Table}
\else
  \newcommand\tablename{Table}
\fi
}
\@ifpackageloaded{float}{}{\usepackage{float}}
\floatstyle{ruled}
\@ifundefined{c@chapter}{\newfloat{codelisting}{h}{lop}}{\newfloat{codelisting}{h}{lop}[chapter]}
\floatname{codelisting}{Listing}

\makeatother
\makeatletter
\@ifpackageloaded{caption}{}{\usepackage{caption}}
\@ifpackageloaded{subcaption}{}{\usepackage{subcaption}}
\makeatother

\ifLuaTeX
  \usepackage{selnolig}  % disable illegal ligatures
\fi
\usepackage[]{natbib}
\usepackage{bookmark}

\IfFileExists{xurl.sty}{\usepackage{xurl}}{} % add URL line breaks if available
\hypersetup{
  pdftitle={Features Extraction Across Multi-Source High-Dimensional Heterogeneous Data},
  pdfauthor={Wenbo Li; Xianli Zeng; Weiming Li; Zeng Li},
  pdfkeywords={3 to 6 keywords, that do not appear in the title},
  colorlinks=true,
  linkcolor={blue},
  filecolor={Maroon},
  citecolor={Blue},
  urlcolor={Blue},
  pdfcreator={LaTeX via pandoc}}

\newcommand{\anon}{1}

\newcommand{\bX}{\mathbf{X}}

\newcommand{\bu}{\mathbf{u}}
\newcommand{\bP}{\mathbf{P}}

\newcommand{\bSigma}{\boldsymbol{\Sigma}}
\newcommand{\sbU}{\widehat{\mathbf{U}}}
\newcommand{\sbu}{\hat{\mathbf{u}}}

\newcommand{\sbP}{\widehat{\mathbf{P}}}

\newcommand{\sbSigma}{\widehat{\boldsymbol{\Sigma}}}
\newcommand{\slambda}{\hat{\lambda}}

\newcommand{\scfi}{\hat{\mathcal{A}}_k^{\mathrm{s}}}

\newtheorem{assumption}{Assumption}

\usepackage{algorithm}
\usepackage{algorithmic}

\ShortHeadings{Features Extraction Across Multi-Source Data}{Li et al.}
\firstpageno{1}

\begin{document}

\def\spacingset#1{\renewcommand{\baselinestretch}%
{#1}\small\normalsize} \spacingset{1}

%%%%%%%%%%%%%%%%%%%%%%%%%%%%%%%%%%%%%%%%%%%%%%%%%%%%%%%%%%%%%%%%%%%%%%%%%%%%%%

\if1\anon
{
  \title{Features Extraction Across Multi-Source Heterogeneous Data}
  \author{\name Wenbo Li \email 12431107@mail.sustech.edu.cn \\
    \addr Department of Statistics and Data Science\\
    Southern University of Science and Technology
    \AND
    \name Xianli Zeng \email xlzeng@xmu.edu.cn \\
    \addr Department of Statistics and Data Science\\
    Xiamen University
    \AND
    \name Weiming Li \email li.weiming@sufe.edu.cn \\
    \addr School of Statistics and Management\\
    Shanghai University of Finance and Economics
    \AND
    \name Zeng Li (corresponding author) \email liz9@sustech.edu.cn \\
    \addr Department of Statistics and Data Science\\
    Southern University of Science and Technology}
  \editor{}
  \makeatletter
  \renewcommand{\@starteditor}{}
  \renewcommand{\@endeditor}{}
  \makeatother
  \maketitle
} \fi

\if0\anon
{
  \bigskip
  \bigskip
  \bigskip
  \begin{center}
    {\LARGE\bf Title}
\end{center}
  \medskip
} \fi

\bigskip
\begin{abstract}%
{\color{black}High-dimensional data from multiple sources often combine shared structure with source-specific variation. We study heterogeneous principal component analysis, which seeks to recover common and private principal subspaces when their dimensions are unknown. {We propose \emph{Feature Extraction (FE)}, a novel data-driven framework with two steps. First, the Eigenvector-based Spike-Count (ESC) estimator determines the number of principal components (PC) in each source using only sample eigenvector information, rather than eigenvalue-based criteria. Second, FE identifies the locations of common features by comparing bias-corrected inner products between sample eigenvectors across sources. These matches yield estimates of the common and source-specific subspaces without iterative optimization. We establish the consistency of our estimators under a spiked covariance model in a high-dimensional regime where the dimension and sample size grow proportionally. Simulations and applications to federated image data and U.S. housing prices demonstrate the performance of FE in rank recovery, subspace estimation, and data reconstruction.} Code is available at \url{https://github.com/liwenbo-research/HeterogeneousPCA}.}
\end{abstract}


% \usepackage{algorithm}      % 提供 algorithm 环境
% \usepackage{algpseudocode}  % 提供伪代码命令，如 \State, \For 等
%%%%% Preprint Options %%%%%
%% The track changes option allows you to mark changes
%% and will produce a list of changes, their line number
%% and page number at the end of the article.
 %\documentclass[preprint,trackchanges]{JASA}

%% NumberedRefs is used for numbered bibliography and citations.
%% Default is Author-Year style.
%\documentclass[preprint,NumberedRefs]{JASA}

%%%%%%%%%%%%%%%%%%%%%%%%%%%%%%%%%%%%%%%%%%%%%%%%%%%%%%%%%%%%%%%%%
%%%%%%%%%%%%%%%%%%%%%%%%%%%%%%%%%%%%%%%%%%%%%%%%%%%%%%%%%%%%%%%%%
%%%%%%% For Reprint
%% For appearance of finished article; 2 columns, 10 pt fonts
%% Also used for estimating print page length.

% \documentclass[reprint]{JASA}

%%%%% Reprint Options %%%%%

%% NumberedRefs is used for numbered bibliography and citations.
%% Default is Author-Year style.
% \documentclass[reprint,NumberedRefs]{JASA}

%% TurnOnLineNumbers
%% Make lines be numbered in reprint style:
% \documentclass[reprint,TurnOnLineNumbers]{JASA}

\section{Introduction}\label{sec:introduction}

Principal component analysis (PCA) is a fundamental tool for recovering low-dimensional structure from high-dimensional observations. Its leading eigenspaces provide interpretable representations for dimension reduction, denoising, visualization, and factor extraction. Modern data are frequently distributed across machines, sensors, institutions, or clients, motivating distributed PCA: local data holders communicate low-dimensional summaries to recover a population-level principal subspace without centralizing raw observations. Early work developed collective PCA for distributed clustering \citep{kargupta2001distributed}; subsequent methods addressed adaptive covariance-eigenvector estimation in sensor networks \citep{BERTRAND2014120}, collaborative dimension reduction over networked agents \citep{7180399}, statistically efficient distributed estimation of principal eigenspaces \citep{fan2019distributed}, and fast distributed PCA for federated-scale data \citep{shen2023fadifastdistributedprincipal}.

However, existing distributed PCA methods typically require samples across clients to come from the same population, or at least from populations with the same leading eigenspace. This assumption can be overly restrictive for modern heterogeneous data. A more general setting allows data from different clients to arise from distinct populations whose principal subspaces share common features while also containing private features specific to each client. The central task is then to use information from all clients to estimate the common subspace and recover the private subspace of each client. This common-and-private formulation has appeared in Personalized PCA (PerPCA) and Knowledge Transfer (KT) across multiple PCA studies \citep{shi2024personalized,li2024knowledgetransfermultipleprincipal}. We distill this formulation into a unified PCA framework, which we call heterogeneous PCA. 

% By distinguishing common structure from client-specific private variation, this framework enables information sharing across heterogeneous populations while preserving the distinctive features of each, providing a foundation for transferable representations, personalized learning, and scientific discovery across related studies.

The first major application direction of heterogeneous PCA combines federated learning with knowledge transfer across related studies. In personalized or federated learning, a representation must transfer across clients while retaining the client-specific directions needed to model heterogeneous local data. PerPCA \citep{shi2024personalized} addresses this goal by jointly estimating a shared basis and source-specific bases through a reconstruction-loss minimization with orthogonality constraints. In transfer learning, KT \citep{li2024knowledgetransfermultipleprincipal} pools information from the leading principal subspaces of related datasets to improve estimation of a target shared projection space. Thus, heterogeneous PCA supports federated learning by sharing common representations while retaining client-specific variation for personalization. For transfer learning, the common subspace enables information pooling across related studies to improve target estimation, while the private subspaces accommodate differences between source and target populations.

The second major direction is the group factor model, in which multiple groups are driven jointly by global factors and separately by group-specific local factors. PCA-based panel-data methods provide a foundation for estimating latent common structure \citep{bai2009panel}, while grouped panel and factor models accommodate structured heterogeneity \citep{bonhomme2015grouped,ando2016panel}. Recent methods estimate the global factor space through complementary spectral constructions: Circularly Projected common factors (CP) uses a circular product of group-specific projection matrices \citep{chen2022circularly}; Generalized Canonical Correlation (GCC) exploits cross-group canonical correlation \citep{lin2023generalised}; and the Aggregated Projection Method (APM) uses the leading eigenspace of an aggregated projection matrix \citep{hu2025aggregated}. These methods, which we compare with below, highlight that global and local factor estimation is intrinsically a common-private subspace problem. From an eigenspace perspective, global factors are common principal components and local factors are private principal components. 

{\color{black}Despite these developments, the literature on heterogeneous PCA is relatively sparse. Representative approaches include PerPCA \citep{shi2024personalized} and KT \citep{li2024knowledgetransfermultipleprincipal}, which formulate shared-subspace estimation as constrained optimization problems using the original data or estimated principal subspaces. These approaches require the relevant ranks, including the dimension of the shared subspace and the dimensions governing the source-specific structure, to be specified in advance. In real applications, these quantities are generally unknown, and their misspecification can substantially affect subspace estimation. Moreover, iterative optimization can incur substantial computational time and memory costs when applied to high-dimensional matrices.}

{To address these limitations, we propose \emph{FE}, a data-driven method for multi-source heterogeneous PCA. The key idea of FE is to exploit the information about covariance spikes encoded in sample eigenvectors. For single-source PCA, we propose the ESC estimator, whose core is to distinguish spike eigenvectors from bulk eigenvectors through the persistence of their directions across independent sample splits. For multi-source heterogeneous PCA, we identify the locations of common features within the ordered sample eigenvectors associated with each population by comparing bias-corrected eigenvector similarities across sources. This matching step determines both which directions are common across the populations and how many common features there are. We then use the matched common features to estimate the shared subspace and the remaining source-specific features to estimate each private subspace, without requiring the relevant ranks to be specified in advance.}

{
\textbf{Contributions.} The contributions of this paper are threefold:
\begin{enumerate}
  \item For single-source data, we develop a novel ESC estimator that estimates the number of PCs using only sample-eigenvector information.
  \item We provide a data-driven estimator of the subspace consisting of common features. The number of common features need not be known in advance; it is estimated directly from the data by exploiting information encoded in the sample eigenvectors. We avoid solving a complex iterative optimization problem by directly averaging the sample subspace of common feature from different data source. 
  \item We identify the positions of common features within the ordered PCs of each source. The procedure therefore reveals which PC in each source corresponds to the shared structure, making the extracted common information more interpretable in real-data applications.
\end{enumerate}}

The rest of this paper is organized as follows. Section~\ref{sec:setup} introduces the heterogeneous PCA model and the common-private principal-component structure. Section~\ref{sec:method} develops FE, including the {ESC} spike-number estimator and the common-feature identification procedure. Section~\ref{sec:asymptotic} establishes the asymptotic properties of the proposed estimators. Section~\ref{sec:simulation} evaluates their finite-sample performance. Section~\ref{sec:real-data} presents applications to group-factor and federated image data.

\section{\label{sec:setup} Model Setup for Heterogeneous Data}

We consider a collection of heterogeneous data matrices 
$\{\mathbf{X}_k, 1\leq k\leq K\}$, where each $\mathbf{X}_k \in \mathbb{R}^{p \times n_k}$ 
is associated with a $p\times p$ population covariance matrix $\boldsymbol{\Sigma}_k$. Throughout this work, we assume that each 
$\boldsymbol{\Sigma}_k$ follows the classical \emph{spiked covariance model} (\cite{Johnstone2001Spikedmodel}), where 
\[
\boldsymbol{\Sigma}_k 
= \sum_{i=1}^p \lambda_{ki}\, \mathbf{u}_{ki}\mathbf{u}_{ki}^\top.
\]

% % Under this model, each covariance matrix admits the spectral decomposition
% \[
% \boldsymbol{\Sigma}_k 
% = \mathbf{U}_k\, \operatorname{diag}(\lambda_{k1},\dots,\lambda_{kr_k},\dots,\lambda_{kp})\, 
% \mathbf{U}_k^\top,
% \] 
{The leading $r_k$ eigenvalues, $\lambda_{k1}\ge\cdots\ge\lambda_{kr_k}$, are relatively large and are referred to as the \emph{spike eigenvalues}. Their associated eigenvectors define the leading population principal component directions. The spikes are well separated from the remaining smaller eigenvalues, $\lambda_{k,r_k+1}\ge\cdots\ge\lambda_{kp}$, which form the \emph{bulk eigenvalues}.}
$\mathbf{U}_k \overset{\Delta}{=} [\mathbf{u}_{k1},\dots,\mathbf{u}_{kp}]_{p\times p}$ is the eigenvectors matrix of $\bSigma_k$.

While each data matrix $\bX_k$ follows different covariance structure $\bSigma_k$, we assume that 
$\{\boldsymbol{\Sigma}_k\}_{k=1}^K$ share a subset of eigenvectors in common, while the 
remaining eigenvectors are specific to each individual data source. 
To formalize this structure, we introduce the following definition of common features.

\begin{definition}[\textbf{Common Feature}]
The covariance matrices 
$\boldsymbol{\Sigma}_1, \dots, \boldsymbol{\Sigma}_K$ 
are said to share a \emph{common feature} if there exists a collection of indices 
$(l_1, \dots, l_K)$ such that the associated eigenvectors coincide \textcolor{black}{up to sign; after compatible sign choices}, i.e., 
$\mathbf{U}_{1}[l_1] = \cdots = \mathbf{U}_{K}[l_K]$, 
where $\mathbf{U}_{k}[i]$ denotes the $i$th column of $\mathbf{U}_{k}$.
\end{definition}

We use $\mathcal{A}_k$ to denote the index set corresponding to the spiked eigenvectors of the covariance matrix $\boldsymbol{\Sigma}_k$, that is, $\mathcal{A}_k = \{1, \dots, r_k\}$. Each set $\mathcal{A}_k$ is partitioned into two subsets: a subset of indices corresponding to the common features, denoted by $\mathcal{A}_k^{\mathrm{s}}$ with cardinality $|\mathcal{A}_k^{\mathrm{s}}| = r_{\mathrm{s}}$, and the remaining subset, denoted by $\mathcal{A}_k^{\mathrm{p}} \overset{\Delta}{=} \mathcal{A}_k \backslash \mathcal{A}_k^{\mathrm{s}}$, which corresponds to the eigenvectors unique to $\boldsymbol{\Sigma}_k$ and is referred to as the set of private features.

Then the common features of the covariance matrices $\{\bSigma_k, 1\leq k \leq K\}$ are given by the eigenvectors 
$\mathbf{U}_k[\mathcal{A}_k^{\mathrm{s}}]$, i.e.
\[
\mathbf{U}_1[\mathcal{A}_1^{\mathrm{s}}] 
= \dots 
= \mathbf{U}_K[\mathcal{A}_K^{\mathrm{s}}] 
\overset{\Delta}{=} \mathbf{U}^{\mathrm{s}}.
\]
The private features of $\boldsymbol{\Sigma}_k$ are denoted by 
$\mathbf{U}_k[\mathcal{A}_k^{\mathrm{p}}]$.
The projection matrix onto the principal subspace of $X_k$ is
\[
\mathbf{P}_k 
= \mathbf{U}^{\mathrm{s}} (\mathbf{U}^{\mathrm{s}})^{\top}
  + \mathbf{U}_k[\mathcal{A}_k^{\mathrm{p}}]\,
    \bigl(\mathbf{U}_k[\mathcal{A}_k^{\mathrm{p}}]\bigr)^{\top}
\;\overset{\Delta}{=}\;
\mathbf{P}^{\mathrm{s}} + \mathbf{P}_k^{\mathrm{p}}.
\]
Our target is to perform PCA for heterogeneous data collected from multiple sources. 
Formally, this amounts to estimating the shared component $\mathbf{P}^{\mathrm{s}}$ and the source-specific components $\mathbf{P}^{\mathrm{p}}_{k}$.

Throughout the paper, $r_k$ denotes the number of spiked principal components in source $k$, $r_{\mathrm{s}}$ denotes the number of common features, and $\mathcal{A}_k^{\mathrm{s}}$ records the locations of these common features within source $k$. The corresponding private index set is $\mathcal{A}_k^{\mathrm{p}}=\mathcal{A}_k\backslash \mathcal{A}_k^{\mathrm{s}}$. In our method, the procedure consists of three main steps:
\begin{enumerate}
  \item To estimate $r_k$, the number of spikes for covariance matrix $\Sigma_k$, denoted by $\hat{r}_k$.
  \item To estimate $\{\mathcal{A}_k^{\mathrm{s}}\}_{k=1}^K$, the locations of the common features within each source, denoted by $\{\hat{\mathcal{A}}_k^{\mathrm{s}}\}_{k=1}^K$.
  \item To estimate $\mathbf{P}^{\mathrm{s}}$ and $\mathbf{P}^{\mathrm{p}}_k$, the projection matrix of the common feature and private feature, denoted by $\sbP^{\mathrm{s}}$ and $\sbP^{\mathrm{p}}_k$.
\end{enumerate}

\section{Methodology}\label{sec:method}

% Importantly, this algorithm requires minimal prior knowledge. In the first step, we estimate the number of principal components for each individual source, whereas in existing related work \citep{li2024knowledgetransfermultipleprincipal,shi2024personalized}, this information is typically assumed to be known. In the second step, we identify the locations of the common features within each data source, a fine-grained procedure for discovering shared principal components that has not been proposed before. In the third step, we integrate the common-feature information across all sources to estimate the shared principal components, and use source-specific information to estimate the private components, thereby completing a comprehensive PCA framework for multi-source heterogeneous data.

\subsection{Estimation of the Number of Spikes for Each Data Source}
\label{subsec:spike}
In the existing literature on multi-source PCA \citep{li2024knowledgetransfermultipleprincipal,shi2024personalized}, it is commonly assumed that the ranks of 
$\mathbf{P}_k$, i.e., the number of spikes, are known in advance. 
However, such information is rarely available in practice, making this assumption unrealistic. 

To address this limitation, our method begins by estimating the ranks of 
$\mathbf{P}_k$, i.e., the number of spikes $\{r_k\}_{k=1}^K$.
Most existing methods for estimating the number of principal components rely on eigenvalue information from the sample covariance or correlation matrix \citep{Yao2012,Yao2014,Ke2023,liu2023asymptotic,Fan03042022}. For example, these methods use consecutive eigenvalue gaps, estimates of the bulk spectral distribution, or thresholds derived from spectral asymptotics to separate spike eigenvalues from the bulk. Their performance can depend on the eigenvalue configuration and on how the bulk spectrum is modeled.

A different line of work uses resampling to assess the stability of estimated principal subspaces. In particular, \citet{songshin2018stability} repeatedly subsample the observations, recompute PCA, and select the number of principal components according to the stability of the resulting eigenspaces. Although effective, this bootstrap-type strategy can be time-consuming because it requires repeated subsampling and eigendecomposition. Thus, existing approaches typically either rely on eigenvalue information or incur the computational cost of a resampling-based stability analysis.

 \begin{wrapfigure}[22]{r}{0.5\textwidth}
    \centering
    \includegraphics[width=\linewidth]{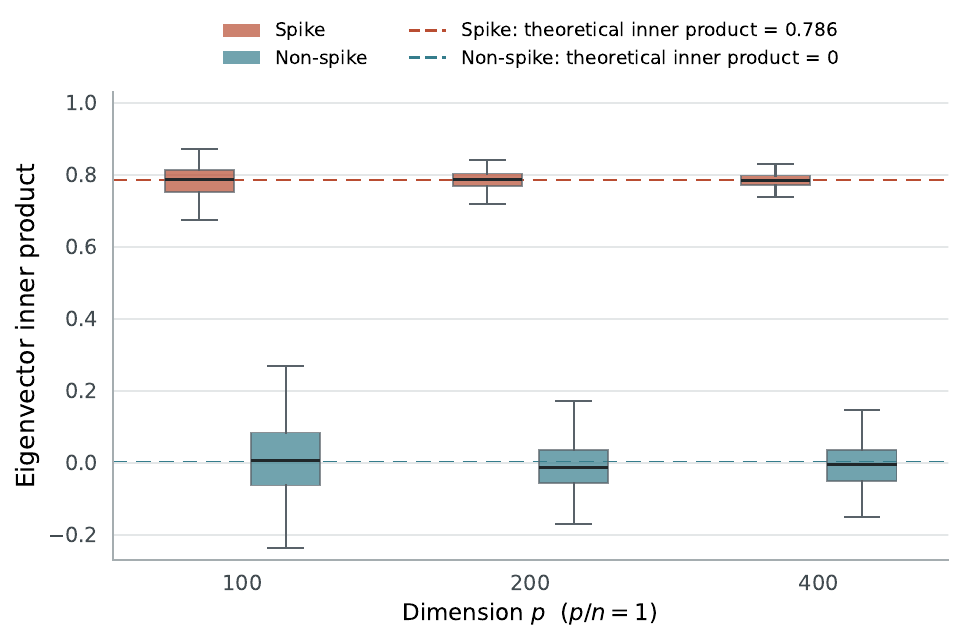}
    \caption{
    Absolute inner products for leading eigenvectors corresponding to the spike
    and signed inner products for second (bulk) eigenvectors from two independent
    Gaussian samples with spectrum
    \(\operatorname{spec}(\boldsymbol{\Sigma})
    =(5,1,\ldots,1,0.75,\ldots,0.75)\), \(p/n=1\), and
    \(p\in\{100,200,400\}\). Boxplots summarize 1000 replications; dashed lines
    mark the theoretical limits \(0.786\) (spike) and \(0\) (bulk).
    }
    \label{fig:innerproduct}
\end{wrapfigure}

Motivated by the spike information encoded in sample eigenvectors, we propose the ESC, which estimates the number of spikes by thresholding the averaged absolute inner products between corresponding sample eigenvectors across independent data splits. The following example illustrates how sample eigenvectors encode information about population spikes.

\textbf{Example.}  To illustrate how inner products between sample eigenvectors can be used to
  identify covariance spikes, we generate two independent data matrices
$  \mathbf{X}_1,\mathbf{X}_2\in\mathbb{R}^{p\times n}, \mathbf{X}_1,\mathbf{X}_2\sim\mathcal{N}(\mathbf{0},\boldsymbol{\Sigma})$,
with \(p/n=1\). The population covariance matrix has one spike and two bulk levels:
$  \operatorname{spec}(\boldsymbol{\Sigma})
  =
  (5,\underbrace{1,\ldots,1}_{p/2-1},
  \underbrace{0.75,\ldots,0.75}_{p/2}).$
  For each realization, let \(\widehat{\mathbf{u}}_{1j}\) and
  \(\widehat{\mathbf{u}}_{2j}\) denote the \(j\)-th leading eigenvectors of the
  sample covariance matrices$  \frac{1}{n}\mathbf{X}_1\mathbf{X}_1^{T}
  \quad\text{and}\quad
  \frac{1}{n}\mathbf{X}_2\mathbf{X}_2^{T},$
  respectively. We repeat the experiment 200 times for each
  \(p=n\in\{100,200,400\}\).

  The boxplots in Figure~\ref{fig:innerproduct} show the inner products between
  the corresponding sample eigenvectors. Since the sign of an eigenvector is
  arbitrary, the spike comparison is displayed as
  \(\left|\langle\widehat{\mathbf{u}}_{11},
  \widehat{\mathbf{u}}_{21}\rangle\right|\). For the non-spiked eigenvectors,
  we display the signed inner product
  \(\langle\widehat{\mathbf{u}}_{12},
  \widehat{\mathbf{u}}_{22}\rangle\). The spike inner products concentrate
  around the nonzero theoretical limit \(0.786\), whereas the non-spike
  inner products fluctuate around zero and become increasingly concentrated as
  \(p\) grows. Thus, a persistent nonzero inner product between corresponding
  sample eigenvectors provides an empirical indication of a population spike,
  while an inner product near zero is characteristic of bulk eigenvectors.
Motivated by these observations, we propose to exploit the information carried by the eigenvectors to estimate the number of spikes. The detailed steps are as follows:

\textbf{Step 1}. For each dataset $\mathbf{X}_k$, randomly split $\mathbf{X}_k$ into two equal-sized subsets of size $\textcolor{black}{h_k=\lfloor n_k/2\rfloor}$, denoted by $\mathbf{X}_k^{(1)}$ and $\mathbf{X}_k^{(2)}$. \textcolor{black}{For odd $n_k$, discard one observation; normalize both subset covariances by $h_k$.} Compute the corresponding sample covariance matrices $\hat{\boldsymbol{\Sigma}}_k^{(1)}$ and $\hat{\boldsymbol{\Sigma}}_k^{(2)}$, eigenvectors $\{\hat{\mathbf{u}}_{kj}^{(1)}, 1\leq j\leq p\}$ and $\{\hat{\mathbf{u}}_{kj}^{(2)}, 1\leq j\leq p\}$, and absolute inner products
  $\{|\langle \hat{\mathbf{u}}_{kj}^{(1)}, \hat{\mathbf{u}}_{kj}^{(2)} \rangle|, 1\leq j\leq p\}$.
        
\textbf{Step 2}. Repeat \textbf{Step}~1 $m$ times, \textcolor{black}{where $m$ is a fixed positive integer.} For each $j$, average the absolute inner products across the $m$ splits:
\[
\bar I_{k,j}\triangleq \frac{1}{m}\sum_{t=1}^m
\left|\left\langle \hat{\mathbf{u}}_{kj}^{(1,t)},\hat{\mathbf{u}}_{kj}^{(2,t)}\right\rangle\right|,
\qquad 1\leq j\leq p.
\]

\textbf{Step 3}. The number of spikes $r_k$ can then be estimated using only the eigenvector information. 
        Specifically, with a threshold $\varepsilon(n) = n^{\varepsilon_0}(n^{-1/2}+\phi^{3/2}(n))$, where $\phi(n)=n^{2/a-1/2}$ \textcolor{black}{and $0<\varepsilon_0<1/2-2/a$.} Our estimator is
        \[
        \hat{r}_k^{\mathrm{{ESC}}} = \min \Bigl\{ l : \bar I_{k,l+1} \bigvee \bar I_{k,l+2} < \varepsilon(n_k)\Bigl\}.
        \]

\begin{algorithm}[ht]
\baselineskip=14pt
\caption{Estimation of the Number of Spikes $\{r_k\}_{k=1}^K$}
\label{alg:estimate}
\begin{algorithmic}[1]
\REQUIRE Data matrices $\{\mathbf{X}_k \in \mathbb{R}^{p \times n_k}\}_{k=1}^K$; threshold $\varepsilon(n_k)$.
\ENSURE $\{\hat{r}_k\}_{k=1}^K$, the estimates of $\{r_k\}_{k=1}^K$

\STATE For each $\mathbf{X}_k$, randomly split it into two equal-sized subsets $\mathbf{X}_k^{(1)}$ and $\mathbf{X}_k^{(2)}$. 
  Compute the corresponding sample covariance matrices $\hat{\boldsymbol{\Sigma}}_k^{(1)}$ and $\hat{\boldsymbol{\Sigma}}_k^{(2)}$, 
       and obtain their eigenvector matrices $\{\hat{\mathbf{u}}_{kj}^{(1)}, j \in [p]\}$ and $\{\hat{\mathbf{u}}_{kj}^{(2)}, j \in [p]\}$. 
       Calculate the inner products 
       \[
       \left\{ \left|\langle \hat{\mathbf{u}}_{kj}^{(1)}, \hat{\mathbf{u}}_{kj}^{(2)} \rangle \right|, j \in [p] \right\}.
       \]

\STATE Repeat Step 1 a total of $m$ times and average the absolute inner products:
       \[
       \bar I_{k,j}=\frac{1}{m}\sum_{t=1}^m
       \left|\left\langle \hat{\mathbf{u}}_{kj}^{(1,t)},\hat{\mathbf{u}}_{kj}^{(2,t)}\right\rangle\right|,
       \qquad j\in[p].
       \]

\STATE Estimate the number of spikes from the averaged eigenvector inner products:
       \[
       \hat{r}_k^{\mathrm{{ESC}}} = \min \Bigl\{ l :
       \bar I_{k,l+1} \bigvee \bar I_{k,l+2} < \varepsilon(n_k) \Bigr\}.
       \]
\end{algorithmic}
\end{algorithm}

\subsection{Identification of Common Features}\label{subsec:common feature}
In Section~\ref{subsec:spike}, we obtained the estimates of $\mathcal{A}_k$ and $r_k$. The goal of this section is to determine $\mathcal{A}_k^{\mathrm{s}}$ and $r_{\mathrm{s}}$, that is, the locations and the number of the common features, with $\mathcal{A}_k$ and $r_k$ are known. To address this problem, we first design an algorithm to identify the locations of common feature for a pair of data source.

\subsubsection{\textcolor{black}{Common-Feature Identification for Two Sources}}\label{subsec:two-source-identification}

As for data pair $(\mathbf{X}_1:p\times n_1, \mathbf{X}_2: p\times n_2)$, we first calculate the eigenvalues $\{\slambda_{kl}; k=1,2,1 ; \leq l\leq p\}$ and eigenvectors $\{\sbu_{kl}; k=1,2; 1\leq l\leq p\}$ of sample covariance matrix $\frac{1}{n_k}\bX_k\bX_k^T$ for each data source $\bX_k$. Our method relies on the following key result (\cite{couillet2022random,bloemendal2016principal}):

% \begin{align}
% \theta_{1l}^{-2} \, \mathbf{e}_a^\top \hat{\mathbf{u}}_{1l} \hat{\mathbf{u}}_{1l}^\top \mathbf{e}_b
% &= \mathbf{e}_a^\top \mathbf{u}_{1l} \mathbf{u}_{1l}^\top \mathbf{e}_b + o_p(1), \label{eq:existingresults-1}\\
% \theta_{2m}^{-2} \, \mathbf{e}_a^\top \hat{\mathbf{u}}_{2m} \hat{\mathbf{u}}_{2m}^\top \mathbf{e}_b
% &= \mathbf{e}_a^\top \mathbf{u}_{2m} \mathbf{u}_{2m}^\top \mathbf{e}_b + o_p(1), \label{eq:existingresults-2}
% \end{align}

% \textcolor{black}{we have
% \begin{align}\label{eq:existingresults-3}
%   \hat{\mathbf{u}}_{1\ell} \hat{\mathbf{u}}_{1\ell}^\top 
% = (\mathbf{u}_{1l}^\top \hat{\mathbf{u}}_{1\ell})^2 \, \mathbf{u}_{1l} \mathbf{u}_{1l}^\top + o_P(1).
% \end{align}}

% Multiplying both sides of \eqref{eq:existingresults-3} by a corresponding eigenvector from the second sample, $\hat{\mathbf{u}}_{2m}$ and combining \eqref{eq:existingresults-1} and \eqref{eq:existingresults-2}, yields
\begin{align}\label{eq:main}
  \vert\hat{\mathbf{u}}_{1\ell}^\top \hat{\mathbf{u}}_{2m}\vert
=\vert\theta_{1l}\theta_{2m}\bu_{1l}^\top \bu_{2m}\vert + o_P(1).
\end{align}
where 
\begin{equation}\label{eq:zeta-estimator}
\theta_{kl} = \sqrt{-\frac{\varphi_{kl}}{\hat{\lambda}_{kl} \phi_{kl}}}, \quad k=1,2,
\end{equation}
\[
\begin{aligned}
\varphi_{kl} &= -\frac{1-(p-\textcolor{black}{\hat r_k})/n_k}{\hat{\lambda}_{kl}} + \frac{1}{n_k} \sum_{j = \textcolor{black}{\hat r_k} +1}^p \textcolor{black}{\frac{1}{\hat{\lambda}_{kj}-\hat{\lambda}_{kl}}},\\
\phi_{kl} &= \textcolor{black}{+}\frac{1-(p-\textcolor{black}{\hat r_k})/n_k}{\hat{\lambda}_{kl}^2} + \frac{1}{n_k} \sum_{j = \textcolor{black}{\hat r_k} +1}^p \frac{1}{(\hat{\lambda}_{kl}-\hat{\lambda}_{kj})^2}.
\end{aligned}
\]

% Finally, replacing $\mathbf{e}_a$ and $\mathbf{e}_b$ by $\mathbf{u}_{1l}$ in \eqref{eq:existingresults-1}
% and by $\mathbf{u}_{2m}$ in \eqref{eq:existingresults-2}, we obtain
% \[
% \frac{ \hat{\mathbf{u}}_{1l}^\top \hat{\mathbf{u}}_{2m} }{ \theta_{1l} \, \theta_{2m} } 
% = \mathbf{u}_{1l}^\top \mathbf{u}_{2m} + o_p(1).
% \]

We have proved that, if $|\bu_{1l}^\top \bu_{2m}| = 1$, 
\[
\left\vert \frac{ \hat{\mathbf{u}}_{1l}^\top \hat{\mathbf{u}}_{2m} }{ \theta_{1l} \, \theta_{2m} } \right\vert - 1 = \textcolor{black}{o_P(\textcolor{black}{\gamma_n})},
\]
\textcolor{black}{where $\gamma_n=n^{\varepsilon_0}\phi$ with $n$ as defined in Assumption~\ref{ass:2}, $\phi=n^{2/a-1/2}$, and $0<\varepsilon_0<1/2-2/a$.}
On the other hand, if $|\bu_{1l}^\top \bu_{2m}| \neq 1$, 
\[
\left\vert \frac{ \hat{\mathbf{u}}_{1l}^\top \hat{\mathbf{u}}_{2m} }{ \theta_{1l} \, \theta_{2m} } \right\vert - 1 = O_P(1).
\]

% the inner product between two sample eigenvectors, 
% after appropriate normalization by the correction factors, 
% converges to the inner product of their corresponding population eigenvectors. 
% Motivated by this observation, we define the following criterion: the sample eigenvectors 
% $\hat{\mathbf{u}}_{1l}$ and $\hat{\mathbf{u}}_{2m}$ are considered to correspond to the same common feature if
% \[
% \left\vert \frac{ \hat{\mathbf{u}}_{1l}^\top \hat{\mathbf{u}}_{2m} }{ \theta_{1l} \, \theta_{2m} } \right\vert > 1 - \epsilon.
% \]
% Therefore we consider design the following criterion $\bu_{1l} = \bu_{2m}$ if
% \begin{equation}\label{eq:cf-criterion}
%  \frac{\left|\langle\widehat{\mathbf{u}}_{1l}, \widehat{\mathbf{u}}_{2m}\rangle\right|}{\theta_{1l} \theta_{2m}}   > 1 - \gamma(\sqrt{\frac{2n_1n_2}{n_1+n_2}})
% \end{equation}
% where $\gamma(n)=n^{\varepsilon_0}n^{2/a-1/2}$.

% Thus, we Apply the above criterion \eqref{eq:cf-criterion}  to the dataset pairs 
% $(\mathbf{X}_1, \mathbf{X}_2)$, $(\mathbf{X}_2, \mathbf{X}_3$, $\dots$, $(\mathbf{X}_{K-1}, \mathbf{X}_K)$, 
Thus we sequentially obtain the corresponding sets of identified common features 
$\{\mathcal{S}_{12}, \mathcal{S}_{23}, \dots, \mathcal{S}_{K-1,K}\}$, where
\[
\mathcal{S}_{k,k+1} = \left\{(l,m) : l \in [\hat{r}_k],\ m \in [\hat{r}_{k+1}],\ \left|\langle\widehat{\mathbf{u}}_{kl}, \widehat{\mathbf{u}}_{k+1,m}\rangle\right| / \textcolor{black}{(\theta_{kl} \theta_{k+1,m})}  \textcolor{black}{>} 1 - \textcolor{black}{\gamma_n}\right\}.
\]

\subsubsection{{Common-Feature Identification for Multiple Sources}}\label{subsec:match-aggregation}
For $K$ data sources, we identify common features sequentially. We first match $\mathbf X_1$ and $\mathbf X_2$ using the two-source criterion, obtaining $\mathcal S_{12}$. Each pair $(j_1,j_2)\in\mathcal S_{12}$ identifies sample eigenvectors $\widehat{\mathbf u}_{1j_1}$ and $\widehat{\mathbf u}_{2j_2}$ corresponding to a candidate shared direction. We then compare $\widehat{\mathbf u}_{2j_2}$ with the spike eigenvectors of $\mathbf X_3$ and retain $(j_1,j_2,j_3)$ only when $(j_2,j_3)\in\mathcal S_{23}$. Thus, only features matched across all three sources remain. We repeat this step with each subsequent source until $\mathbf X_K$ is included, discarding any candidate that cannot be matched to the next source.

Equivalently, the retained index tuples satisfy
\[
(j_1,\ldots,j_K):\qquad
(j_k,j_{k+1})\in\mathcal S_{k,k+1},\quad k=1,\ldots,K-1.
\]

 % end revision marking
We summarize the entire procedure in the following algorithm.

\begin{algorithm}[ht]
\begingroup\color{black}
\baselineskip=14pt
\caption{Identifying common features across $K$ data sources.}
\label{alg:identification}
\begin{algorithmic}[1]
\REQUIRE Data matrices $\mathbf{X}_k \in \mathbb{R}^{p \times n_k}$, $1\leq k\leq K$, $K\geq2$; estimated spike counts $\{\hat r_k\}_{k=1}^K$; threshold $\gamma_n$.
\ENSURE Estimated common feature index sets $\{\scfi\}_{k=1}^K$ and common rank $\hat r_{\mathrm{s}}$.
\FOR{$k = 1$ \TO $K$}
  \STATE $\sbSigma_k = \frac{1}{n_k}\bX_k\bX_k^\top$.
  \STATE $\sbSigma_k = \sbU_k\operatorname{diag}\{\slambda_{k1},\dots,\slambda_{kp}\}\sbU_k^\top$, where $\sbU_k=[\sbu_{k1},\dots,\sbu_{kp}]$ and $\slambda_{k1}\geq\cdots\geq\slambda_{kp}$.
  \STATE Compute $\{\theta_{kl}\}_{l=1}^{\hat r_k}$ using \eqref{eq:zeta-estimator}, applied to source $k$.
\ENDFOR
\FOR{$k = 1$ \TO $K-1$}
  \STATE Calculate $\mathcal S_{k,k+1}=\left\{(l,m):l\in[\hat r_k],\ m\in[\hat r_{k+1}],\ \frac{|\langle\widehat{\mathbf u}_{kl},\widehat{\mathbf u}_{k+1,m}\rangle|}{\theta_{kl}\theta_{k+1,m}}>1-\gamma_n\right\}$.
\ENDFOR
\STATE Aggregate $\{\mathcal S_{k,k+1}\}_{k=1}^{K-1}$ sequentially and retain only index tuples matched across all $K$ sources.
\STATE Extract the common feature index sets $\{\scfi\}_{k=1}^K$ from the retained tuples and set $\hat r_{\mathrm{s}}=|\scfi|$ for any $k\in[K]$.
\end{algorithmic}
\endgroup
\end{algorithm}

\subsection{Estimation of the Target Projection Matrix}\label{subsec:projection matrices}
{\color{black}
After Sections~\ref{subsec:spike} and~\ref{subsec:common feature}, we have obtained the estimated common feature index sets $\scfi$. We use these sets to estimate both the projection matrix of common features, $\mathbf{P}^{\mathrm{s}}$,
and the source-specific (private) projection matrices, $\mathbf{P}_k^{\mathrm{p}}$.

} % end revision marking
The projection matrix of the common features is estimated as 
$\hat{\mathbf{P}}^{\mathrm{s}}$, obtained by taking the leading $\hat{r}_{\mathrm{s}}$ eigenvectors of
\[
\frac{1}{K} \sum_{k=1}^{K} \textcolor{black}{\widehat{\mathbf{U}}_k[\widehat{\mathcal{A}}_k^{\mathrm{s}}]} \textcolor{black}{\widehat{\mathbf{U}}_k[\widehat{\mathcal{A}}_k^{\mathrm{s}}]}^T.
\]

{\color{black}
For the private features, let $\hat r_k^{\mathrm p}=\hat r_k-\hat r_{\mathrm s}$ and define the residualized covariance
\begin{equation}\label{eq:private-residual-covariance}
\widehat{\boldsymbol\Sigma}_k^{\perp}
=(\mathbf I_p-\widehat{\mathbf P}^{\mathrm s})\widehat{\boldsymbol\Sigma}_k
(\mathbf I_p-\widehat{\mathbf P}^{\mathrm s}).
\end{equation}
Let $\widehat{\mathbf V}_k^{\mathrm p}$ contain the leading $\hat r_k^{\mathrm p}$ orthonormal eigenvectors of this matrix restricted to $\operatorname{range}(\mathbf I_p-\widehat{\mathbf P}^{\mathrm s})$. The private projector is
\begin{equation}\label{eq:private-residual-projector}
\widehat{\mathbf P}_k^{\mathrm p}
=\widehat{\mathbf V}_k^{\mathrm p}(\widehat{\mathbf V}_k^{\mathrm p})^\top.
\end{equation}
For $\hat r_k^{\mathrm p}=0$, take an empty private basis and set $\widehat{\mathbf P}_k^{\mathrm p}=\mathbf0$; if $\hat r_{\mathrm s}=0$, set $\widehat{\mathbf P}^{\mathrm s}=\mathbf0$. The notation $\widehat{\mathbf V}_k^{\mathrm p}$ distinguishes the final private basis from columns of the original sample eigenvector matrix $\widehat{\mathbf U}_k$: the private directions are re-estimated after removing the aggregated common space. In particular,
\[
\widehat{\mathbf P}^{\mathrm s}\widehat{\mathbf P}_k^{\mathrm p}=\mathbf0,
\]
so the common and private reconstructions are orthogonal.

} % end revision marking
The overall projection matrix for the $k$-th source is computed as
\[
\hat{\mathbf{P}}_k = \hat{\mathbf{P}}^{\mathrm{s}} + \hat{\mathbf{P}}_k^{\mathrm{p}}.
\]
Finally, the complete \emph{FE} procedure for multi-source heterogeneous PCA is presented in Algorithm~\ref{alg:all}.

\begin{algorithm}[ht]
\baselineskip=14pt
\caption{\textcolor{black}{FE across $K$ data sources.}}
\label{alg:all}
\begin{algorithmic}[1]
\REQUIRE  Data matrices $\{\mathbf{X}_k \in \mathbb{R}^{p \times n_k};1\leq k\leq K\}$; threshold $\textcolor{black}{\gamma_n}$
\ENSURE $\{\sbP_k; 1\leq k\leq K\}$: the estimate of $\{\bP_k;1\leq k\leq K\}$.
\STATE Apply Algorithm \ref{alg:estimate} to obtain $\{\hat{r}_k;1\leq k\leq K\}$;
\STATE Apply Algorithm \ref{alg:identification} to obtain $\{\scfi;1\leq k\leq K\}$ and $\hat{r}_{\mathrm{s}}$;
\STATE Obtain $\sbP^{\mathrm{s}}$: the estimate of projection space of common feature $\bP^{\mathrm{s}}$ by taking the leading $\hat{r}_{\mathrm{s}}$ eigenvectors of
\[
\frac{1}{K} \sum_{k=1}^{K} \textcolor{black}{\widehat{\mathbf{U}}_k[\widehat{\mathcal{A}}_k^{\mathrm{s}}]} \textcolor{black}{\widehat{\mathbf{U}}_k[\widehat{\mathcal{A}}_k^{\mathrm{s}}]}^T.
\]
\STATE \textcolor{black}{For each $k$, form $\widehat{\boldsymbol\Sigma}_k^\perp$ as in \eqref{eq:private-residual-covariance}, extract its leading $\hat r_k-\hat r_{\mathrm s}$ orthonormal directions in the common-space complement as $\widehat{\mathbf V}_k^{\mathrm p}$, and set $\widehat{\mathbf P}_k^{\mathrm p}=\widehat{\mathbf V}_k^{\mathrm p}(\widehat{\mathbf V}_k^{\mathrm p})^\top$ (zero for an empty private basis).}
\STATE Obtain $\{\sbP_k\}_{k=1}^K$ by 
\[
\sbP_k = \sbP^{\mathrm{s}} + \sbP^{\mathrm{p}}_k\]
\end{algorithmic}
\end{algorithm}

\section{Asymptotic Properties}\label{sec:asymptotic}

\subsection{\textcolor{black}{Preliminaries}}\label{subsec:preliminary}
{
For a $p\times p$ real symmetric matrix $A_p$ with eigenvalues $\lambda_1(A_p),\ldots,\lambda_p(A_p)$, its empirical spectral distribution (ESD) is the probability measure
\[
F^{A_p}(x)=\frac{1}{p}\sum_{j=1}^{p}\mathbf{1}\{\lambda_j(A_p)\le x\},
\qquad x\in\mathbb R,
\]
where eigenvalues are counted with multiplicity.  A probability distribution $F$ is called the limiting spectral distribution (LSD) of $\{A_p\}$ if the ESD converges weakly to $F$, denoted by $F^{A_p}\Rightarrow F$, as $p\to\infty$.

For a probability distribution $F$ on $\mathbb R$, the Stieltjes transform is defined as
\[
m(z)=\int\frac{1}{x-z}\,dF(x),
\qquad z\in\mathbb C_+=\{z\in\mathbb C:\operatorname{Im}z>0\}.
\]
}

\subsection{Consistency of spike number estimation}

{We establish the consistency of the ESC spike-count estimator $\hat r_k^{\mathrm{ESC}}$ for each source $k=1,\ldots,K$.}

\begin{assumption}\label{ass:1}
The distributed data consist of $K$ sub-datasets
\[
\mathbf{X}_k = \boldsymbol{\Sigma}_k^{1/2} \mathbf{W}_k, \quad k = 1, \ldots, K,
\]
where $\mathbf{W}_k = (w_{kij}) \in \mathbb{R}^{p \times n_k}$, with $\sum_{k=1}^K n_k = N$, 
and the collection $\{w_{kij} : 1 \le k \le K,\, 1 \le i \le p,\, 1 \le j \le n_k\}$ consists of i.i.d.\ random variables satisfying
\[
\mathbb{E}(w_{kij}) = 0, \quad 
\mathbb{E}(w_{kij}^2) = 1, \quad 
\mathbb{E}\lvert w_{kij} \rvert^{a} < \infty,
\]
for some constant $a > 4$.

\end{assumption}

\begin{assumption}\label{ass:2}
The dimension $p$ and the sample sizes $\{n_1, \ldots, n_K\}$ tend to infinity such that
\[
n_k = n_k(p),\quad c_{n_k} \triangleq \frac{p}{n_k} \to c_k, 
\quad \text{and} \quad 
\max_{1 \le k \le K} c_{n_k} < c_0 < \infty,
\]
for some positive constants $\{c_k\}$ and $c_0$.
Let $n=\min_{1\le k\le K}n_k$.
\end{assumption}

\begin{assumption}\label{ass:3}
The $K$ covariance matrices $\boldsymbol{\Sigma}_k$'s are positive definite with bounded spectral norms. 
For each $\boldsymbol{\Sigma}_k$, its empirical spectral distribution $H_{p,k}$ converges weakly to a probability distribution $H_k$ with compact support $\Gamma_{H_k} \subset \mathbb{R}^+$. 
The $r_k$ largest eigenvalues of $\boldsymbol{\Sigma}_k$, denoted by \textcolor{black}{$\lambda_{k1},\ldots,\lambda_{kr_k}$, are spike eigenvalues}, with $r_k$ being a finite integer.
They satisfy
\begin{equation}\label{eq:ass3}
\min_{1 \le j \le r_k} \bigl( \lambda_{k j} - \lambda_{k, j+1} \bigr) > d_0,\quad
\psi_{2c_0, H_k}'(\lambda_{k r_k})
\triangleq
1 - 2c_0 \int \frac{t^2}{(\lambda_{k r_k} - t)^2} \, dH_k(t)
> d_0,
\end{equation}
for some constant $d_0 > 0$, uniformly in $k\in\{1,\ldots,K\}$. 
Moreover, the $(r_k+1)$th largest eigenvalue $\lambda_{k,r_k+1}$ belongs to the bulk in the sense that
\[
\operatorname{dist}\!\bigl(\lambda_{k,r_k+1}, \Gamma_{H_k}\bigr) = o(1), \qquad p \to \infty.
\]

\begingroup\color{black}\hypersetup{linkcolor=black,citecolor=black,urlcolor=black}
In addition, assume that, for some constant $\tau>0$ and all sufficiently large $p$, uniformly in $k$,
\begin{equation}
  1+\lambda_{k,r_k+1}\underline m_{0,k,n_k}(\lambda_{+,k,n_k})\ge\tau.
  \label{eq:dingassump}
\end{equation}

Here $\underline m_{0,k,n_k}$ is the companion Stieltjes transform associated with the population bulk spectral distribution $(p-r_k)^{-1}\sum_{i=r_k+1}^{p}\delta_{\lambda_{ki}}$ and aspect ratio $(p-r_k)/n_k$. It is the unique solution in $\mathbb C_+$ to the this equation
\[
\begin{aligned}
z=-\frac{1}{\underline m_{0,k,n_k}(z)}
+\frac{1}{n_k}\sum_{i=r_k+1}^{p}
\frac{\lambda_{ki}}{1+\lambda_{ki}\underline m_{0,k,n_k}(z)},
\qquad z\in\mathbb C_+.
\end{aligned}
\]
The quantity $\lambda_{+,k,n_k}$ is the right endpoint of the support of the probability distribution whose Stieltjes transform is $\underline m_{0,k,n_k}$. 
\endgroup
\end{assumption}

\begin{remark}
Assumptions~\ref{ass:1} and~\ref{ass:2} specify a high-dimensional framework for the observed data matrices $\mathbf{X}_k$'s, in which the population exhibits a linear dependence structure with finite $a$th moments. The dimension $p$ and the sample sizes grow proportionally to infinity, while the number of sources $K$ remains fixed.    
\end{remark}

\begin{remark}
Assumption~\ref{ass:3} characterizes a spiked population model \citep{Johnstone2001Spikedmodel} for the covariance matrix $\boldsymbol{\Sigma}_k$. 
The conditions in~\eqref{eq:ass3}, adopted from \citet{Bai2012Sample}, guarantee that the population covariance matrix $\boldsymbol{\Sigma}_k$, as well as its sample counterpart, admits exactly $r_k$ simple leading eigenvalues that are separated from the bulk spectrum.
These spike eigenvalues, together with their associated eigenvectors, form the basis for recovering the principal eigenspace of the population covariance matrices.
\textcolor{black}{Condition~\eqref{eq:dingassump}, adapted from the regularity assumptions of \citet{Ding2021}, imposes uniform regularity on the population eigenvalues of $\boldsymbol{\Sigma}_k$ near the bulk spectral edge. It is used in the proof of Lemma~\ref{lem:4}.}
\end{remark}

\begin{theorem}\label{th1}
\textcolor{black}{Let $m$ be a fixed positive integer.}
Let $\phi(n_k)=n_k^{2/a-1/2}$ and $\varepsilon(n_k) = n_k^{\varepsilon_0}(n_k^{-1/2}+\phi(n_k)^{3/2})$ for \textcolor{black}{$0<\varepsilon_0<1/2-2/a$}.
Under Assumptions~\ref{ass:1}--\ref{ass:3}, we have
\[
\mathbb{P}\!\left(\hat r_k^{\mathrm{{ESC}}}=r_k\right)\to 1
\quad k=1,\ldots,K.
\]
\end{theorem}

\begin{remark}
The consistency of $\hat{r}_k^{\mathrm{{ESC}}}$ is driven by the distinct split-sample behavior of the spike and bulk eigenvectors.
Under the spiked covariance model, the leading sample eigenvectors exhibit non-vanishing inner products across independent splits, whereas the bulk eigenvectors do not.
%; moreover, the inner products between spike and bulk eigenvectors vanish asymptotically.
Consequently, any threshold sequence $\varepsilon(n_k) \to 0$ that satisfies $\varepsilon(n_k) >n_k^{\varepsilon_0}(n_k^{-1/2}+\phi(n_k)^{3/2})$ produces a consistent separation between the spike and bulk regimes.
\end{remark}

\subsection{Consistent identification of common spike eigenvectors}

{We establish the consistency of the common-feature count estimator $\hat r_{\mathrm s}$ and the common-feature location estimators $\widehat{\mathcal A}_k^{\mathrm s}$ for $k=1,\ldots,K$.}

\begin{assumption}\label{ass:A4}
There exists $\eta \in (0,1)$ such that for any $k\neq \ell$, $j\in\mathcal S_k$, $m\in\mathcal S_\ell$, we have either $\mathbf u_{k,j} = \pm\mathbf u_{\ell,m}$ or $|\langle \mathbf u_{k,j},\mathbf u_{\ell,m}\rangle| \le 1-\eta$.
\end{assumption}

% \begin{theorem}\label{th2}
% \textcolor{black}{Let $n=\min_k n_k$, $\phi=n^{2/a-1/2}$, and $\textcolor{black}{\gamma}=n^{\varepsilon_0}\phi$} for \textcolor{black}{$0<\varepsilon_0<1/2-2/a$}.
% Suppose that Assumptions~\ref{ass:1}-~\ref{ass:A4} hold.
% Then, as $p\to\infty$ and $\min_{1\le k\le K}n_k\to\infty$,
% \[
% \mathbb P\!\left(\widehat{\mathcal E}_{k}={\mathcal E}_{k}^{\star}\ \text{for all }k=1,\ldots,K-1\right)\to 1.
% \]
% \end{theorem}

\begingroup

\begin{theorem}[Consistency of the common-feature rank and locations]\label{th3}
Let $\phi=n^{2/a-1/2}$ and $\gamma_n=n^{\varepsilon_0}\phi$ for $0<\varepsilon_0<1/2-2/a$.
Under Assumptions~\ref{ass:1}--\ref{ass:A4}, we have
\[
\mathbb P\!\left(\hat r_{\mathrm s}=r_{\mathrm s}\right)\to 1.
\]
Moreover,
\[
\mathbb P\!\left(\widehat{\mathcal A}_k^{\mathrm s}=\mathcal A_k^{\mathrm s}
\text{ for all }k=1,\ldots,K\right)\to 1.
\]
\end{theorem}
\endgroup

\section{Simulation Studies}\label{sec:simulation}

In this section, we conduct extensive simulation studies to evaluate the effectiveness of our proposed method.

\subsection{Spike-Number Estimation for Single-Source Data}

We generate samples according to
\[
\mathbf{X} = \boldsymbol{\Sigma}^{1/2}\mathbf{Z},
\]
where $\mathbf{Z}=[\mathbf{z}_{1},\dots,\mathbf{z}_{n}]_{p\times n}$ contains $n$ i.i.d. samples. We consider two distributions for the entries $\mathbf{z}_{i}=(z_{ij})_{j=1}^p$: $z_{ij}\sim N(0,1)$ and $z_{ij}\sim \mathrm{exp}(1)-1$, $1\leq j\leq p$. Since this subsection focuses on single-source data, we suppress the source subscript.

For the population covariance matrix $\bSigma$, we consider the following structure:
\[
\bSigma = \mathrm{diag}\Bigg(
    \frac{\lambda_{1} - 1/4}{8}\mathbf{1}_{8}\mathbf{1}_{8}^\top + \frac{1}{4}\mathbf{I}_{8}, \ 
    \frac{\lambda_2 - 1/4}{8}\mathbf{1}_{8}\mathbf{1}_{8}^\top + \frac{1}{4}\mathbf{I}_{8}, \ 
    \boldsymbol{\Sigma}_{\mathrm{bulk}}
\Bigg),
\]
which contains two spikes. The spiked eigenvalues and corresponding eigenvectors are
\[
\lambda_1(\bSigma) = \lambda_{1}, \quad
\lambda_2(\bSigma) = \lambda_{2}, \quad
\mathbf{u}_{1}^\top = \Big(\frac{1}{2\sqrt{2}}\mathbf{1}_{8}^\top, \mathbf{0}_{p-8}^\top\Big), \quad
\mathbf{u}_{2}^\top = \Big(\mathbf{0}_{8}^\top, \frac{1}{2\sqrt{2}}\mathbf{1}_{8}^\top, \mathbf{0}_{p-16}^\top\Big),
\]

% Highlight the revised scenario definitions in Section 5.1.
\begingroup\color{black}
\textbf{Scenario 1.  Diagonal bulk.}
\[
(\lambda_1,\lambda_2)=(8,6),
\qquad
\boldsymbol{\Sigma}_{\mathrm{bulk}}
=
\operatorname{diag}\!\left(
1.5\,\mathbf I_{p/2-8},
1.2\,\mathbf I_{p/2-8}
\right).
\]

\textbf{Scenario 2. AR(1) bulk.}
We set $(\lambda_1,\lambda_2)=(8,6)$ and let the residual bulk be an AR(1)
covariance matrix with unit marginal variance and correlation $\rho=0.3$:
\[
\boldsymbol{\Sigma}_{\mathrm{bulk}}
=\left(0.3^{|i-j|}\right)_{i,j=1}^{p-16}.
\]

\textbf{Scenario 3. Gamma bulk.}
We set $(\lambda_1,\lambda_2)=(8,6)$ and use independent Gamma-distributed
residual variances:
\[
\boldsymbol{\Sigma}_{\mathrm{bulk}}=\boldsymbol{\Lambda}_{\Gamma},\qquad
\boldsymbol{\Lambda}_{\Gamma}
=\operatorname{diag}(\Lambda_{\Gamma,1},\ldots,\Lambda_{\Gamma,p-16}),
\quad \Lambda_{\Gamma,j}\stackrel{\mathrm{i.i.d.}}{\sim}
\operatorname{Gamma}(10,0.1),
\]
where the second parameter is the scale. 

\textbf{Scenario 4. Gamma bulk with an AR(1) residual block.}
We set $(\lambda_1,\lambda_2)=(6,4)$ and use a Gamma diagonal bulk together
with a 40-dimensional AR(1) residual block:
\[
\boldsymbol{\Sigma}_{\mathrm{bulk}}
=\operatorname{diag}\!\left(
\boldsymbol{\Lambda}_{\Gamma},
0.15\left(0.30^{|i-j|}\right)_{i,j=1}^{40}
\right),\qquad m=p-56,
\]
where $\boldsymbol{\Lambda}_{\Gamma}=\operatorname{diag}
(\Lambda_{\Gamma,1},\ldots,\Lambda_{\Gamma,m})$ and
$\Lambda_{\Gamma,j}\stackrel{\mathrm{i.i.d.}}{\sim}
\operatorname{Gamma}(10,0.1)$, with shape $10$ and scale $0.1$.
In all four scenarios, the two $8$-dimensional spike blocks and the bulk
blocks together have total dimension $p$.
\endgroup
% We consider the following two scenarios for our simulations:
% \begin{itemize}
%   \item \textbf{Scenario 1 (weak spikes)}: 
%         $\lambda_{1} = 6$, $\lambda_{2} = 4$, 
%         $\sigma_1^2 = 1$, $\sigma_2^2 = 0.5$.
%   \item \textbf{Scenario 2 (strong spikes)}: 
%         $\lambda_{1} = 10$, $\lambda_{2} = 8$, 
%         $\sigma_1^2 = 1.5$, $\sigma_2^2 = 1$.
% \end{itemize}

\textbf{Metric.} 
We fix the ratio $p/n\in\{0.8,1,1.2\}$ and consider $p\in\{200,300,400\}$. In our setting, the true number of spikes is $r=2$. The estimation accuracy of a given estimator $\hat r$ is computed from {1000} independent repetitions as the fraction for which $\hat r=r$. The innovations are either standard normal or standardized $\operatorname{Exp}(1)-1$.

\textbf{Baseline methods.}
Denote the eigenvalues of the sample covariance matrix $\sbSigma = \frac{1}{n}\bX\bX^\top$ and the sample correlation matrix $\hat{\mathbf{R}} = \mathrm{diag}(\sbSigma)^{-1/2}\sbSigma\,\mathrm{diag}(\sbSigma)^{-1/2}$ by $\{\slambda_j: 1\leq j\leq p\}$ and $\{\hat{\alpha}_j: 1\leq j\leq p\}$, respectively. We compare our proposed {ESC} method with PY \citep{Yao2014}, BEMA \citep{Ke2023}, EDB \citep{liu2023asymptotic}, and ACT \citep{Fan03042022}.

\textbf{PY \cite{Yao2014}.} The estimator of the number of spikes is defined as
\[
\hat{r}^{\mathrm{PY}} = \min \Bigl\{ j : \delta_{n,j} < d(n) \ \text{and} \ \delta_{n,j+1} < d(n) \Bigr\}\textcolor{black}{-1},
\]
where $d(n)$ is a pre-specified threshold and 
\[
\delta_{n,j} = \hat{\lambda}_{j} - \hat{\lambda}_{j+1}, \quad j = 1,\dots,\textcolor{black}{p-1}.
\]

\begingroup\color{black}
\textbf{BEMA \cite{Ke2023}.} We use the general Gamma-bulk version of BEMA (Algorithm~2), which models residual variances by $\operatorname{Gamma}(\theta,\sigma^2/\theta)$, with shape $\theta$ and scale $\sigma^2/\theta$. The parameters $(\theta,\sigma^2)$ are fitted by matching the middle sample eigenvalues to the corresponding model quantiles. The fitted model is then used to simulate the largest sample eigenvalue under a model without spikes. Writing $\hat t_{1-\beta}$ for its estimated $(1-\beta)$ quantile, the estimated spike count is
\[
\hat r^{\mathrm{BEMA}}
=\#\{1\leq j\leq \min(p,n):\hat\lambda_j>\hat t_{1-\beta}\}.
\]
The implementation uses bulk trimming parameter $\alpha=0.2$, $\beta=0.1$, 10 Monte Carlo samples to approximate each bulk quantile curve, and 100 samples to estimate the null threshold; it centers each simulated null matrix as it centers the observed data. This general version differs from the scalar-noise BEMA0 procedure in Algorithm~1.

\textbf{EDB \cite{liu2023asymptotic}.} Let $q=\min(p,n)$ and retain the ordered sample covariance eigenvalues $\hat\lambda_1,\ldots,\hat\lambda_q$. Following the original EDB criterion when $p<n$ and its pseudo-EDB extension when $p>n$, we use
\[
\hat r^{\mathrm{EDB}}
=\arg\min_{1\leq j\leq w}
\left\{-\log(p)(\hat\lambda_1-\hat\lambda_{j+1})
+(q-j-1)\log\widetilde\theta_{q,j}
+\frac{p j\log n}{n}\right\},
\]
where
\[
\widetilde\theta_{q,j}
=\frac{1}{q-j-1}\sum_{i=j+1}^{q-1}
\exp\{2(\hat\lambda_i-\hat\lambda_{i+1})\},
\qquad
w=\min\{\lfloor6n^{0.1}\rfloor,q-2\}.
\]
The summand uses the squared exponential gap indexed by $i$. When $p>n$, the reduced spectrum of length $q$ enters the bulk term, while the penalty retains $p$.

\textbf{ACT \cite{Fan03042022}.} ACT applies its threshold to bias-corrected eigenvalues of the sample correlation matrix. Denoting these adjusted values by $\hat\alpha_j^{\mathrm{ACT}}$, the estimator is
\[
\hat r^{\mathrm{ACT}}
=\max\left(\{j:1\leq j\leq p-1,\ 
\hat\alpha_j^{\mathrm{ACT}}>1+\sqrt{p/(n-1)}\}\cup\{0\}\right).
\]
Specifically, for $j\leq p-1$, set $\rho_j=(p-j)/(n-1)$ and
\[
m_j=\frac{1}{p-j}\left[\sum_{\ell=j+1}^{p}\frac{1}{\hat\alpha_\ell-\hat\alpha_j}
+\frac{4}{\hat\alpha_{j+1}-\hat\alpha_j}\right],\qquad
\hat\alpha_j^{\mathrm{ACT}}=\left[\frac{1-\rho_j}{\hat\alpha_j}-\rho_j m_j\right]^{-1}.
\]
Thus the threshold is applied to the bias-corrected, rather than raw, correlation eigenvalues.
\endgroup

% Simulation-1: updated spike-number estimation results from the tabels/ CSV files.
% Duplicate (p,n) records retain the row with the largest epsilon0.

\begin{table}[!htbp]

\centering
\caption{Scenario 1: exact spike-number recovery with $(\lambda_1,\lambda_2)=(8,6)$ and diagonal bulk $\boldsymbol{\Sigma}_{\mathrm{bulk}}=\operatorname{diag}(1.5\mathbf I_{p/2-8},1.2\mathbf I_{p/2-8})$. Entries are proportions of 1000 repetitions with $\hat r=2$, under standard-normal or standardized $\operatorname{Exp}(1)-1$ innovations.}
\label{tab:simulation1-scenario1}
\setlength{\tabcolsep}{2pt}\renewcommand{\arraystretch}{1.05}\resizebox{\textwidth}{!}{%
\begin{tabular}{@{}lrrr|rrr|rrr|rrr|rrr|rrr@{}}
\toprule
\rowcolor{gray!15} \textbf{Method} & \multicolumn{9}{c|}{\textbf{Normal}} & \multicolumn{9}{c}{\textbf{$\mathrm{exp}(1)-1$}} \\
\cmidrule(lr){2-10} \cmidrule(lr){11-19}
\rowcolor{gray!8} \textbf{$p/n$} & \multicolumn{3}{c|}{\textbf{0.80}} & \multicolumn{3}{c|}{\textbf{1.00}} & \multicolumn{3}{c|}{\textbf{1.20}} & \multicolumn{3}{c|}{\textbf{0.80}} & \multicolumn{3}{c|}{\textbf{1.00}} & \multicolumn{3}{c}{\textbf{1.20}} \\
\rowcolor{gray!4} \textbf{$p$} & 200 & 300 & 400 & 200 & 300 & 400 & 200 & 300 & 400 & 200 & 300 & 400 & 200 & 300 & 400 & 200 & 300 & 400 \\
\midrule
\rowcolor{highlightColor} \textbf{ESC} & \textbf{1.00} & \textbf{1.00} & \textbf{1.00} & \textbf{0.99} & \textbf{1.00} & \textbf{1.00} & \textbf{1.00} & \textbf{1.00} & \textbf{1.00} & 0.99 & 0.99 & \textbf{1.00} & \textbf{1.00} & 0.99 & 0.99 & 0.95 & \textbf{1.00} & \textbf{1.00} \\
\textbf{PY} & 0.82 & 0.90 & 0.87 & 0.87 & 0.77 & 0.85 & 0.87 & 0.87 & 0.82 & 0.80 & 0.87 & 0.85 & 0.72 & 0.82 & 0.87 & 0.73 & 0.84 & 0.83 \\
\textbf{EDB} & \textbf{1.00} & \textbf{1.00} & \textbf{1.00} & \textbf{0.99} & \textbf{1.00} & \textbf{1.00} & \textbf{1.00} & \textbf{1.00} & \textbf{1.00} & \textbf{1.00} & \textbf{1.00} & \textbf{1.00} & 0.99 & \textbf{1.00} & \textbf{1.00} & 0.95 & \textbf{1.00} & 0.99 \\
\textbf{BEMA} & \textbf{1.00} & \textbf{1.00} & \textbf{1.00} & \textbf{0.99} & 0.99 & \textbf{1.00} & 0.99 & \textbf{1.00} & \textbf{1.00} & 0.98 & \textbf{1.00} & \textbf{1.00} & 0.94 & 0.99 & \textbf{1.00} & 0.93 & 0.99 & \textbf{1.00} \\
\textbf{ACT} & 0.93 & 0.98 & 0.91 & 0.96 & 0.96 & 0.96 & 0.93 & 0.95 & 0.93 & 0.96 & 0.98 & 0.95 & 0.96 & 0.95 & 0.96 & \textbf{0.97} & 0.95 & 0.95 \\
\bottomrule
\end{tabular}
}
\end{table}

Table~\ref{tab:simulation1-scenario1} shows high exact-recovery rates under the diagonal bulk. Across both innovation distributions, ESC attains $0.95$--$1.00$, comparable to EDB ($0.95$--$1.00$) and BEMA ($0.93$--$1.00$); PY ranges from $0.72$ to $0.90$.

\begin{table}[!htbp]

\centering
\caption{Scenario 2: exact spike-number recovery with $(\lambda_1,\lambda_2)=(8,6)$ and a pure $(p-16)$-dimensional AR(1) bulk with unit marginal variance and correlation $\rho=0.3$. Entries are proportions of 1000 repetitions with $\hat r=2$, under standard-normal or standardized $\operatorname{Exp}(1)-1$ innovations.}
\label{tab:simulation1-scenario2}
\setlength{\tabcolsep}{2pt}\renewcommand{\arraystretch}{1.05}\resizebox{\textwidth}{!}{%
\begin{tabular}{@{}lrrr|rrr|rrr|rrr|rrr|rrr@{}}
\toprule
\rowcolor{gray!15} \textbf{Method} & \multicolumn{9}{c|}{\textbf{Normal}} & \multicolumn{9}{c}{\textbf{$\mathrm{exp}(1)-1$}} \\
\cmidrule(lr){2-10} \cmidrule(lr){11-19}
\rowcolor{gray!8} \textbf{$p/n$} & \multicolumn{3}{c|}{\textbf{0.80}} & \multicolumn{3}{c|}{\textbf{1.00}} & \multicolumn{3}{c|}{\textbf{1.20}} & \multicolumn{3}{c|}{\textbf{0.80}} & \multicolumn{3}{c|}{\textbf{1.00}} & \multicolumn{3}{c}{\textbf{1.20}} \\
\rowcolor{gray!4} \textbf{$p$} & 200 & 300 & 400 & 200 & 300 & 400 & 200 & 300 & 400 & 200 & 300 & 400 & 200 & 300 & 400 & 200 & 300 & 400 \\
\midrule
\rowcolor{highlightColor} \textbf{ESC} & \textbf{1.00} & \textbf{1.00} & \textbf{1.00} & 0.98 & 0.98 & \textbf{1.00} & 0.99 & \textbf{1.00} & \textbf{1.00} & 0.97 & 0.97 & 0.96 & 0.98 & \textbf{1.00} & 0.99 & 0.97 & 0.98 & \textbf{1.00} \\
\textbf{PY} & 0.81 & 0.79 & 0.87 & 0.81 & 0.79 & 0.86 & 0.86 & 0.78 & 0.86 & 0.74 & 0.76 & 0.82 & 0.74 & 0.81 & 0.83 & 0.77 & 0.74 & 0.81 \\
\textbf{EDB} & \textbf{1.00} & \textbf{1.00} & \textbf{1.00} & 0.99 & \textbf{1.00} & \textbf{1.00} & 0.99 & \textbf{1.00} & \textbf{1.00} & \textbf{1.00} & \textbf{1.00} & \textbf{1.00} & \textbf{1.00} & 0.99 & \textbf{1.00} & \textbf{1.00} & \textbf{0.99} & \textbf{1.00} \\
\textbf{BEMA} & \textbf{1.00} & \textbf{1.00} & \textbf{1.00} & \textbf{1.00} & \textbf{1.00} & \textbf{1.00} & \textbf{1.00} & \textbf{1.00} & \textbf{1.00} & \textbf{1.00} & \textbf{1.00} & \textbf{1.00} & 0.97 & \textbf{1.00} & \textbf{1.00} & 0.92 & 0.98 & \textbf{1.00} \\
\textbf{ACT} & 0.18 & 0.08 & 0.03 & 0.26 & 0.15 & 0.07 & 0.37 & 0.25 & 0.10 & 0.14 & 0.11 & 0.05 & 0.30 & 0.17 & 0.12 & 0.35 & 0.31 & 0.15 \\
\bottomrule
\end{tabular}%
}
\end{table}

{Table~\ref{tab:simulation1-scenario2} reports the pure AR(1) bulk experiment. ESC attains $0.96$--$1.00$; EDB and BEMA attain $0.99$--$1.00$ and $0.92$--$1.00$. ACT is less accurate, ranging from $0.03$ to $0.37$.}

\begin{table}[!htbp]

\centering
\caption{Scenario 3: exact spike-number recovery with $(\lambda_1,\lambda_2)=(8,6)$ and a pure $(p-16)$-dimensional diagonal Gamma bulk with independent $\operatorname{Gamma}(10,0.1)$ variances (shape $10$, scale $0.1$). Entries are proportions of 1000 repetitions with $\hat r=2$, under standard-normal or standardized $\operatorname{Exp}(1)-1$ innovations.}
\label{tab:simulation1-scenario3}
\setlength{\tabcolsep}{2pt}\renewcommand{\arraystretch}{1.05}\resizebox{\textwidth}{!}{%
\begin{tabular}{@{}lrrr|rrr|rrr|rrr|rrr|rrr@{}}
\toprule
\rowcolor{gray!15} \textbf{Method} & \multicolumn{9}{c|}{\textbf{Normal}} & \multicolumn{9}{c}{\textbf{$\mathrm{exp}(1)-1$}} \\
\cmidrule(lr){2-10} \cmidrule(lr){11-19}
\rowcolor{gray!8} \textbf{$p/n$} & \multicolumn{3}{c|}{\textbf{0.80}} & \multicolumn{3}{c|}{\textbf{1.00}} & \multicolumn{3}{c|}{\textbf{1.20}} & \multicolumn{3}{c|}{\textbf{0.80}} & \multicolumn{3}{c|}{\textbf{1.00}} & \multicolumn{3}{c}{\textbf{1.20}} \\
\rowcolor{gray!4} \textbf{$p$} & 200 & 300 & 400 & 200 & 300 & 400 & 200 & 300 & 400 & 200 & 300 & 400 & 200 & 300 & 400 & 200 & 300 & 400 \\
\midrule
\rowcolor{highlightColor} \textbf{ESC} & \textbf{0.99} & 0.99 & 0.98 & 0.97 & 0.99 & \textbf{1.00} & \textbf{1.00} & \textbf{1.00} & 0.98 & 0.95 & 0.99 & \textbf{0.99} & 0.97 & \textbf{0.98} & 0.98 & \textbf{0.98} & \textbf{0.99} & \textbf{0.98} \\
\textbf{PY} & 0.77 & 0.73 & 0.82 & 0.77 & 0.78 & 0.85 & 0.90 & 0.85 & 0.81 & 0.71 & 0.65 & 0.61 & 0.73 & 0.70 & 0.73 & 0.78 & 0.82 & 0.77 \\
\textbf{EDB} & \textbf{0.99} & 0.99 & \textbf{1.00} & \textbf{1.00} & \textbf{1.00} & \textbf{1.00} & \textbf{1.00} & \textbf{1.00} & \textbf{1.00} & 0.96 & 0.98 & \textbf{0.99} & 0.97 & \textbf{0.98} & \textbf{0.99} & \textbf{0.98} & \textbf{0.99} & \textbf{0.98} \\
\textbf{BEMA} & 0.90 & 0.87 & 0.84 & 0.91 & 0.84 & 0.84 & 0.95 & 0.89 & 0.86 & 0.90 & 0.92 & 0.91 & 0.89 & 0.84 & 0.88 & 0.92 & 0.88 & 0.89 \\
\textbf{ACT} & \textbf{0.99} & 0.98 & 0.99 & 0.99 & \textbf{1.00} & 0.99 & \textbf{1.00} & \textbf{1.00} & \textbf{1.00} & \textbf{0.99} & 0.97 & 0.98 & \textbf{1.00} & 0.97 & 0.98 & 0.97 & 0.98 & 0.93 \\
\bottomrule
\end{tabular}%
}
\end{table}

\begin{table}[ht]
\centering
\caption{Scenario 4: exact spike-number recovery with $(\lambda_1,\lambda_2)=(6,4)$, a Gamma diagonal block of dimension $p-56$ with independent $\operatorname{Gamma}(10,0.1)$ variances, and a 40-dimensional AR(1) block with marginal variance $0.15$ and correlation $\rho=0.30$. Entries are proportions of 1000 repetitions with $\hat r=2$, under standard-normal or standardized $\operatorname{Exp}(1)-1$ innovations.}
\label{tab:simulation1-scenario4}
\setlength{\tabcolsep}{2pt}\renewcommand{\arraystretch}{1.05}\resizebox{\textwidth}{!}{%
\begin{tabular}{@{}lrrr|rrr|rrr|rrr|rrr|rrr@{}}
\toprule
\rowcolor{gray!15} \textbf{Method} & \multicolumn{9}{c|}{\textbf{Normal}} & \multicolumn{9}{c}{\textbf{$\mathrm{exp}(1)-1$}} \\
\cmidrule(lr){2-10} \cmidrule(lr){11-19}
\rowcolor{gray!8} \textbf{$p/n$} & \multicolumn{3}{c|}{\textbf{0.80}} & \multicolumn{3}{c|}{\textbf{1.00}} & \multicolumn{3}{c|}{\textbf{1.20}} & \multicolumn{3}{c|}{\textbf{0.80}} & \multicolumn{3}{c|}{\textbf{1.00}} & \multicolumn{3}{c}{\textbf{1.20}} \\
\rowcolor{gray!4} \textbf{$p$} & 200 & 300 & 400 & 200 & 300 & 400 & 200 & 300 & 400 & 200 & 300 & 400 & 200 & 300 & 400 & 200 & 300 & 400 \\
\midrule
\rowcolor{highlightColor} \textbf{ESC} & \textbf{1.00} & \textbf{1.00} & \textbf{1.00} & \textbf{0.98} & \textbf{1.00} & \textbf{0.99} & \textbf{0.95} & \textbf{0.98} & 0.97 & \textbf{0.95} & \textbf{0.98} & \textbf{0.99} & \textbf{0.92} & \textbf{0.93} & \textbf{0.96} & \textbf{0.84} & \textbf{0.87} & 0.87 \\
\textbf{PY} & 0.85 & 0.72 & 0.76 & 0.89 & 0.84 & 0.77 & 0.91 & 0.84 & 0.86 & 0.67 & 0.75 & 0.71 & 0.72 & 0.72 & 0.76 & 0.70 & 0.75 & 0.77 \\
\textbf{EDB} & 0.99 & 0.99 & 0.99 & 0.95 & 0.99 & 0.95 & 0.88 & 0.87 & 0.89 & 0.94 & 0.96 & 0.98 & 0.86 & 0.86 & 0.84 & 0.74 & 0.67 & 0.68 \\
\textbf{BEMA} & 0.07 & 0.83 & 0.83 & 0.07 & 0.61 & 0.78 & 0.04 & 0.70 & 0.79 & 0.19 & 0.49 & 0.74 & 0.16 & 0.28 & 0.72 & 0.17 & 0.46 & 0.75 \\
\textbf{ACT} & 0.64 & 0.66 & 0.67 & 0.79 & 0.74 & 0.73 & 0.81 & 0.88 & 0.84 & 0.68 & 0.63 & 0.67 & 0.79 & 0.74 & 0.77 & 0.75 & 0.86 & 0.86 \\
\bottomrule
\end{tabular}%
}
\par\vspace{1.5\baselineskip}
\includegraphics[width=\textwidth]{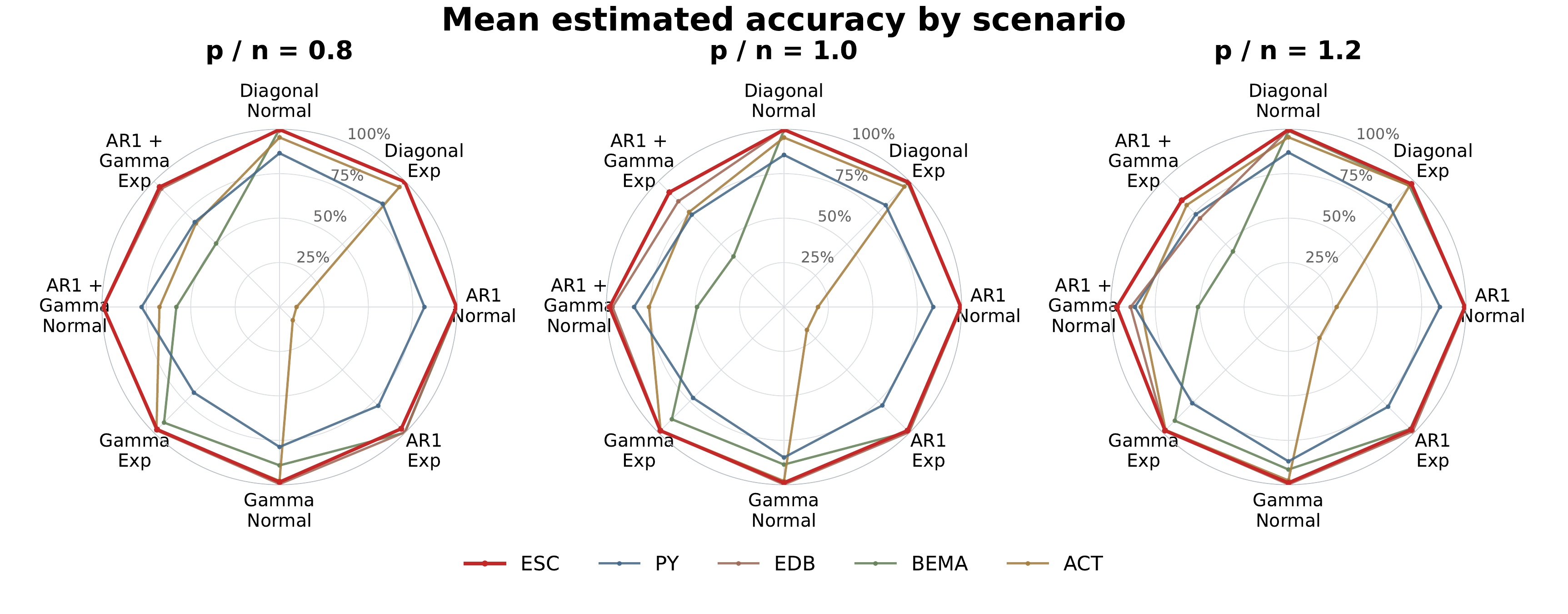}
\captionof{figure}{Mean exact spike-count recovery rates for the four bulk scenarios under Normal and standardized $\operatorname{Exp}(1)-1$ innovations, averaged over $p\in\{200,300,400\}$ at each aspect ratio $p/n\in\{0.8,1,1.2\}$. Each radial axis represents a scenario--distribution pair.}
\label{fig:spike-correctness-radar}
\end{table}

For the pure Gamma bulk in Table~\ref{tab:simulation1-scenario3}, ESC attains $0.95$--$1.00$; EDB and ACT attain $0.96$--$1.00$ and $0.93$--$1.00$. BEMA ranges from $0.84$ to $0.95$, while PY ranges from $0.61$ to $0.90$.

{Figure~\ref{fig:spike-correctness-radar} summarizes the recovery rates in Tables~\ref{tab:simulation1-scenario1}--\ref{tab:simulation1-scenario4}. ESC has mean rates close to one for the diagonal, AR(1), and Gamma bulks, with lower rates for the mixed Gamma--AR(1) bulk under exponential innovations. ACT performs poorly for the pure AR(1) bulk, whereas BEMA is less accurate for the mixed bulk.}

{Table~\ref{tab:simulation1-scenario4} reports the mixed Gamma and AR(1) bulk experiment with weaker spikes $(6,4)$. Under Normal innovations, ESC attains $0.95$--$1.00$; under exponential innovations, ESC attains $0.84$--$0.99$. BEMA ranges from $0.04$ to $0.83$ across the two distributions.}

\subsection{Common-Feature Identification for Multi-Source Data}  \label{subsec:simulationcommonfeature}

We generate multi-source samples according to
\[
\mathbf{X}_k = \boldsymbol{\Sigma}_k^{1/2}\mathbf{Z}_k, \quad k = 1,\ldots,K,
\]
where $\mathbf{Z}_k=[\mathbf{z}_{k1},\dots,\mathbf{z}_{kn_k}]_{p\times n_k}$ contains $n_k$ i.i.d. samples. We consider two distributions for the entries $\mathbf{z}_{ki}=(z_{kij})_{j=1}^p$: $z_{kij}\sim N(0,1)$ and $z_{kij}\sim \mathrm{exp}(1)-1$, $1\leq j\leq p$.

\begingroup\color{black}
For the population covariance matrices $\boldsymbol{\Sigma}_k$, $1\leq k\leq K$, we consider the following spiked structure:
\[
\boldsymbol{\Sigma}_k
= \operatorname{diag}\left(
B_{k1}, B_{k2}, B_{k3},
\sigma_1^2 \mathbf{I}_{\lceil b/2\rceil},
\sigma_2^2 \mathbf{I}_{\lfloor b/2\rfloor}
\right).
\]
Each $\boldsymbol{\Sigma}_k$ has $r_k=3$ spikes, two of which correspond to common features shared across all $K$ sources; hence $r_{\mathrm{s}}=2$.
The two 8-dimensional blocks associated with these common features are
\[
B_{ki}=\frac{\lambda_{ki}-\frac{1}{4}}{8}\mathbf{1}_8\mathbf{1}_8^\top
+\frac{1}{4}\mathbf{I}_8,\qquad i=1,2.
\]
Their spike eigenvalues are $\lambda_{k1}$ and $\lambda_{k2}$, with corresponding eigenvectors
\[
\mathbf{u}_{k i}^{\top} = (\mathbf{0}_{8(i-1)}^\top, \frac{1}{2\sqrt{2}}\mathbf{1}_8^\top, \mathbf{0}_{p-8i}^\top), \quad 1 \leq i \leq 2,
\]
which have the same directions for every source.
The $(K+1)$-dimensional block associated with the private feature of source $k$ is
\[
B_{k3}=(\lambda_{k3}-\tfrac{1}{4})\mathbf{v}_k\mathbf{v}_k^\top
+\tfrac{1}{4}\mathbf{I}_{K+1},\qquad
\mathbf{v}_k\in\mathbb{R}^{K+1},\quad \|\mathbf{v}_k\|_2=1.
\]
It has spike eigenvalue $\lambda_{k3}$ and corresponding eigenvector
\[
\mathbf{u}_{k3}^\top = (\mathbf{0}_{16}^\top, \mathbf{v}_k^\top, \mathbf{0}_{b}^\top).
\]
Finally, $b=p-(K+17)>0$ is the remaining bulk dimension after the two common blocks and the private block; its coordinates are split between the $\sigma_1^2$ and $\sigma_2^2$ identity blocks.
\endgroup

We use the following unified parameter setting throughout the simulation study. We set $K=10$ data sources, $r_k=3$ for every source, $r_{\mathrm{s}}=2$, $\sigma_1^2=1$, and $\sigma_2^2=\textcolor{black}{0.75}$. For each source $k$, the $r_k$ spike eigenvalues $(\lambda_{k1},\ldots,\lambda_{kr_k})$ are a permutation of $(\lambda_1,\lambda_2,\lambda_3)$, so all sources share the same three spike strengths while their assignment to feature directions may differ. We set $n_k=p$ for every source, so that $p/n_k=1$ for all $k$, and use the prescribed values of $p$ for each experiment.

For the spike eigenvalues, we consider three signal-strength scenarios:
\begin{itemize}
  \item \textbf{S1 (Strong-Signal Regime):} $(\lambda_1,\lambda_2,\lambda_3)=(15,8,4)$;
  \item \textbf{S2 (Intermediate-Signal Regime):} $(\lambda_1,\lambda_2,\lambda_3)=(15,8,3)$;
  \item \textbf{S3 (Weak-Signal Regime):} $(\lambda_1,\lambda_2,\lambda_3)=(15,5,2.5)$.
\end{itemize}

We consider two private-feature configurations, denoted by C1 and C2.

\textbf{C1 (Orthonormal Private Features).} The private-feature block has dimension $K+1$. Let $\{\mathbf{e}_0,\mathbf{e}_1,\ldots,\mathbf{e}_K\}$ denote its canonical orthonormal basis, where $\mathbf{e}_0$ is reserved for the nonorthogonal construction. In configuration C1, source $k$ uses
\[
\mathbf{v}_k=\mathbf{e}_k,\qquad k=1,\ldots,K.
\]
Thus, $\|\mathbf{v}_k\|_2=1$ and $\mathbf{v}_k^{\mathsf T}\mathbf{v}_\ell=0$ for $k\neq\ell$. 

\textbf{C2 (Controlled Nonorthogonal Private Features).} Using the same orthonormal basis, the code constructs each private direction from a common anchor $\mathbf{e}_0$ and a source-specific basis vector:
\[
\mathbf{v}_k=\sqrt{\rho}\,\mathbf{e}_0+\sqrt{1-\rho}\,\mathbf{e}_k,
\quad k=1,\ldots,K.
\]
The experiments use $\rho=0.2$. Consequently, $\|\mathbf{v}_k\|_2=1$ and $\mathbf{v}_k^{\mathsf T}\mathbf{v}_\ell=\rho$ for every $k\neq\ell$, so all private directions have the same controlled pairwise overlap.

\textbf{Baseline methods.} We compare FE with two classes of baselines. The first class consists of \textbf{PerPCA} \citep{shi2024personalized} and \textbf{KT} \citep{li2024knowledgetransfermultipleprincipal}. Both methods are supplied with the true dimensions $r_k=3$ and $r_{\mathrm{s}}=2$ in this experiment; these fixed/oracle baselines are denoted PerPCA(r) and KT(r) in the tables below. PerPCA estimates a shared basis $\mathbf U^{\mathrm{s}}$ and source-specific bases $\mathbf U_k^{\mathrm p}$ by minimizing
\[
\min_{\mathbf U^{\mathrm{s}},\{\mathbf U_k^{\mathrm p}\}}
\frac{1}{2}\sum_{k=1}^{K}\frac{1}{n_k}
\left\|\mathbf X_k-\mathbf U^{\mathrm{s}}(\mathbf U^{\mathrm{s}})^{\mathsf T}\mathbf X_k
-\mathbf U_k^{\mathrm p}(\mathbf U_k^{\mathrm p})^{\mathsf T}\mathbf X_k\right\|_F^2,
\]
subject to orthonormality and $\mathbf U_k^{\mathrm p\mathsf T}\mathbf U^{\mathrm{s}}=0$. KT estimates the shared basis from the leading $r_{\mathrm{s}}$ eigenvectors of the weighted average of source-specific projectors,
\[
\widehat{\mathbf U}^{\mathrm{s}}
=\operatorname{eig}_{r_{\mathrm{s}}}\left(
\sum_{k=1}^{K}\frac{n_k}{\sum_{\ell=1}^{K}n_\ell}
\widehat{\mathbf U}_k[\mathcal A_k]\widehat{\mathbf U}_k[\mathcal A_k]^{\mathsf T}
\right),
\]
and then obtains the target private component from the residual after removing the estimated shared projection. In the reported comparisons, both the oracle dimensions and FE-estimated dimensions are supported where applicable.

The second class consists of group-factor methods: the \textbf{APM} \citep{hu2025aggregated}, \textbf{CP} \citep{chen2022circularly}, and \textbf{GCC} \citep{lin2023generalised}. We use their group-factor notation, under which the data structure is
\[
\mathbf X_k=\mathbf G\boldsymbol\Gamma_k^{\mathsf T}
+\mathbf F_k\boldsymbol\Lambda_k^{\mathsf T}+\mathbf E_k,
\qquad k=1,\ldots,K,
\]
where $\mathbf G\in\mathbb R^{p\times r_0}$ and $\mathbf F_k\in\mathbb R^{p\times r_k}$ are respectively the global and group-specific factor matrices, $\boldsymbol\Gamma_k\in\mathbb R^{n_k\times r_0}$ and $\boldsymbol\Lambda_k\in\mathbb R^{n_k\times r_k}$ are the corresponding loading matrices, and $\mathbf E_k$ is the idiosyncratic error matrix. 
{\color{black}
For FE, PerPCA, and KT, the common and private projectors are constructed from their estimated bases as follows:
\[
\widehat{\mathbf P}^{\mathrm{s}}=\widehat{\mathbf U}^{\mathrm{s}}(\widehat{\mathbf U}^{\mathrm{s}})^{\mathsf T},\quad \widehat{\mathbf P}_{k}^{\mathrm p}=\widehat{\mathbf V}_k^{\mathrm p}(\widehat{\mathbf V}_k^{\mathrm p})^{\mathsf T}.
\]
For FE, $\widehat{\mathbf V}_k^{\mathrm p}$ is obtained from the residualized covariance as in \eqref{eq:private-residual-covariance}--\eqref{eq:private-residual-projector}; for PerPCA and KT, it denotes the orthonormal private basis returned by the respective method.

} % end revision marking
For APM, CP, and GCC, the group-factor estimates are mapped to the heterogeneous-PCA projectors by
\[
\widehat{\mathbf P}^{\mathrm{s}}=\widehat{\mathbf G}(\widehat{\mathbf G}^\top\widehat{\mathbf G})^{\dagger} \widehat{\mathbf G}^\top, \quad \widehat{\mathbf P}_{k}^{\mathrm p}=\widehat{\mathbf F}_k(\widehat{\mathbf F}_k^\top\widehat{\mathbf F}_k)^{\dagger}\widehat{\mathbf F}_k^{\top}.
\]
% The loading estimates are used only for the factor representation. For FE, PerPCA, and KT, the corresponding projectors are formed directly from their estimated common and private bases.

\textbf{Metric.} Each configuration is evaluated over 1000 independent repetitions. For Tables~\ref{tab:simulation3-rank-combined} and~\ref{tab:simulation3-rank-combined-nonorth}, the rank metrics are
\[
\hat r_{\mathrm{s}},\qquad
\Pr(\hat r_{\mathrm{s}}>r_{\mathrm{s}}),\qquad
\Pr(\hat r_{\mathrm{s}}<r_{\mathrm{s}}),
\qquad r_{\mathrm{s}}=2.
\]
The subspace metrics use the true projectors $\mathbf P^{\mathrm{s}}=\mathbf U^{\mathrm{s}}(\mathbf U^{\mathrm{s}})^{\mathsf T}$ and $\mathbf P_k^{\mathrm p}=\mathbf U_k[\mathcal A_k^{\mathrm p}]\mathbf U_k[\mathcal A_k^{\mathrm p}]^{\mathsf T}$. Writing the estimated projectors as $\widehat{\mathbf P}^{\mathrm{s}}$ and $\widehat{\mathbf P}_k^{\mathrm p}$, we report
\[
\mathrm{Global}=\left\|\widehat{\mathbf P}^{\mathrm{s}}-\mathbf P^{\mathrm{s}}\right\|_F^2, \quad 
\mathrm{Local}=\frac{1}{K}\sum_{k=1}^K
\left\|\widehat{\mathbf P}_k^{\mathrm p}-\mathbf P_k^{\mathrm p}\right\|_F^2,\quad
\mathrm{Total}=\mathrm{Global}+\mathrm{Local}.
\]
The table entries are averages over the repetitions; smaller values indicate more accurate recovery.

For the focused common-feature identification experiment, we fix the true
spike rank at $r_k=3$ for every source and apply FE to estimate only the
common-feature index sets.  We report the client-average exact recovery rate,
the true-positive rate (TPR), and the false-discovery rate (FDR).  For source
$k$, let
\[
 \mathrm{TP}_k=|\widehat{\mathcal A}_k^{\mathrm{s}}
   \cap \mathcal A_k^{\mathrm{s}}|,\quad
 \mathrm{FP}_k=|\widehat{\mathcal A}_k^{\mathrm{s}}\setminus
   \mathcal A_k^{\mathrm{s}}|,\quad
 \mathrm{FN}_k=|\mathcal A_k^{\mathrm{s}}\setminus
   \widehat{\mathcal A}_k^{\mathrm{s}}|.
\]
The client-average exact recovery rate is
\[
 \frac{1}{K}\sum_{k=1}^K
 \mathbf{1}\{\widehat{\mathcal A}_k^{\mathrm{s}}
 =\mathcal A_k^{\mathrm{s}}\},
\]
and the pooled TPR and FDR are respectively
\[
 \mathrm{TPR}=\frac{\sum_{k=1}^K \mathrm{TP}_k}{K r_{\mathrm{s}}},
 \qquad
 \mathrm{FDR}=\frac{\sum_{k=1}^K \mathrm{FP}_k}
 {\max\{\sum_{k=1}^K|\widehat{\mathcal A}_k^{\mathrm{s}}|,1\}}.
\]
Each configuration uses 1,000 independent repetitions, with
$p\in\{100,300,500\}$ and $p/n\in\{0.8,1,1.2\}$. Table~\ref{tab:simulation3-common-identification}
reports rates averaged over the three aspect ratios, separately for
orthogonal (C1) and nonorthogonal (C2) private directions and for Gaussian and
centered exponential noise. Under weak signals (S3), increasing $p$ from 100
to 500 raises exact recovery from 74.6\% to 96.5\% for C1 and from 73.8\% to
96.2\% for C2 under Gaussian noise; under centered exponential noise, it
rises from 28.5\% to 86.2\% and from 27.5\% to 85.5\%, respectively. Even
in the latter setting, FDR is at most 0.4\%. Under strong signals (S1),
Gaussian-noise exact recovery is already at least 99.7\% at $p=100$ in both
private-direction settings.

\begin{table}[!htb]
\centering
\caption{\textcolor{black}{Common-feature index recovery by FE with $K=10$, $r_k=3$, and true common rank $r_{\mathrm{s}}=2$. The tuning parameters are $(a,\varepsilon_0)=(10,0.1)$. Entries are percentages averaged over 1,000 repetitions and $p/n\in\{0.8,1,1.2\}$ for each setting. Exact recovery is averaged over clients; TPR and FDR are pooled across clients (definitions in the text).}}
\label{tab:simulation3-common-identification}
\setlength{\tabcolsep}{4pt}
\renewcommand{\arraystretch}{1.08}
\resizebox{\textwidth}{!}{%
\begin{tabular}{@{}c c *{6}{c}@{}}
\toprule
\rowcolor{gray!8}
\multicolumn{2}{c}{} & \multicolumn{3}{c}{\textbf{Normal}} & \multicolumn{3}{c}{\textbf{$\mathrm{exp}(1)-1$}} \\
\cmidrule(lr){3-5} \cmidrule(lr){6-8}
\textbf{Signal} & \textbf{$p$} & \textbf{Exact recovery (\%)} & \textbf{TPR (\%)} & \textbf{FDR (\%)} & \textbf{Exact recovery (\%)} & \textbf{TPR (\%)} & \textbf{FDR (\%)} \\
\midrule
\rowcolor{gray!15} \multicolumn{8}{c}{\textbf{C1: Orthogonal private directions}} \\
\multirow{3}{*}{\textbf{S1}} & 100 & 99.7 & 99.8 & 0.1 & 89.6 & 94.7 & 2.3 \\
 & 300 & 100 & 100 & 0.0 & 99.7 & 99.8 & 0.1 \\
 & 500 & 100 & 100 & 0.0 & 100 & 100 & 0.0 \\
\midrule
\addlinespace[2pt]
\multirow{3}{*}{\textbf{S2}} & 100 & 96.7 & 98.4 & 0.1 & 66.9 & 82.9 & 1.8 \\
 & 300 & 99.8 & 99.9 & 0.0 & 97.3 & 98.7 & 0.1 \\
 & 500 & 100 & 100 & 0.0 & 99.9 & 99.9 & 0.0 \\
\midrule
\addlinespace[2pt]
\multirow{3}{*}{\textbf{S3}} & 100 & 74.6 & 87.3 & 0.1 & 28.5 & 64.2 & 0.4 \\
 & 300 & 90.8 & 95.4 & 0.0 & 68.9 & 84.4 & 0.0 \\
 & 500 & 96.5 & 98.2 & 0.0 & 86.2 & 93.1 & 0.0 \\
\midrule
\rowcolor{gray!15} \multicolumn{8}{c}{\textbf{C2: Nonorthogonal private directions ($\rho=0.2$)}} \\
\multirow{3}{*}{\textbf{S1}} & 100 & 99.8 & 99.9 & 0.1 & 90.1 & 95.0 & 1.8 \\
 & 300 & 100 & 100 & 0.0 & 99.7 & 99.9 & 0.1 \\
 & 500 & 100 & 100 & 0.0 & 100 & 100 & 0.0 \\
\midrule
\addlinespace[2pt]
\multirow{3}{*}{\textbf{S2}} & 100 & 96.2 & 98.1 & 0.0 & 67.9 & 83.5 & 1.7 \\
 & 300 & 99.8 & 99.9 & 0.0 & 97.1 & 98.5 & 0.1 \\
 & 500 & 100 & 100 & 0.0 & 99.8 & 99.9 & 0.0 \\
\midrule
\addlinespace[2pt]
\multirow{3}{*}{\textbf{S3}} & 100 & 73.8 & 86.9 & 0.1 & 27.5 & 63.7 & 0.3 \\
 & 300 & 90.6 & 95.3 & 0.0 & 69.4 & 84.7 & 0.0 \\
 & 500 & 96.2 & 98.1 & 0.0 & 85.5 & 92.8 & 0.0 \\
\midrule
\bottomrule
\end{tabular}
}
\end{table}

{Tables~\ref{tab:simulation3-rank-combined} and~\ref{tab:simulation3-rank-combined-nonorth} show the clearest advantage of FE under weak signals. At $p/n=1.20$, S3, and $p=100$, FE estimates the common rank as $1.85$ in C1 and $1.84$ in C2, with zero observed over-estimation; APM, CP, and GCC each estimate $1.00$ in both settings. FE also has zero observed over-estimation throughout the tables, while its mean estimate remains between $1.84$ and $2.00$ across all reported settings.}

\begin{table}[!htb]
\centering
\caption{\textcolor{black}{Common-rank estimation with orthogonal private directions (C1), $K=10$, $r_k=3$, and true common rank $r_{\mathrm{s}}=2$. Observations are standard Gaussian. Tuning parameters $(a,\varepsilon_0)$ are $(10,0.1)$ for S1--S2 and $(6,0.1)$ for S3. Each entry reports the mean estimate, over-estimation rate, and under-estimation rate, in that order, over 1,000 repetitions.}}
\label{tab:simulation3-rank-combined}
\fontsize{10.5}{11.5}\selectfont
\setlength{\tabcolsep}{2pt}
\renewcommand{\arraystretch}{1.20}
\resizebox{\textwidth}{!}{%
\begin{tabular}{@{}l|ccc|ccc|ccc@{}}
\toprule
\rowcolor{gray!8} \multirow{2}{*}{\textbf{Method}} & \multicolumn{3}{c|}{\textbf{S1C1}} & \multicolumn{3}{c|}{\textbf{S2C1}} & \multicolumn{3}{c}{\textbf{S3C1}} \\
\cmidrule(lr){2-4} \cmidrule(lr){5-7} \cmidrule(lr){8-10}
\rowcolor{gray!8} & $p=50$ & $p=100$ & $p=200$ & $p=50$ & $p=100$ & $p=200$ & $p=50$ & $p=100$ & $p=200$ \\
\midrule
\rowcolor{gray!15} \multicolumn{10}{c}{\textbf{$p/n=0.80$}} \\
\midrule
\rowcolor{gray!4} APM & \shortstack{1.99\\(0.00, 0.01)} & \shortstack{1.87\\(0.00, 0.13)} & \shortstack{1.03\\(0.00, 0.97)} & \shortstack{1.31\\(0.00, 0.69)} & \shortstack{1.00\\(0.00, 1.00)} & \shortstack{1.00\\(0.00, 1.00)} & \shortstack{1.00\\(0.00, 1.00)} & \shortstack{1.00\\(0.00, 1.00)} & \shortstack{1.00\\(0.00, 1.00)} \\[4pt]
CP & \shortstack{2.00\\(0.00, 0.00)} & \shortstack{2.00\\(0.00, 0.00)} & \shortstack{2.00\\(0.00, 0.00)} & \shortstack{2.00\\(0.00, 0.00)} & \shortstack{1.99\\(0.00, 0.01)} & \shortstack{1.67\\(0.00, 0.33)} & \shortstack{1.84\\(0.00, 0.15)} & \shortstack{1.02\\(0.00, 0.98)} & \shortstack{1.00\\(0.00, 1.00)} \\[4pt]
GCC & \shortstack{2.00\\(0.00, 0.00)} & \shortstack{2.00\\(0.00, 0.00)} & \shortstack{2.00\\(0.00, 0.00)} & \shortstack{1.87\\(0.00, 0.13)} & \shortstack{1.73\\(0.00, 0.27)} & \shortstack{1.63\\(0.00, 0.37)} & \shortstack{1.40\\(0.00, 0.60)} & \shortstack{1.07\\(0.00, 0.93)} & \shortstack{1.00\\(0.00, 1.00)} \\[4pt]
\rowcolor{highlightColor} \textbf{FE} & \shortstack{2.00\\(0.00, 0.00)} & \shortstack{2.00\\(0.00, 0.00)} & \shortstack{2.00\\(0.00, 0.00)} & \shortstack{1.99\\(0.00, 0.01)} & \shortstack{2.00\\(0.00, 0.00)} & \shortstack{2.00\\(0.00, 0.00)} & \shortstack{1.97\\(0.00, 0.03)} & \shortstack{1.98\\(0.00, 0.02)} & \shortstack{2.00\\(0.00, 0.01)} \\[4pt]
\midrule
\rowcolor{gray!15} \multicolumn{10}{c}{\textbf{$p/n=1.00$}} \\
\midrule
\rowcolor{gray!4} APM & \shortstack{1.89\\(0.00, 0.11)} & \shortstack{1.43\\(0.00, 0.57)} & \shortstack{1.00\\(0.00, 1.00)} & \shortstack{1.09\\(0.00, 0.91)} & \shortstack{1.00\\(0.00, 1.00)} & \shortstack{1.00\\(0.00, 1.00)} & \shortstack{1.00\\(0.00, 1.00)} & \shortstack{1.00\\(0.00, 1.00)} & \shortstack{1.00\\(0.00, 1.00)} \\[4pt]
CP & \shortstack{2.00\\(0.00, 0.00)} & \shortstack{2.00\\(0.00, 0.00)} & \shortstack{1.91\\(0.00, 0.10)} & \shortstack{1.98\\(0.00, 0.02)} & \shortstack{1.59\\(0.00, 0.41)} & \shortstack{1.01\\(0.00, 0.99)} & \shortstack{1.47\\(0.00, 0.53)} & \shortstack{1.00\\(0.00, 1.00)} & \shortstack{1.00\\(0.00, 1.00)} \\[4pt]
GCC & \shortstack{2.00\\(0.00, 0.00)} & \shortstack{2.00\\(0.00, 0.00)} & \shortstack{2.00\\(0.00, 0.00)} & \shortstack{1.70\\(0.00, 0.30)} & \shortstack{1.39\\(0.00, 0.61)} & \shortstack{1.20\\(0.00, 0.80)} & \shortstack{1.21\\(0.00, 0.80)} & \shortstack{1.01\\(0.00, 0.99)} & \shortstack{1.00\\(0.00, 1.00)} \\[4pt]
\rowcolor{highlightColor} \textbf{FE} & \shortstack{1.99\\(0.00, 0.01)} & \shortstack{2.00\\(0.00, 0.00)} & \shortstack{2.00\\(0.00, 0.00)} & \shortstack{1.96\\(0.00, 0.04)} & \shortstack{1.98\\(0.00, 0.02)} & \shortstack{2.00\\(0.00, 0.01)} & \shortstack{1.92\\(0.00, 0.08)} & \shortstack{1.93\\(0.00, 0.07)} & \shortstack{1.98\\(0.00, 0.03)} \\[4pt]
\midrule
\rowcolor{gray!15} \multicolumn{10}{c}{\textbf{$p/n=1.20$}} \\
\midrule
\rowcolor{gray!4} APM & \shortstack{1.45\\(0.00, 0.55)} & \shortstack{1.03\\(0.00, 0.97)} & \shortstack{1.00\\(0.00, 1.00)} & \shortstack{1.01\\(0.00, 0.99)} & \shortstack{1.00\\(0.00, 1.00)} & \shortstack{1.00\\(0.00, 1.00)} & \shortstack{1.00\\(0.00, 1.00)} & \shortstack{1.00\\(0.00, 1.00)} & \shortstack{1.00\\(0.00, 1.00)} \\[4pt]
CP & \shortstack{2.00\\(0.00, 0.00)} & \shortstack{1.93\\(0.00, 0.07)} & \shortstack{1.27\\(0.00, 0.73)} & \shortstack{1.94\\(0.00, 0.06)} & \shortstack{1.14\\(0.00, 0.86)} & \shortstack{1.00\\(0.00, 1.00)} & \shortstack{1.21\\(0.00, 0.79)} & \shortstack{1.00\\(0.00, 1.00)} & \shortstack{1.00\\(0.00, 1.00)} \\[4pt]
GCC & \shortstack{1.98\\(0.00, 0.02)} & \shortstack{2.00\\(0.00, 0.01)} & \shortstack{2.00\\(0.00, 0.00)} & \shortstack{1.50\\(0.00, 0.50)} & \shortstack{1.19\\(0.00, 0.81)} & \shortstack{1.04\\(0.00, 0.96)} & \shortstack{1.11\\(0.00, 0.89)} & \shortstack{1.00\\(0.00, 0.99)} & \shortstack{1.00\\(0.00, 1.00)} \\[4pt]
\rowcolor{highlightColor} \textbf{FE} & \shortstack{1.97\\(0.00, 0.03)} & \shortstack{1.99\\(0.00, 0.01)} & \shortstack{2.00\\(0.00, 0.00)} & \shortstack{1.86\\(0.00, 0.14)} & \shortstack{1.92\\(0.00, 0.08)} & \shortstack{1.99\\(0.00, 0.01)} & \shortstack{1.86\\(0.00, 0.14)} & \shortstack{1.85\\(0.00, 0.15)} & \shortstack{1.92\\(0.00, 0.08)} \\[4pt]
\midrule
\bottomrule
\end{tabular}
}
\end{table}

\begin{table}[!htb]
\centering
\caption{\textcolor{black}{Common-rank estimation with nonorthogonal private directions (C2), whose pairwise inner products are $0.2$. We use $K=10$, $r_k=3$, true common rank $r_{\mathrm{s}}=2$, and standard Gaussian observations. Tuning parameters $(a,\varepsilon_0)$ are $(10,0.1)$ for S1--S2 and $(6,0.1)$ for S3. Each entry reports the mean estimate, over-estimation rate, and under-estimation rate, in that order, over 1,000 repetitions.}}
\label{tab:simulation3-rank-combined-nonorth}
\fontsize{10.5}{11.5}\selectfont
\setlength{\tabcolsep}{2pt}
\renewcommand{\arraystretch}{1.20}
\resizebox{\textwidth}{!}{%
\begin{tabular}{@{}l|ccc|ccc|ccc@{}}
\toprule
\rowcolor{gray!8} \multirow{2}{*}{\textbf{Method}} & \multicolumn{3}{c|}{\textbf{S1C2}} & \multicolumn{3}{c|}{\textbf{S2C2}} & \multicolumn{3}{c}{\textbf{S3C2}} \\
\cmidrule(lr){2-4} \cmidrule(lr){5-7} \cmidrule(lr){8-10}
\rowcolor{gray!8} & $p=50$ & $p=100$ & $p=200$ & $p=50$ & $p=100$ & $p=200$ & $p=50$ & $p=100$ & $p=200$ \\
\midrule
\rowcolor{gray!15} \multicolumn{10}{c}{\textbf{$p/n=0.80$}} \\
\midrule
\rowcolor{gray!4} APM & \shortstack{1.99\\(0.00, 0.01)} & \shortstack{1.87\\(0.00, 0.13)} & \shortstack{1.03\\(0.00, 0.97)} & \shortstack{1.31\\(0.00, 0.69)} & \shortstack{1.00\\(0.00, 1.00)} & \shortstack{1.00\\(0.00, 1.00)} & \shortstack{1.00\\(0.00, 1.00)} & \shortstack{1.00\\(0.00, 1.00)} & \shortstack{1.00\\(0.00, 1.00)} \\[4pt]
CP & \shortstack{2.00\\(0.00, 0.00)} & \shortstack{2.00\\(0.00, 0.00)} & \shortstack{2.00\\(0.00, 0.00)} & \shortstack{2.00\\(0.00, 0.00)} & \shortstack{1.99\\(0.00, 0.01)} & \shortstack{1.67\\(0.00, 0.34)} & \shortstack{1.85\\(0.00, 0.15)} & \shortstack{1.02\\(0.00, 0.98)} & \shortstack{1.00\\(0.00, 1.00)} \\[4pt]
GCC & \shortstack{2.00\\(0.00, 0.00)} & \shortstack{2.00\\(0.00, 0.00)} & \shortstack{2.00\\(0.00, 0.00)} & \shortstack{1.73\\(0.00, 0.27)} & \shortstack{1.41\\(0.00, 0.59)} & \shortstack{1.14\\(0.00, 0.86)} & \shortstack{1.23\\(0.00, 0.77)} & \shortstack{1.00\\(0.00, 0.99)} & \shortstack{1.00\\(0.00, 1.00)} \\[4pt]
\rowcolor{highlightColor} \textbf{FE} & \shortstack{2.00\\(0.00, 0.00)} & \shortstack{2.00\\(0.00, 0.00)} & \shortstack{2.00\\(0.00, 0.00)} & \shortstack{1.99\\(0.00, 0.01)} & \shortstack{2.00\\(0.00, 0.00)} & \shortstack{2.00\\(0.00, 0.00)} & \shortstack{1.97\\(0.00, 0.03)} & \shortstack{1.98\\(0.00, 0.02)} & \shortstack{2.00\\(0.00, 0.00)} \\[4pt]
\midrule
\rowcolor{gray!15} \multicolumn{10}{c}{\textbf{$p/n=1.00$}} \\
\midrule
\rowcolor{gray!4} APM & \shortstack{1.89\\(0.00, 0.11)} & \shortstack{1.43\\(0.00, 0.57)} & \shortstack{1.00\\(0.00, 1.00)} & \shortstack{1.09\\(0.00, 0.91)} & \shortstack{1.00\\(0.00, 1.00)} & \shortstack{1.00\\(0.00, 1.00)} & \shortstack{1.00\\(0.00, 1.00)} & \shortstack{1.00\\(0.00, 1.00)} & \shortstack{1.00\\(0.00, 1.00)} \\[4pt]
CP & \shortstack{2.00\\(0.00, 0.00)} & \shortstack{2.00\\(0.00, 0.00)} & \shortstack{1.91\\(0.00, 0.10)} & \shortstack{1.99\\(0.00, 0.01)} & \shortstack{1.60\\(0.00, 0.40)} & \shortstack{1.01\\(0.00, 0.99)} & \shortstack{1.47\\(0.00, 0.53)} & \shortstack{1.00\\(0.00, 1.00)} & \shortstack{1.00\\(0.00, 1.00)} \\[4pt]
GCC & \shortstack{1.99\\(0.00, 0.01)} & \shortstack{2.00\\(0.00, 0.00)} & \shortstack{2.00\\(0.00, 0.00)} & \shortstack{1.52\\(0.00, 0.47)} & \shortstack{1.15\\(0.00, 0.85)} & \shortstack{1.02\\(0.00, 0.98)} & \shortstack{1.10\\(0.00, 0.90)} & \shortstack{1.00\\(0.00, 1.00)} & \shortstack{1.00\\(0.00, 1.00)} \\[4pt]
\rowcolor{highlightColor} \textbf{FE} & \shortstack{1.99\\(0.00, 0.01)} & \shortstack{2.00\\(0.00, 0.00)} & \shortstack{2.00\\(0.00, 0.00)} & \shortstack{1.95\\(0.00, 0.05)} & \shortstack{1.98\\(0.00, 0.02)} & \shortstack{2.00\\(0.00, 0.01)} & \shortstack{1.93\\(0.00, 0.07)} & \shortstack{1.93\\(0.00, 0.07)} & \shortstack{1.98\\(0.00, 0.03)} \\[4pt]
\midrule
\rowcolor{gray!15} \multicolumn{10}{c}{\textbf{$p/n=1.20$}} \\
\midrule
\rowcolor{gray!4} APM & \shortstack{1.45\\(0.00, 0.55)} & \shortstack{1.03\\(0.00, 0.97)} & \shortstack{1.00\\(0.00, 1.00)} & \shortstack{1.01\\(0.00, 0.99)} & \shortstack{1.00\\(0.00, 1.00)} & \shortstack{1.00\\(0.00, 1.00)} & \shortstack{1.00\\(0.00, 1.00)} & \shortstack{1.00\\(0.00, 1.00)} & \shortstack{1.00\\(0.00, 1.00)} \\[4pt]
CP & \shortstack{2.00\\(0.00, 0.00)} & \shortstack{1.93\\(0.00, 0.07)} & \shortstack{1.27\\(0.00, 0.73)} & \shortstack{1.93\\(0.00, 0.07)} & \shortstack{1.14\\(0.00, 0.86)} & \shortstack{1.00\\(0.00, 1.00)} & \shortstack{1.21\\(0.00, 0.79)} & \shortstack{1.00\\(0.00, 1.00)} & \shortstack{1.00\\(0.00, 1.00)} \\[4pt]
GCC & \shortstack{1.96\\(0.00, 0.04)} & \shortstack{1.96\\(0.00, 0.04)} & \shortstack{1.98\\(0.00, 0.02)} & \shortstack{1.32\\(0.00, 0.68)} & \shortstack{1.06\\(0.00, 0.94)} & \shortstack{1.00\\(0.00, 1.00)} & \shortstack{1.05\\(0.00, 0.95)} & \shortstack{1.00\\(0.00, 1.00)} & \shortstack{1.00\\(0.00, 1.00)} \\[4pt]
\rowcolor{highlightColor} \textbf{FE} & \shortstack{1.97\\(0.00, 0.03)} & \shortstack{2.00\\(0.00, 0.00)} & \shortstack{2.00\\(0.00, 0.00)} & \shortstack{1.87\\(0.00, 0.13)} & \shortstack{1.92\\(0.00, 0.08)} & \shortstack{1.99\\(0.00, 0.01)} & \shortstack{1.87\\(0.00, 0.13)} & \shortstack{1.84\\(0.00, 0.16)} & \shortstack{1.92\\(0.00, 0.08)} \\[4pt]
\midrule
\bottomrule
\end{tabular}
}
\end{table}
\renewcommand{\arraystretch}{1}

\renewcommand{\arraystretch}{1}

Table~\ref{tab:simulation3-subspace-error} shows that FE recovers both common and private spaces substantially better than APM, CP, and GCC. FE's Total error decreases from $0.793$ to $0.752$ as $p$ increases in the orthogonal case and is nearly identical in the nonorthogonal case. It is close to KT(r) and PerPCA(r), although PerPCA(r) is slightly better at some dimensions because it is supplied with the true common rank. Relative to APM/GCC and CP, FE reduces Total error by roughly $70\%$--$72\%$. FE obtains this performance while estimating the ranks from the data.

\begin{table}[!htb]
\centering
\caption{\textcolor{black}{Subspace recovery under the weak-signal setting $(\lambda_1,\lambda_2,\lambda_3)=(15,5,2.5)$. Global Error and Local Error report $\|\widehat{\mathbf P}^{\mathrm{s}}-\mathbf P^{\mathrm{s}}\|_F^2$ and $K^{-1}\sum_{k=1}^K\|\widehat{\mathbf P}_k^{\mathrm p}-\mathbf P_k^{\mathrm p}\|_F^2$, respectively; Total Error is their sum. Here $r_k=3$ is the total number of spike features in each source, $r_{\mathrm{s}}=2$ is the shared rank, and $r_k-r_{\mathrm{s}}=1$ is the private rank. Entries are averages over 1000 repetitions; smaller is better.}}
\label{tab:simulation3-subspace-error}
\footnotesize
\renewcommand{\arraystretch}{0.92}
\setlength{\tabcolsep}{4pt}
\begin{tabular*}{\textwidth}{@{\extracolsep{\fill}}l*{9}{r}@{}}
\toprule
\multicolumn{10}{@{}l}{\textbf{S3C1 (Orthogonal)}} \\
\cmidrule(lr){1-10}
\textbf{Method} & \multicolumn{3}{c}{\textbf{Global Error}} & \multicolumn{3}{c}{\textbf{Local Error}} & \multicolumn{3}{c}{\textbf{Total Error}} \\
\cmidrule(lr){2-4} \cmidrule(lr){5-7} \cmidrule(lr){8-10}
& $p=200$ & $500$ & $800$ & $p=200$ & $500$ & $800$ & $p=200$ & $500$ & $800$ \\
\midrule
APM & 1.0212 & 1.0222 & 1.0224 & 1.5874 & 1.5919 & 1.6086 & 2.6086 & 2.6141 & 2.6310 \\
CP & 1.1198 & 1.1223 & 1.1228 & 1.5991 & 1.5995 & 1.6119 & 2.7189 & 2.7218 & 2.7348 \\
GCC & 1.0197 & 1.0206 & 1.0208 & 1.5873 & 1.5919 & 1.6086 & 2.6070 & 2.6125 & 2.6294 \\
KT & 0.1297 & 0.1414 & 0.1431 & 0.5860 & 0.5911 & 0.6084 & 0.7157 & 0.7324 & 0.7515 \\
PerPCA & 0.0999 & 0.1075 & 0.1094 & 0.6142 & 0.6522 & 0.6966 & 0.7142 & 0.7597 & 0.8059 \\
\rowcolor{highlightColor} \textbf{FE} & 0.1671 & 0.1602 & 0.1436 & 0.6261 & 0.6112 & 0.6084 & 0.7933 & 0.7714 & 0.7520 \\
\midrule
\addlinespace
\multicolumn{10}{@{}l}{\textbf{S3C2 (Nonorthogonal)}} \\
\cmidrule(lr){1-10}
\textbf{Method} & \multicolumn{3}{c}{\textbf{Global Error}} & \multicolumn{3}{c}{\textbf{Local Error}} & \multicolumn{3}{c}{\textbf{Total Error}} \\
\cmidrule(lr){2-4} \cmidrule(lr){5-7} \cmidrule(lr){8-10}
& $p=200$ & $500$ & $800$ & $p=200$ & $500$ & $800$ & $p=200$ & $500$ & $800$ \\
\midrule
APM & 1.0212 & 1.0222 & 1.0224 & 1.5872 & 1.5919 & 1.6087 & 2.6084 & 2.6142 & 2.6311 \\
CP & 1.1201 & 1.1223 & 1.1228 & 1.6016 & 1.5992 & 1.6122 & 2.7217 & 2.7215 & 2.7350 \\
GCC & 1.0197 & 1.0206 & 1.0208 & 1.5871 & 1.5919 & 1.6087 & 2.6068 & 2.6125 & 2.6294 \\
KT & 0.1298 & 0.1415 & 0.1431 & 0.5859 & 0.5912 & 0.6086 & 0.7157 & 0.7326 & 0.7517 \\
PerPCA & 0.1001 & 0.1077 & 0.1094 & 0.6228 & 0.6514 & 0.6813 & 0.7228 & 0.7591 & 0.7907 \\
\rowcolor{highlightColor} \textbf{FE} & 0.1673 & 0.1602 & 0.1435 & 0.6260 & 0.6112 & 0.6085 & 0.7934 & 0.7714 & 0.7521 \\
\midrule
\bottomrule
\end{tabular*}
\end{table}

\subsection{Visual Illustration of Multi-Source Heterogeneous PCA}
We consider four grayscale images containing different icons (triangle, disk, cross, and cloud) over a shared background. Let $\bX_k\in\mathbb{R}^{p\times p}$ denote the image matrix for source $k$, which can be written as
\[
\bX_k=\bX^{\mathrm{background}}+\bX_k^{\mathrm{icon}},\qquad k=1,\ldots,4.
\]
The background component is shared across sources, whereas $X_k^{\mathrm{icon}}$ represents the source-specific icon.

\textbf{Baselines}. We compare FE with the APM of \citet{hu2025aggregated}, CP of \citet{chen2022circularly}, GCC of \citet{lin2023generalised}, PerPCA of \citet{shi2024personalized}, and KT of \citet{li2024knowledgetransfermultipleprincipal}. {For PerPCA and KT, the suffix (r) uses a fixed common rank of $3$ with ESC-estimated total ranks; (e) uses the common rank estimated by FE and the same ESC-estimated total ranks.} For each method, the common and source-specific projection matrices are estimated from the four image matrices, and the resulting global and total reconstructions are evaluated using the same image scale.

\textbf{Metric.}
We use the same notation as in Section~\ref{subsec:simulationcommonfeature}: $\widehat{\mathbf P}^{\mathrm{s}}$ denotes the estimated common projection matrix and $\widehat{\mathbf P}_k^{\mathrm{p}}$ the estimated source-specific projection matrix for source $k$. The global reconstruction is
\[
\widehat{\mathbf X}_k^{\mathrm{global}}=\widehat{\mathbf P}^{\mathrm{s}}\mathbf X_k,
\]
whereas the global-plus-local reconstruction is denoted by
\[
\widehat{\mathbf X}_k^{\mathrm{total}}=\left(\widehat{\mathbf P}^{\mathrm{s}}+\widehat{\mathbf P}_k^{\mathrm{p}}\right)\mathbf X_k.
\]
The two reconstruction errors are computed against their corresponding target components: the gloabl error (GE) is
\[
\operatorname{GE}
 =\frac{1}{4}\sum_{k=1}^{4}
 \frac{1}{p^2}\left\|\widehat{\mathbf X}_k^{\mathrm{global}}-\mathbf X^{\mathrm{background}}\right\|_F^2,
\]
The reconstruction error (RE) is
\[
\operatorname{RE}
 =\frac{1}{4}\sum_{k=1}^{4}
 \frac{1}{p^2}\left\|\widehat{\mathbf X}_k^{\mathrm{total}}-\mathbf X_k\right\|_F^2.
\]
Here $\mathbf X_k=\mathbf X^{\mathrm{background}}+\mathbf X_k^{\mathrm{icon}}$. 

\begin{table}[htbp]
\centering
\caption{\textcolor{black}{Reconstruction errors for the four $400\times400$ images. Global Error reports $\operatorname{MSE}_{\mathrm{global}}$ for the shared background; Total Error reports $\operatorname{MSE}_{\mathrm{total}}$ for reconstruction using both common and private projectors. FE uses $a=7$ and $\varepsilon_0=0.01$ to estimate the common rank. ESC estimates total ranks $(12,10,5,9)$ using $a=6$ and $\varepsilon_0=0.1$. PerPCA(r) and KT(r) use common rank $3$; PerPCA(e) and KT(e) use the FE-estimated common rank $2$.}}
\label{tab:table4-summary}
\small
\setlength{\tabcolsep}{4pt}
\renewcommand{\arraystretch}{1.18}
\begin{tabular}{@{}lrrrrrrrr@{}}
\toprule
\rowcolor{gray!15} \textbf{Methods} & \textbf{FE} & \textbf{APM} & \textbf{CP} & \textbf{GCC} & \textbf{PerPCA(r)} & \textbf{KT(r)} & \textbf{PerPCA(e)} & \textbf{KT(e)} \\
\midrule
\rowcolor{gray!4} \textcolor{black}{GE} & 0.1323 & 0.1296 & 0.1431 & 0.1296 & 0.1421 & 0.1390 & 0.1308 & 0.1296 \\
\textcolor{black}{RE} & 0.0033 & 0.0032 & 0.0042 & 0.0032 & 0.0051 & 0.0033 & 0.0050 & 0.0032 \\
\bottomrule
\end{tabular}
\end{table}
\renewcommand{\arraystretch}{1}

Table~\ref{tab:table4-summary} shows how the common-rank choice affects reconstruction. ESC estimates total ranks $(12,10,5,9)$, while FE estimates a common rank of $2$. With the common rank fixed at $3$, PerPCA(r) and KT(r) have Global/Total MSEs $(0.1421,0.0051)$ and $(0.1390,0.0033)$, respectively. Using the FE-estimated common rank, PerPCA(e) has $(0.1308,0.0050)$ and KT(e) has $(0.1296,0.0032)$. The estimated common rank lowers both global and total error for these two methods. FE's own Global/Total MSEs are $(0.1323,0.0033)$.

\begin{figure}[htbp]
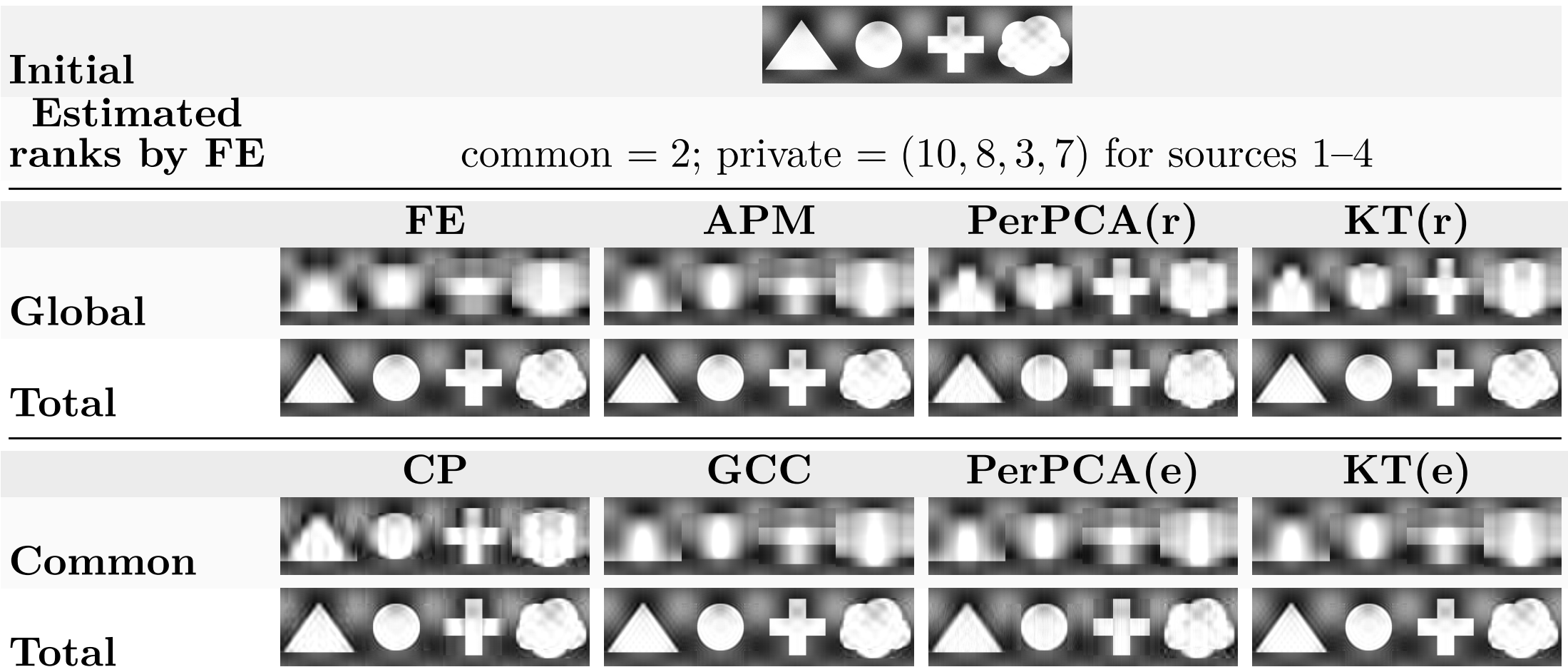

\centering
% Highlight revisions made in the current editing pass.
\begingroup\color{black}
\setlength{\tabcolsep}{2pt}
\renewcommand{\arraystretch}{0.9}
\newcommand{\fourimages}[2]{%
  \includegraphics[width=0.047\textwidth]{#1/#2-1.png}%
  \includegraphics[width=0.047\textwidth]{#1/#2-2.png}%
  \includegraphics[width=0.047\textwidth]{#1/#2-3.png}%
  \includegraphics[width=0.047\textwidth]{#1/#2-4.png}}
\begin{tabular}{@{}lcccc@{}}
\rowcolor{gray!10} \shortstack{\textbf{Initial}} & \multicolumn{4}{c}{\fourimages{Plots/Table4-400-pathfe-a7-ranka6-e0p1-fixed-common-3}{original}} \\
\rowcolor{gray!4} \shortstack{\textbf{Estimated}\\\textbf{ranks by FE}} & \multicolumn{4}{c}{common $=2$; private $=(10,8,3,7)$ for sources $1$--$4$} \\
\midrule
\rowcolor{gray!15} & \textbf{FE} & \textbf{APM} & \textbf{PerPCA(r)} & \textbf{KT(r)} \\
\rowcolor{gray!4} \textbf{Global} & \fourimages{Plots/Table4-400-pathfe-a7-ranka6-e0p1-fixed-common-3}{pathfe-global} & \fourimages{Plots/Table4-400-pathfe-a7-ranka6-e0p1-fixed-common-3}{apm-global} & \fourimages{Plots/Table4-400-pathfe-a7-ranka6-e0p1-fixed-common-3}{ppca-global} & \fourimages{Plots/Table4-400-pathfe-a7-ranka6-e0p1-fixed-common-3}{kt-global} \\
\textbf{Total} & \fourimages{Plots/Table4-400-pathfe-a7-ranka6-e0p1-fixed-common-3}{pathfe-global_plus_local} & \fourimages{Plots/Table4-400-pathfe-a7-ranka6-e0p1-fixed-common-3}{apm-global_plus_local} & \fourimages{Plots/Table4-400-pathfe-a7-ranka6-e0p1-fixed-common-3}{ppca-global_plus_local} & \fourimages{Plots/Table4-400-pathfe-a7-ranka6-e0p1-fixed-common-3}{kt-global_plus_local} \\
\midrule
\rowcolor{gray!15} & \textbf{CP} & \textbf{GCC} & \textbf{PerPCA(e)} & \textbf{KT(e)} \\
\rowcolor{gray!4} \textbf{Common} & \fourimages{Plots/Table4-400-pathfe-a7-ranka6-e0p1-estimated-common}{cp-global} & \fourimages{Plots/Table4-400-pathfe-a7-ranka6-e0p1-estimated-common}{gcc-global} & \fourimages{Plots/Table4-400-pathfe-a7-ranka6-e0p1-estimated-common}{ppca-global} & \fourimages{Plots/Table4-400-pathfe-a7-ranka6-e0p1-estimated-common}{kt-global} \\
\textbf{Total} & \fourimages{Plots/Table4-400-pathfe-a7-ranka6-e0p1-estimated-common}{cp-global_plus_local} & \fourimages{Plots/Table4-400-pathfe-a7-ranka6-e0p1-estimated-common}{gcc-global_plus_local} & \fourimages{Plots/Table4-400-pathfe-a7-ranka6-e0p1-estimated-common}{ppca-global_plus_local} & \fourimages{Plots/Table4-400-pathfe-a7-ranka6-e0p1-estimated-common}{kt-global_plus_local}
\end{tabular}
\caption{Reconstructions of the four $400\times400$ images. Global shows $\widehat{\mathbf X}_k^{\mathrm{global}}=\widehat{\mathbf P}^{\mathrm{s}}\mathbf X_k$; Total shows $\widehat{\mathbf X}_k^{\mathrm{total}}=(\widehat{\mathbf P}^{\mathrm{s}}+\widehat{\mathbf P}_k^{\mathrm p})\mathbf X_k$. FE uses $a=7$ and $\varepsilon_0=0.01$ and estimates common rank $2$. ESC estimates total ranks $(12,10,5,9)$ using $a=6$ and $\varepsilon_0=0.1$, giving private ranks $(10,8,3,7)$ at common rank $2$. PerPCA(r) and KT(r) use common rank $3$; PerPCA(e) and KT(e) use FE's estimated common rank $2$.}
\label{fig:four-picture-fe-ppca}
\endgroup
\end{figure}
\renewcommand{\arraystretch}{1}

Figure~\ref{fig:four-picture-fe-ppca} compares the initial images with the global and global-plus-local reconstructions from all eight method variants. The (r) and (e) variants differ in their common-rank inputs: the former fix it at $3$, while the latter use the FE estimate of $2$; both use the same ESC-estimated total ranks.

Table~\ref{tab:table4-summary} quantifies the effect of this rank choice. In this run, using the estimated common rank reduces both global and total reconstruction errors for PerPCA and KT relative to fixing the common rank at $3$.

\section{Real Data Application}\label{sec:real-data}

{This section reports two real-data applications of FE. We first evaluate reconstruction on FEMNIST, then study global and state-specific variation in United States housing-price data. For each application, we describe the data construction, competing methods, rank inputs, and evaluation criteria before reporting the numerical results.}

\subsection{Federated Image Data}
{We apply our algorithm to FEMNIST (Caldas et al., 2019).} FEMNIST is a popular dataset in federated analytics. It consists of grayscale images of handwritten digits and English letters contributed by 3550 different writers. Each image has $28 \times 28 = 784$
pixels. Different writers have different writing styles. Thus, the dataset is inherently heterogeneous. Our task is to learn a few principal components (PCs) that can represent the dataset. On average, each client has 89 images. We represent each image as a vector in $\mathbb{R}^{784}$. {We use a 70\%/15\%/15\% training/validation/test split for the 20 clients.}

\iffalse % CIFAR-10 experiment temporarily omitted.
CIFAR10 is a multiclass image dataset consisting of 60000 images from 10 classes. To simulate a heterogeneous setting, we divide the CIFAR10 training and test sets into 20 parts, such that each part contains images from only two classes. We then assign these parts to 20 clients. This data partitioning scheme is consistent with the federated learning literature (McMahan et al., 2017). We then apply similar data preprocessing procedures to vectorize the images on each client and construct the dataset
$\left\{Y^{(i)}\right\}_{i=1}^{20}.$
\fi

\textbf{Baseline methods.} We compare FE with the same methods as in Section~\ref{subsec:simulationcommonfeature}: PerPCA, KT, APM, CP, and GCC. PerPCA and KT estimate common and client-specific principal spaces, whereas APM, CP, and GCC estimate global and local factor spaces. All methods are fitted on the training observations using the dimensions required by their respective procedures.

{\color{black}
\textbf{Metric.} The entries reported in {Table~\ref{tab:femnist-reconstruction}} are global-plus-local reconstruction errors. For each source $k$, a method estimates the common and private projection matrices $\widehat{\mathbf P}^{\mathrm{s}}$ and $\widehat{\mathbf P}_k^{\mathrm{p}}$. These are constructed from its orthonormal common and private bases, denoted by $\widehat{\mathbf U}^{\mathrm{s}}$ and $\widehat{\mathbf V}_k^{\mathrm p}$, as
\[
\widehat{\mathbf P}^{\mathrm{s}}
=\widehat{\mathbf U}^{\mathrm{s}}
\bigl(\widehat{\mathbf U}^{\mathrm{s}}\bigr)^{\top},
\qquad
\widehat{\mathbf P}_k^{\mathrm{p}}
=\widehat{\mathbf V}_k^{\mathrm p}
\bigl(\widehat{\mathbf V}_k^{\mathrm p}\bigr)^{\top}.
\]
For FE, the private basis is re-estimated from \eqref{eq:private-residual-covariance} according to \eqref{eq:private-residual-projector}, so it is orthogonal to the estimated common space. The other methods use their respective estimated private bases.

} % end revision marking
The global-plus-local reconstruction of $\mathbf X_k$ is
\[
\widehat{\mathbf X}_k
=\left(\widehat{\mathbf P}^{\mathrm{s}}
+\widehat{\mathbf P}_k^{\mathrm{p}}\right)\mathbf X_k.
\]
% The global-only and private-only reconstructions are
% \[
% \widehat{\mathbf X}_k^{\mathrm{global}}=\widehat{\mathbf P}^{\mathrm{s}}\mathbf X_k,
% \qquad
% \widehat{\mathbf X}_k^{\mathrm{local}}=\widehat{\mathbf P}_k^{\mathrm{p}}\mathbf X_k.
% \]
% For $a\in\{\mathrm{global},\mathrm{local},\mathrm{total}\}$, with
% $\widehat{\mathbf X}_k^{\mathrm{total}}=\widehat{\mathbf X}_k$, 
The reconstruction error(RE) is
\[
\mathrm{RE} = {\displaystyle\sum_{k=1}^{K}
\left\|\mathbf X_k-\widehat{\mathbf X}_k\right\|_F^2}/
{\displaystyle\sum_{k=1}^{K}n_k},
\]
with the matrices understood as either the training or test collection. Thus, ``global+local'' means that both projectors are used in one reconstruction; it is not the sum of two separate error metrics. Each method is fitted using only the training data. The Train column evaluates this formula on the fitted data, whereas the Test column evaluates the same estimated projection matrices on held-out observations without refitting.

\begin{figure*}[ht]
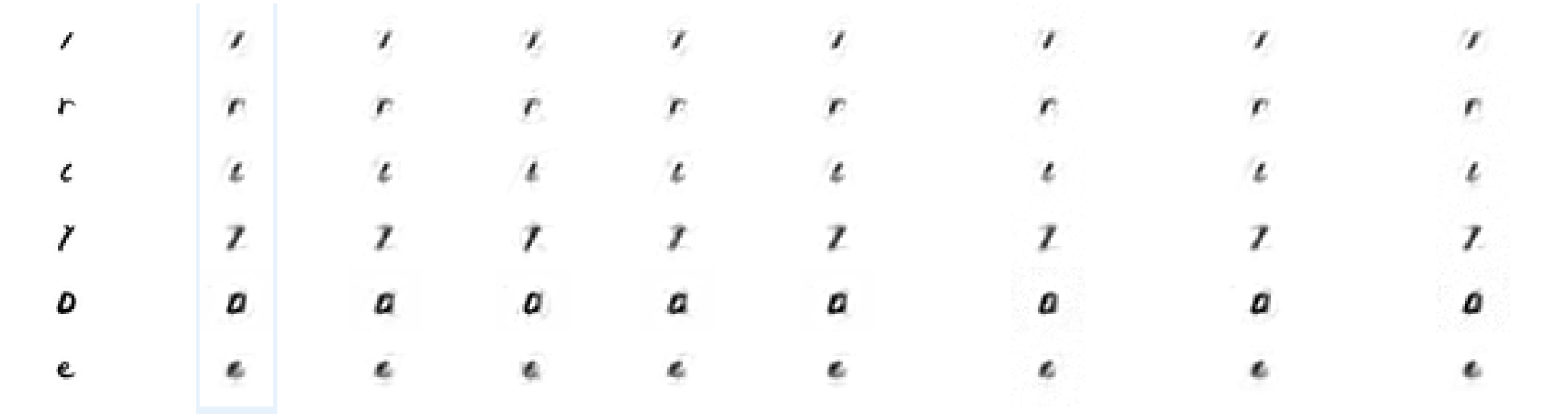

\centering
\footnotesize
\setlength{\tabcolsep}{1pt}
\newcommand{\FEMNISToriginal}[2]{\includegraphics[width=0.065\textwidth]{Plots/FEMNIST-a5p00-e0p01-n20/femnist/original-c#1-i#2.png}}
\newcommand{\FEMNISTreconstruction}[4]{\includegraphics[width=0.065\textwidth]{Plots/FEMNIST-a5p00-e0p01-n20/femnist/#1-#2-c#3-i#4.png}}
\renewcommand{\arraystretch}{0.9}%
\begin{tabular*}{\textwidth}{@{\extracolsep{\fill}}c>{\columncolor{highlightColor}}c*{7}{c}@{}}
\toprule
\rowcolor{gray!15} \textbf{Original} & \cellcolor{highlightColor}\textbf{FE} & \textbf{APM} & \textbf{CP} & \textbf{GCC} & \textbf{KT(r)} & \textbf{PerPCA(r)} & \textbf{KT(e)} & \textbf{PerPCA(e)} \\
% Removed: separate common-component reconstructions.
\rowcolor{gray!4} {\textbf{$r_{\mathrm{s}}$}} & \cellcolor{highlightColor}{10} & {3} & {9} & {1} & {5} & {5} & {10} & {10} \\
{\textbf{$r_k-r_{\mathrm{s}}$}} & \cellcolor{highlightColor}{3--7} & {10--14} & {4--8} & {12--16} & {8--12} & {8--12} & {3--7} & {3--7} \\
\midrule
\FEMNISToriginal{018}{000} & \FEMNISTreconstruction{pathfe}{global_plus_local}{018}{000} & \FEMNISTreconstruction{apm}{global_plus_local}{018}{000} & \FEMNISTreconstruction{cp}{global_plus_local}{018}{000} & \FEMNISTreconstruction{gcc}{global_plus_local}{018}{000} & \FEMNISTreconstruction{kt-fixed-5}{global_plus_local}{018}{000} & \FEMNISTreconstruction{ppca-fixed-5}{global_plus_local}{018}{000} & \FEMNISTreconstruction{kt}{global_plus_local}{018}{000} & \FEMNISTreconstruction{ppca}{global_plus_local}{018}{000} \\[-5pt]
\FEMNISToriginal{000}{001} & \FEMNISTreconstruction{pathfe}{global_plus_local}{000}{001} & \FEMNISTreconstruction{apm}{global_plus_local}{000}{001} & \FEMNISTreconstruction{cp}{global_plus_local}{000}{001} & \FEMNISTreconstruction{gcc}{global_plus_local}{000}{001} & \FEMNISTreconstruction{kt-fixed-5}{global_plus_local}{000}{001} & \FEMNISTreconstruction{ppca-fixed-5}{global_plus_local}{000}{001} & \FEMNISTreconstruction{kt}{global_plus_local}{000}{001} & \FEMNISTreconstruction{ppca}{global_plus_local}{000}{001} \\[-5pt]
\FEMNISToriginal{002}{001} & \FEMNISTreconstruction{pathfe}{global_plus_local}{002}{001} & \FEMNISTreconstruction{apm}{global_plus_local}{002}{001} & \FEMNISTreconstruction{cp}{global_plus_local}{002}{001} & \FEMNISTreconstruction{gcc}{global_plus_local}{002}{001} & \FEMNISTreconstruction{kt-fixed-5}{global_plus_local}{002}{001} & \FEMNISTreconstruction{ppca-fixed-5}{global_plus_local}{002}{001} & \FEMNISTreconstruction{kt}{global_plus_local}{002}{001} & \FEMNISTreconstruction{ppca}{global_plus_local}{002}{001} \\[-5pt]
\FEMNISToriginal{005}{000} & \FEMNISTreconstruction{pathfe}{global_plus_local}{005}{000} & \FEMNISTreconstruction{apm}{global_plus_local}{005}{000} & \FEMNISTreconstruction{cp}{global_plus_local}{005}{000} & \FEMNISTreconstruction{gcc}{global_plus_local}{005}{000} & \FEMNISTreconstruction{kt-fixed-5}{global_plus_local}{005}{000} & \FEMNISTreconstruction{ppca-fixed-5}{global_plus_local}{005}{000} & \FEMNISTreconstruction{kt}{global_plus_local}{005}{000} & \FEMNISTreconstruction{ppca}{global_plus_local}{005}{000} \\[-5pt]
\FEMNISToriginal{010}{001} & \FEMNISTreconstruction{pathfe}{global_plus_local}{010}{001} & \FEMNISTreconstruction{apm}{global_plus_local}{010}{001} & \FEMNISTreconstruction{cp}{global_plus_local}{010}{001} & \FEMNISTreconstruction{gcc}{global_plus_local}{010}{001} & \FEMNISTreconstruction{kt-fixed-5}{global_plus_local}{010}{001} & \FEMNISTreconstruction{ppca-fixed-5}{global_plus_local}{010}{001} & \FEMNISTreconstruction{kt}{global_plus_local}{010}{001} & \FEMNISTreconstruction{ppca}{global_plus_local}{010}{001} \\[-5pt]
\FEMNISToriginal{003}{000} & \FEMNISTreconstruction{pathfe}{global_plus_local}{003}{000} & \FEMNISTreconstruction{apm}{global_plus_local}{003}{000} & \FEMNISTreconstruction{cp}{global_plus_local}{003}{000} & \FEMNISTreconstruction{gcc}{global_plus_local}{003}{000} & \FEMNISTreconstruction{kt-fixed-5}{global_plus_local}{003}{000} & \FEMNISTreconstruction{ppca-fixed-5}{global_plus_local}{003}{000} & \FEMNISTreconstruction{kt}{global_plus_local}{003}{000} & \FEMNISTreconstruction{ppca}{global_plus_local}{003}{000} \\
\bottomrule
\end{tabular*}
\caption{{Full reconstructions of six FEMNIST test images from clients 18, 0, 2, 5, 10, and 3, selected for low FE full-reconstruction image MSE and distinct character shapes. FE uses $a=5$ and $\varepsilon_0=0.01$ for these reconstructions. The FE-estimated total ranks $r_k$ are $(16,15,16,13,17,14)$. The rank rows report each method's common rank $r_{\mathrm{s}}$ and private-rank range $r_k-r_{\mathrm{s}}$ across these six clients.}}
\label{fig:femnist-reconstruction-comparison}
\end{figure*}

{Figure~\ref{fig:femnist-reconstruction-comparison} compares full reconstructions of six selected FEMNIST test images from clients 18, 0, 2, 5, 10, and 3, ordered by increasing FE full-reconstruction image MSE in the new run. Each method uses both the estimated shared and client-specific projections.}

\begin{table}[ht]
\centering
\caption{{FEMNIST global-plus-local reconstruction error, $\sum_k\|(\widehat{\mathbf P}^{\mathrm{s}}+\widehat{\mathbf P}_k^{\mathrm p})\mathbf X_k-\mathbf X_k\|_F^2/\sum_k n_k$, on the training and test sets. FE uses $a=5$ and $\varepsilon_0=0.01$; FE rank estimation uses ESC with $a=20$ and $\varepsilon_0=0.001$. The rank rows report each method's common rank $r_{\mathrm{s}}$ and private-rank range $r_k-r_{\mathrm{s}}$ across the 20 clients. KT(r) and PerPCA(r) use a fixed common rank of $5$; KT(e) and PerPCA(e) use FE-estimated ranks. Smaller is better.}}
\label{tab:femnist-reconstruction}

\small
\setlength{\tabcolsep}{3pt}
\begin{tabular}{@{}lrrrrrrrr@{}}
\toprule
\rowcolor{gray!15}
\textbf{Methods} & \cellcolor{highlightColor}\textbf{FE} & \textbf{APM} & \textbf{CP} & \textbf{GCC} & \textbf{KT(r)} & \textbf{PerPCA(r)} & \textbf{KT(e)} & \textbf{PerPCA(e)} \\
\midrule
\rowcolor{gray!4} \textbf{$r_{\mathrm{s}}$} & \cellcolor{highlightColor}10 & 3 & 9 & 1 & 5 & 5 & 10 & 10 \\
\textbf{$r_k-r_{\mathrm{s}}$} & \cellcolor{highlightColor}2--9 & 9--16 & 3--10 & 11--18 & 7--14 & 7--14 & 2--9 & 2--9 \\
\rowcolor{gray!4} \textbf{RE(train)} & \cellcolor{highlightColor}4.9617 & 4.6047 & 5.2227 & 4.5385 & 4.6878 & 5.3723 & 4.9468 & 5.4507 \\
\rowcolor{gray!4} \textbf{RE(test)} & \cellcolor{highlightColor}5.5678 & 5.8605 & 6.0110 & 5.9333 & 5.7902 & 6.3307 & 5.5765 & 5.8561 \\
\bottomrule
\end{tabular}
\end{table}

{On FEMNIST, FE has the lowest test error ($5.5678$), followed closely by KT(e) ($5.5765$). GCC has the lowest training error ($4.5385$), while KT(e) has the lowest validation error ($5.6871$). KT(e) and PerPCA(e) use the same FE-estimated ranks as FE, allowing a direct comparison under matched dimensions; their test errors are $5.5765$ and $5.8561$, respectively.}

\subsection{United States Housing-Price Data}

We illustrate the group-factor interpretation using monthly United States housing-price data. The raw data are Zillow Home Value Index (ZHVI) series for homes with one, two, three, four, and five or more bedrooms across regions. The sample spans January 2000 to April 2023. After removing regions with missing observations and retaining states with at least 90 regions, the data contain $M=16$ states, $N=2241$ regions, and $T=279$ monthly observations. For region $i$ in state $m$, the monthly growth rate is
\[
x_{m,i,t}=\log(z_{m,i,t}/z_{m,i,t-1})\times 100,
\]
where $z_{m,i,t}$ is the ZHVI level. Following the notation used throughout the paper, collect the regional series in state $m$ as the columns of
\[
\mathbf X_m\in\mathbb R^{T\times N_m}.
\]
Each state is treated as one source: shared temporal directions are global factors, while state-specific temporal directions are local factors.

\textcolor{black}{We apply FE to the collection $\{\mathbf{X}_m\}_{m=1}^M$ and estimate each state's total rank $r_m$ from the data. To identify common temporal directions, FE matches adjacent states with threshold $\epsilon_{m,m+1}=n_{\mathrm{eff}}^{1/2-2/a-\varepsilon_0}$, where $n_{\mathrm{eff}}=\min(N_m,N_{m+1})$, $a=10$, and $\varepsilon_0=0.1$, in accordance with the simulation setting.}

\textcolor{black}{\textbf{Metric.} For every method, we calculate each region's explained fraction and then average these fractions within its state. Let $\mathbf x_{m,i}$ be the $i$th column of $\mathbf X_m$. With each method's common and private projectors, the reported ratios are}
\[
\begin{aligned}
\widehat{\mathrm{RIG}}_m
&=\frac{1}{N_m}\sum_{i=1}^{N_m}
\frac{\|\widehat{\mathbf P}^{\mathrm s}\mathbf x_{m,i}\|^2}{\|\mathbf x_{m,i}\|^2},\\
\widehat{\mathrm{RIF}}_m
&=\frac{1}{N_m}\sum_{i=1}^{N_m}
\frac{\|\widehat{\mathbf P}^{\mathrm p}_m\mathbf x_{m,i}\|^2}{\|\mathbf x_{m,i}\|^2},\\
\widehat{\mathrm{RIE}}_m
&=1-\widehat{\mathrm{RIG}}_m-\widehat{\mathrm{RIF}}_m.
\end{aligned}
\]
\textcolor{black}{For FE, the private basis is extracted from $(\mathbf I_T-\widehat{\mathbf P}^{\mathrm s})\widehat{\boldsymbol\Sigma}_m(\mathbf I_T-\widehat{\mathbf P}^{\mathrm s})$, where $\widehat{\boldsymbol\Sigma}_m=\mathbf X_m\mathbf X_m^\top/N_m$. Its private projector is zero when the estimated private rank is zero. The final W-Mean weights each state's ratios by $N_m$.}

\begin{figure}[H]
\centering
\includegraphics[width=0.96\textwidth]{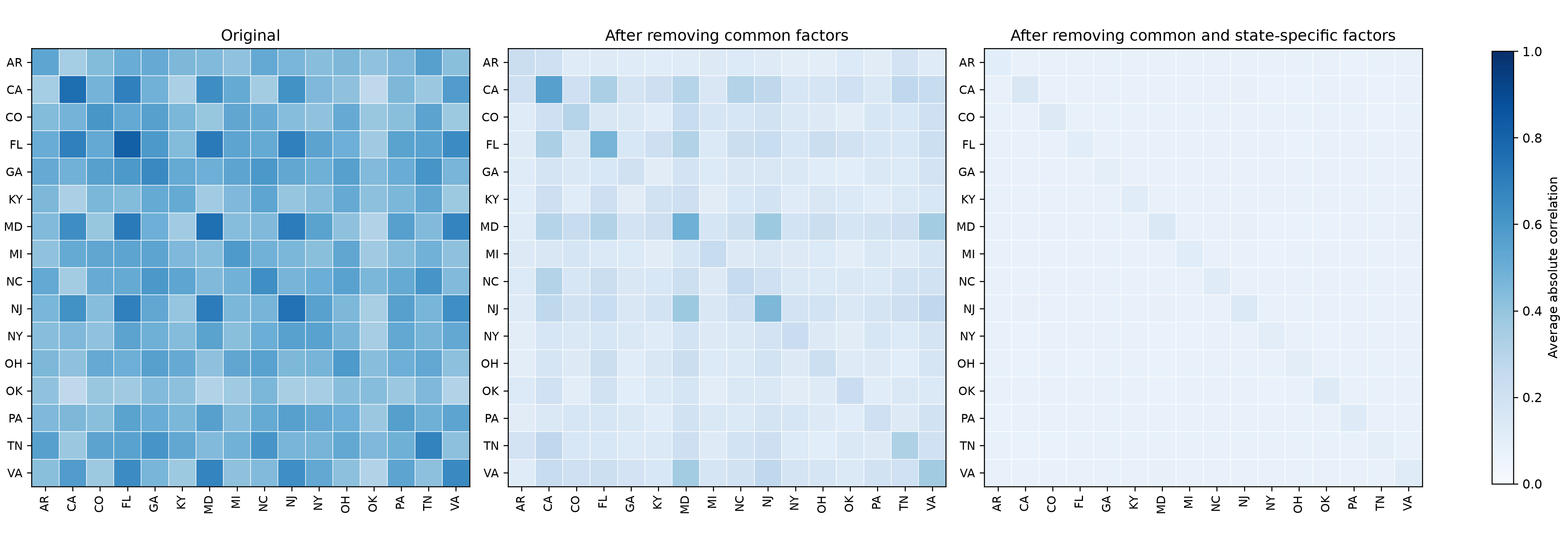}
\caption{Residual cross-state correlations after removing the estimated global and local components using \textcolor{black}{FE}.}
\label{fig:side-by-side-heatmaps}
\end{figure}

\begin{table}[H]
\centering
\scriptsize
\setlength{\tabcolsep}{2pt}
\renewcommand{\arraystretch}{1.05}
\resizebox{\textwidth}{!}{%
\begin{tabular}{@{}l*{16}{r}@{}}
\toprule
\rowcolor{gray!15}
\textbf{Rank} & \textbf{AR} & \textbf{CA} & \textbf{CO} & \textbf{FL} & \textbf{GA} & \textbf{KY} & \textbf{MD} & \textbf{MI} & \textbf{NC} & \textbf{NJ} & \textbf{NY} & \textbf{OH} & \textbf{OK} & \textbf{PA} & \textbf{TN} & \textbf{VA} \\
\midrule
\rowcolor{gray!4} \textcolor{black}{\textbf{$r_k$}} & 5 & 19 & 7 & 18 & 15 & 3 & 9 & 9 & 5 & 7 & 31 & 8 & 3 & 6 & 10 & 10 \\
\textcolor{black}{\textbf{$r_{\mathrm{s}}$}} & 1 & 1 & 1 & 1 & 1 & 1 & 1 & 1 & 1 & 1 & 1 & 1 & 1 & 1 & 1 & 1 \\
\rowcolor{gray!4} \textcolor{black}{\textbf{$r_k-r_{\mathrm{s}}$}} & 4 & 18 & 6 & 17 & 14 & 2 & 8 & 8 & 4 & 6 & 30 & 7 & 2 & 5 & 9 & 9 \\
\bottomrule
\end{tabular}
}
\caption{\textcolor{black}{FE-estimated ranks for the 16 states in the housing-price application, using $a=5$ and $\varepsilon_0=0.01$. Here $r_k$ is the total number of spike features in state $k$, $r_{\mathrm{s}}$ is the number shared across states, and $r_k-r_{\mathrm{s}}$ is the number private to state $k$.}}
\label{tab:application-housing-ranks}
\end{table}

Figure~\ref{fig:side-by-side-heatmaps} shows the residual cross-state correlations after removing the estimated global and local components using \textcolor{black}{FE}. The weak remaining off-diagonal correlations support the interpretation that the extracted common and state-specific spaces capture systematic temporal variation rather than merely fitting isolated states. The corresponding state-level rank estimates are listed in Table~\ref{tab:application-housing-ranks}; every state has one common feature, while the estimated private rank ranges from 2 to 30.

\begin{sidewaystable}[p]
\centering
% Highlight revisions made in the current editing pass.
\caption{Component shares in the housing-price application. Here $r_k$ is the total spike count in state $k$, $r_{\mathrm{s}}$ is the shared count, and $r_k-r_{\mathrm{s}}$ is the private count. For KT and PerPCA, (r) uses FE-estimated total state ranks with the common rank fixed at 3; (e) uses FE-estimated total and common ranks.}
\label{tab:application-housing}
\begingroup
\small
\setlength{\tabcolsep}{1pt}
\renewcommand{\arraystretch}{1.35}
\resizebox{\textheight}{!}{%
\begin{tabular}{@{}l*{24}{r}@{}}
\toprule
\rowcolor{gray!15}
& \multicolumn{8}{c}{Global Component (RIG)} & \multicolumn{8}{c}{Local Component (RIF)} & \multicolumn{8}{c}{Error Part (RIE)} \\
\cmidrule(lr){2-9}\cmidrule(lr){10-17}\cmidrule(lr){18-25}
\textbf{State} & \textbf{FE} & \textbf{APM} & \textbf{CP} & \textbf{GCC} & \textbf{KT(r)} & \textbf{PerPCA(r)} & \textbf{KT(e)} & \textbf{PerPCA(e)} & \textbf{FE} & \textbf{APM} & \textbf{CP} & \textbf{GCC} & \textbf{KT(r)} & \textbf{PerPCA(r)} & \textbf{KT(e)} & \textbf{PerPCA(e)} & \textbf{FE} & \textbf{APM} & \textbf{CP} & \textbf{GCC} & \textbf{KT(r)} & \textbf{PerPCA(r)} & \textbf{KT(e)} & \textbf{PerPCA(e)} \\
\midrule
OH & 0.601 & 0.590 & 0.515 & 0.590 & 0.645 & 0.648 & 0.596 & 0.587 & 0.190 & 0.202 & 0.276 & 0.202 & 0.137 & 0.090 & 0.196 & 0.176 & 0.209 & 0.208 & 0.209 & 0.208 & 0.218 & 0.262 & 0.208 & 0.238 \\
CO & 0.619 & 0.616 & 0.496 & 0.616 & 0.701 & 0.647 & 0.621 & 0.619 & 0.245 & 0.247 & 0.367 & 0.247 & 0.149 & 0.184 & 0.242 & 0.226 & 0.136 & 0.137 & 0.137 & 0.137 & 0.150 & 0.169 & 0.137 & 0.155 \\
VA & 0.599 & 0.560 & 0.552 & 0.559 & 0.597 & 0.663 & 0.553 & 0.510 & 0.297 & 0.336 & 0.347 & 0.337 & 0.292 & 0.213 & 0.343 & 0.370 & 0.104 & 0.104 & 0.101 & 0.104 & 0.112 & 0.124 & 0.104 & 0.121 \\
MD & 0.606 & 0.553 & 0.526 & 0.551 & 0.599 & 0.734 & 0.544 & 0.489 & 0.306 & 0.360 & 0.386 & 0.361 & 0.305 & 0.140 & 0.368 & 0.389 & 0.088 & 0.088 & 0.088 & 0.088 & 0.096 & 0.127 & 0.088 & 0.122 \\
AR & 0.581 & 0.616 & 0.557 & 0.616 & 0.656 & 0.643 & 0.612 & 0.621 & 0.150 & 0.111 & 0.173 & 0.110 & 0.055 & 0.034 & 0.117 & 0.084 & 0.269 & 0.273 & 0.270 & 0.273 & 0.290 & 0.323 & 0.272 & 0.295 \\
KY & 0.559 & 0.581 & 0.529 & 0.582 & 0.614 & 0.635 & 0.584 & 0.590 & 0.120 & 0.100 & 0.148 & 0.100 & 0.000 & 0.000 & 0.098 & 0.080 & 0.321 & 0.319 & 0.323 & 0.318 & 0.386 & 0.365 & 0.318 & 0.330 \\
GA & 0.661 & 0.651 & 0.591 & 0.651 & 0.709 & 0.695 & 0.655 & 0.641 & 0.205 & 0.215 & 0.274 & 0.215 & 0.154 & 0.151 & 0.212 & 0.212 & 0.134 & 0.134 & 0.135 & 0.133 & 0.137 & 0.154 & 0.133 & 0.146 \\
TN & 0.672 & 0.718 & 0.641 & 0.719 & 0.745 & 0.735 & 0.720 & 0.732 & 0.179 & 0.134 & 0.209 & 0.132 & 0.101 & 0.081 & 0.132 & 0.101 & 0.148 & 0.149 & 0.149 & 0.149 & 0.154 & 0.184 & 0.148 & 0.167 \\
MI & 0.545 & 0.505 & 0.419 & 0.505 & 0.613 & 0.572 & 0.511 & 0.493 & 0.260 & 0.302 & 0.387 & 0.302 & 0.180 & 0.196 & 0.296 & 0.274 & 0.195 & 0.193 & 0.194 & 0.193 & 0.207 & 0.232 & 0.193 & 0.233 \\
CA & 0.538 & 0.445 & 0.365 & 0.443 & 0.664 & 0.692 & 0.446 & 0.388 & 0.433 & 0.526 & 0.606 & 0.528 & 0.303 & 0.255 & 0.526 & 0.573 & 0.029 & 0.029 & 0.029 & 0.029 & 0.033 & 0.053 & 0.029 & 0.040 \\
OK & 0.468 & 0.500 & 0.431 & 0.500 & 0.538 & 0.565 & 0.498 & 0.514 & 0.170 & 0.139 & 0.202 & 0.139 & 0.000 & 0.000 & 0.141 & 0.099 & 0.362 & 0.361 & 0.367 & 0.361 & 0.462 & 0.435 & 0.361 & 0.387 \\
PA & 0.635 & 0.633 & 0.574 & 0.632 & 0.662 & 0.678 & 0.628 & 0.609 & 0.171 & 0.177 & 0.234 & 0.178 & 0.129 & 0.082 & 0.181 & 0.179 & 0.193 & 0.190 & 0.191 & 0.190 & 0.209 & 0.240 & 0.191 & 0.212 \\
FL & 0.700 & 0.626 & 0.585 & 0.625 & 0.717 & 0.793 & 0.625 & 0.565 & 0.250 & 0.324 & 0.365 & 0.325 & 0.231 & 0.135 & 0.325 & 0.371 & 0.050 & 0.050 & 0.050 & 0.050 & 0.052 & 0.072 & 0.050 & 0.064 \\
NC & 0.641 & 0.674 & 0.642 & 0.675 & 0.706 & 0.713 & 0.676 & 0.680 & 0.153 & 0.119 & 0.154 & 0.118 & 0.068 & 0.052 & 0.118 & 0.104 & 0.206 & 0.207 & 0.204 & 0.207 & 0.226 & 0.236 & 0.207 & 0.215 \\
NY & 0.617 & 0.609 & 0.566 & 0.609 & 0.642 & 0.651 & 0.609 & 0.583 & 0.322 & 0.330 & 0.373 & 0.331 & 0.295 & 0.276 & 0.331 & 0.348 & 0.061 & 0.061 & 0.061 & 0.061 & 0.063 & 0.072 & 0.060 & 0.069 \\
NJ & 0.649 & 0.600 & 0.545 & 0.598 & 0.661 & 0.710 & 0.595 & 0.541 & 0.261 & 0.310 & 0.364 & 0.312 & 0.234 & 0.142 & 0.315 & 0.324 & 0.090 & 0.090 & 0.091 & 0.090 & 0.105 & 0.147 & 0.090 & 0.135 \\
Mean & 0.606 & 0.592 & 0.533 & 0.592 & 0.654 & 0.673 & 0.592 & 0.573 & 0.232 & 0.246 & 0.304 & 0.246 & 0.165 & 0.127 & 0.246 & 0.244 & 0.162 & 0.162 & 0.163 & 0.162 & 0.181 & 0.200 & 0.162 & 0.183 \\
W-Mean & 0.617 & 0.602 & 0.544 & 0.602 & 0.663 & 0.680 & 0.603 & 0.583 & 0.235 & 0.250 & 0.308 & 0.250 & 0.175 & 0.138 & 0.249 & 0.250 & 0.148 & 0.148 & 0.148 & 0.148 & 0.162 & 0.182 & 0.148 & 0.167 \\
\bottomrule
\end{tabular}
}
\endgroup
\vspace{1ex}

\begin{minipage}{\linewidth}
\scriptsize
\textit{Note.} Mean is the unweighted average across states; W-Mean is weighted by the number of regions $N_m$ in each state.
\end{minipage}
\end{sidewaystable}
\renewcommand{\arraystretch}{1}

For the housing data, FE's weighted-average shares are $0.617$ for the common component, $0.235$ for the local component, and $0.148$ for the error component. In AR, the FE rank estimate has 5 total features, including one common and four private features. Table~\ref{tab:application-housing} also shows the effect of rank specification for KT and PerPCA. Using FE-estimated total state ranks with common rank fixed at 3 gives weighted-average triples $(\mathrm{RIG},\mathrm{RIF},\mathrm{RIE})=(0.663,0.175,0.162)$ for KT(r) and $(0.680,0.138,0.182)$ for PerPCA(r). With FE-estimated common and total state ranks, the corresponding triples for KT(e) and PerPCA(e) are $(0.603,0.249,0.148)$ and $(0.583,0.250,0.167)$. These ratios describe the empirical decomposition and are not, by themselves, evidence of correct common and private subspace recovery.

\section{Conclusion}

We studied the problem of heterogeneousPCA for multi-source data. To address it, we proposed FE, which estimates source-specific spike counts from eigenvector stability, identifies shared directions by matching spike eigenvectors across sources, and uses the identified features to construct common and private subspaces. We established consistency of the proposed spike-count estimators and common-feature identification procedure under a spiked-covariance model. Simulations show that FE estimates the common rank and recovers common features more accurately than APM, CP, and GCC, particularly under weak signals, while also achieving competitive subspace recovery; experiments on federated images and housing data further demonstrate its practical advantages. Extending the theory to growing numbers of sources and more general partially shared subspaces remains an important direction for future work.

% Reconstruction-error table for the FEMNIST application.

\appendix
\section{Proofs}

\begingroup\color{black}\hypersetup{linkcolor=black,citecolor=black,urlcolor=black}
\subsection{Finite-dimensional regularity}\label{app:regularity}
For $h_k=\lfloor n_k/2\rfloor$ and $s\in\{n_k,h_k\}$, define
\[
H_{\mathcal B,k}=\frac1{p-r_k}\sum_{i=r_k+1}^{p}\delta_{\lambda_{ki}},\qquad c_{s,r_k}=\frac{p-r_k}{s}.
\]

\begin{definition}[Stochastic domination in probability]
Let $\{X_n(a_n)\}$ and $\{Y_n(a_n)\}$ be sequences of nonnegative random variables 
that may depend on a parameter sequence $(a_n)$. 
We say that $X_n(a_n)$ is \emph{stochastically dominated in probability} by $Y_n(a_n)$, 
uniformly in the parameter, and write $X_n(a_n) \prec_{\mathrm{p}} Y_n(a_n)$, 
if for every $\varepsilon > 0$,
\[
\sup_{a_n} \mathbb{P}\!\left( X_n(a_n) > n^{\varepsilon} Y_n(a_n) \right) \to 0, 
\qquad n \to \infty.
\]
If $|Z_n| \prec_{\mathrm{p}} Y_n$ for some quantity $Z_n$, we also write $Z_n = O_{\prec_{\mathrm{p}}}(Y_n)$.
\end{definition}

\begin{remark}
The notation $``\prec_{\mathrm{p}}"$ denotes a weak form of stochastic domination \citep{Erdos2013}. It relaxes the usual polynomial tail bound in random matrix theory to convergence in probability.    
\end{remark}

\endgroup
\subsection{Auxiliary Lemmas and their proofs}
For notational simplicity, we focus on a single source dataset $k$ throughout this section and suppress its subscript when it is unambiguous. Let the population and sample covariance matrices admit the spectral decompositions
\[
\boldsymbol{\Sigma} = \sum_{j=1}^p \lambda_j\,\mathbf{u}_j \mathbf{u}_j^\top,
\qquad
\widehat{\boldsymbol{\Sigma}} = \sum_{j=1}^p \hat{\lambda}_j \,\hat{\mathbf{u}}_j \hat{\mathbf{u}}_j^\top,
\]
with $\lambda_1 \ge \cdots \ge \lambda_p$ and $\hat{\lambda}_1 \ge \cdots \ge \hat{\lambda}_p$ denoting the ordered population and sample eigenvalues, respectively, and
$\{\mathbf{u}_j\}_{j=1}^p$ and $\{\hat{\mathbf{u}}_j\}_{j=1}^p$ the corresponding orthonormal eigenvectors. Let $r$ denote the number of spikes and define the spike and bulk index sets
\[
\mathcal{S} = \{1,\ldots,r\}, \qquad \mathcal{B} = \{r+1,\ldots,p\}.
\]
It is well known from random matrix theory that the bulk eigenvalues of $\widehat{\boldsymbol{\Sigma}}$ converge weakly to a deterministic distribution, known as the generalized Mar\v{c}henko-Pastur law \citep{marchenko1967distribution,MR1370408}, while the leading $r$ eigenvalues form outliers whose limits lie outside the support of the bulk spectral distribution \citep{Bai2012Sample}. 
We collect these results in the following two lemmas.

\textcolor{black}{Write $c_{n,r}=(p-r)/n$ for the bulk aspect ratio.} To describe the bulk eigenvalues, let

\[
H_{\mathcal B}=\frac{1}{p-r}\sum_{\ell\in \mathcal B}\delta_{\lambda_\ell}
\quad\text{and}\quad
F_{\mathcal B}=\frac{1}{p-r}\sum_{\ell\in \mathcal B}\delta_{\hat{\lambda}_\ell}
\]
denote the empirical spectral distributions of the bulk eigenvalues of $\boldsymbol{\Sigma}$ and $\widehat{\boldsymbol{\Sigma}}$, respectively. 
Let $m_{\mathcal B}(z)$ be the Stieltjes transform of $F_{\mathcal B}$, defined by
\[
m_{\mathcal B}(z)=\int \frac{1}{x-z}\,dF_{\mathcal B}(x), \qquad z\in\mathbb{C}_+,
\]
where $\mathbb{C}_+$ denotes the upper complex plane.
\begin{lemma}\label{lem:1}
Under Assumptions~\ref{ass:1}--\ref{ass:3}, almost surely,
\[
m_{\mathcal B}(z) - m_0(z) \to 0, \qquad \forall z \in \mathbb{C}_+,
\]
where $m_0(z)$ is the unique solution to
\[
m(z) = \int \frac{1}{t\bigl(1-\textcolor{black}{c_{n,r}}-\textcolor{black}{c_{n,r}} z\, m(z)\bigr)-z}\, dH_{\mathcal B}(t),
\]
on the set $\{\textcolor{black}{m \in \mathbb{C}} : -(1-\textcolor{black}{c_{n,r}})/z + \textcolor{black}{c_{n,r}} \textcolor{black}{m} \in \mathbb{C}_+\}$.
\end{lemma}
\begin{proof}
This lemma follows from Theorem~1.1 in \citet{MR1370408} and Theorem~A.43 in \citet{baibook}.
\end{proof}

\begin{remark}\label{f0}
By the inversion formula for the Stieltjes transform, the function $m_0(z)$ defined in Lemma~\ref{lem:1} induces a sequence of deterministic probability measures, denoted by $\{F_{\textcolor{black}{c_{n,r}}, H_{\mathcal B}}\}_{n\ge1}.$
These measures provide deterministic approximations to the empirical bulk distribution in the sense that, almost surely, for all continuity points $x$ of $F_{\textcolor{black}{c_{n,r}}, H_{\mathcal B}}$, 
\[
F_{\mathcal B}(x) - F_{\textcolor{black}{c_{n,r}}, H_{\mathcal B}}(x) \to 0 .
\]
The support of $F_{\textcolor{black}{c_{n,r}}, H_{\mathcal B}}$ can be determined using the method of \citet{choi1995}.
\end{remark}

To describe the spike eigenvalues, define
\[
\psi_{\textcolor{black}{c_{n,r}}, H_{\mathcal B}}(x)
\triangleq x + \textcolor{black}{c_{n,r}} x \int \frac{t}{x - t}\, dH_{\mathcal B}(t)
= x + \textcolor{black}{\frac{x}{n}}\sum_{\ell\in \mathcal B}\frac{\lambda_\ell}{x-\lambda_\ell}.
\]
\begin{lemma}\label{lem:2}
Under Assumptions~\ref{ass:1}--\ref{ass:3}, almost surely, for any fixed integer $\omega>r$,
\[
\hat{\lambda}_j=
\begin{cases}
\psi_{\textcolor{black}{c_{n,r}},H_{\mathcal B}}(\lambda_j)+o(1), & j\in\mathcal S,\\[3pt]
\lambda_+ + o(1), & r<j\le \omega,
\end{cases}
\]
where $\lambda_+$ denotes the right endpoint of the support of $F_{\textcolor{black}{c_{n,r}},H_{\mathcal B}}$.
\end{lemma}
\begin{proof}
This lemma follows from the main results of \citet{Bai2012Sample} and \citet{bai98}. 
\end{proof}

A more precise characterization of the spectrum can be obtained through large deviation bounds for sample eigenvalues and eigenvectors, a direction that has been extensively studied in the literature (see, e.g., \citet{Knowles2014, bloemendal2016principal, Knowles2017, Yangfan2019, Ding2021, Bao2022} and references therein). In particular, \citet{Ding2021} develops a general theory for separable covariance models, which includes the present setting as a special case.
We summarize below simplified versions of their results, specialized to our framework.
Lemmas \ref{lem:1} and \ref{lem:2} provide the foundation for the stronger conclusions that follow.

\begin{lemma}\label{lem:3}
Under Assumptions~\ref{ass:1}--\ref{ass:3}, let $\phi_n = n^{2/a-1/2}$. 
For each fixed $j \in \mathcal{S}$,
\[
\bigl|\hat{\lambda}_j-\psi_{\textcolor{black}{c_{n,r}},H_{\mathcal B}}(\lambda_j)\bigr|
\prec_{\mathrm{p}} n^{-1/2}+\phi_n.
\]
Moreover, for any fixed integer $\omega>r$, we have
\[
\bigl|\hat{\lambda}_j-\lambda_+\bigr|
\prec_{\mathrm{p}} n^{-2/3}+\phi_n^2,
\qquad r< j \le \omega,
\]
where $\lambda_+$ is the right edge of the support of $F_{\textcolor{black}{c_{n,r}},H_{\mathcal B}}$ defined in Lemma \ref{lem:2}.
\end{lemma}
\begin{proof}
\begingroup\color{black}\hypersetup{linkcolor=black,citecolor=black,urlcolor=black}
This lemma is a direct consequence of Theorem~3.6 and Corollary~3.19 of \citet{Ding2021}.
\endgroup

\end{proof}

\begin{lemma}\label{lem:4}
Under Assumptions~\ref{ass:1}--\ref{ass:3}, let $\phi_n = n^{2/a-1/2}$. 
For each fixed $j \in \mathcal{S}$, define the following deterministic rank-one matrix
\[
\mathcal{Z}_j \triangleq 
\textcolor{black}{\zeta_j}\,\mathbf{u}_j \mathbf{u}_j^\top,
\qquad \textcolor{black}{\zeta_j\triangleq}\frac{1}{\lambda_j}\,
\frac{g_{\textcolor{black}{c_{n,r}}}^{\prime}\!\left(-\lambda_j^{-1}\right)}{g_{\textcolor{black}{c_{n,r}}}\!\left(-\lambda_j^{-1}\right)},
\]
where $g_{\textcolor{black}{c_{n,r}}}(x)$ is the inverse function of $\underline m_0(x)=-(1-\textcolor{black}{c_{n,r}})/x+\textcolor{black}{c_{n,r}} m_0(x):(\lambda_+,\infty)\to (\underline m_0(\lambda_+),0)$.
Then, for any deterministic \textcolor{black}{unit} vector $\mathbf{v} \in \mathbb{R}^p$, we have
\begin{align*}
\bigl|
\langle \mathbf{v}, \hat{\mathbf{u}}_j \rangle^2
-
\langle \mathbf{v}, \mathcal{Z}_j \mathbf{v} \rangle
\bigr|
\prec_{\mathrm{p}}\,
& v_j^2 \bigl(\phi_n + n^{-1/2}\bigr)
+ \bigl(\phi_n + n^{-1/2}\bigr)^2 \\
& + \langle \mathbf{v}, \mathcal{Z}_j \mathbf{v} \rangle^{1/2}
\left( \sum_{\textcolor{black}{i\ne j}} v_i^2 \right)^{1/2}
\bigl(\phi_n + n^{-1/2} \bigr),\quad j\in\mathcal S,
\end{align*}
where $v_\ell = \langle \mathbf{u}_\ell, \mathbf{v} \rangle$, $\ell=1,\ldots,p$.

Moreover, for any fixed integer $\omega>r$ and any deterministic \textcolor{black}{unit} vector $\mathbf{v} \in \mathbb{R}^p$, we have
\[
\langle \mathbf{v}, \hat{\mathbf{u}}_j \rangle^2
\prec_{\mathrm{p}}\, n^{-1} + \phi_n^3,\quad r<j\leq \omega.
\]
\end{lemma}
\begin{proof}
\begingroup\color{black}\hypersetup{linkcolor=black,citecolor=black,urlcolor=black}
Write $\underline m_{0,k,s}$ and $\lambda_{+,k,s}$ for the companion transform and right edge of $F_{c_{s,r_k},H_{\mathcal B,k}}$, respectively. The edge parameter $\alpha_{k,s}=-1/\underline m_{0,k,s}(\lambda_{+,k,s})>\lambda_{k,r_k+1}$ satisfies $c_{s,r_k}\int t^2/(\alpha_{k,s}-t)^2\,dH_{\mathcal B,k}(t)=1$ and increases with $c_{s,r_k}$. Thus Assumption~\ref{ass:3} also gives $1-\lambda_{k,r_k+1}/\alpha_{k,h_k}\ge\tau$ for the half samples. For identity column covariance, the remaining edge condition in Assumption~2.6 of \citet{Ding2021} holds automatically.

The fixed population gap permits uniform application of weak convergence near the spikes. Since $h_k=\lfloor n_k/2\rfloor$ and $c_{n_k}<c_0$, we have $c_{s,r_k}\le 2c_0+o(1)$ for $s\in\{n_k,h_k\}$. Thus Assumption~\ref{ass:3} yields, uniformly in $k,s$ and for some $\delta>0$,
\[
\psi'_{c_{s,r_k},H_{\mathcal B,k}}(\lambda_{kr_k})\ge d_0-o(1),
\qquad
\psi_{c_{s,r_k},H_{\mathcal B,k}}(\lambda_{kr_k})-\lambda_{+,k,s}\ge\delta
\]
for all sufficiently large $p$. The second bound follows by integrating the positive derivative over a fixed interval to the left of the weakest spike, whose existence follows from the uniformly bounded second derivatives there.

This lemma is a direct consequence of Theorems~3.10 and~3.14 and Corollary~3.19 of \citet{Ding2021}. The upper bounds on failure probabilities are uniform over deterministic unit vectors.
\endgroup

\end{proof}

\subsection{Proof of Theorem~\ref{th1}}

Without loss of generality, we focus on a single dataset $k$ and suppress the subscript $k$.
For this dataset, write $\phi_k=\phi(n_k)=n_k^{2/a-1/2}$ and
$\varepsilon(n_k)=n_k^{\varepsilon_0}(n_k^{-1/2}+\phi_k^{3/2})$, with
\textcolor{black}{$0<\varepsilon_0<1/2-2/a$}, as in the theorem statement.

Recall that we split the sample into two independent halves, each of size \textcolor{black}{$h_k=\lfloor n_k/2\rfloor$}, and denote the corresponding sample covariance matrices by $\widehat{\boldsymbol{\Sigma}}^{(1)}$ and $\widehat{\boldsymbol{\Sigma}}^{(2)}$, with eigenvectors $\hat{\mathbf{u}}_j^{(1)}$ and $\hat{\mathbf{u}}_j^{(2)}$, $j=1,\dots,p$. 
\begingroup\color{black}\hypersetup{linkcolor=black,citecolor=black,urlcolor=black}
Write $c_{h_k,r}=(p-r)/h_k$ for the half samples.
\endgroup
We analyze the inner products $\langle \hat{\mathbf{u}}_j^{(1)}, \hat{\mathbf{u}}_j^{(2)} \rangle$ separately for spike and bulk indices.

\noindent\textbf{Spike indices ($j\in\mathcal S$).}  
Let
\[
\zeta_j
=
\frac{1}{\lambda_j}\,
\frac{\textcolor{black}{g_{c_{h_k,r}}}^{\prime}\!\left(-\lambda_j^{-1}\right)}{\textcolor{black}{g_{c_{h_k,r}}}\!\left(-\lambda_j^{-1}\right)},
\qquad j\in\mathcal S.
\]
By definition (Lemma~\ref{lem:4}), the function $\textcolor{black}{g_{c_{h_k,r}}}$ is positive and monotonically increasing on its domain. 
Consequently, $\zeta_j>0$ for all $j\in\mathcal{S}$.
\begingroup\color{black}\hypersetup{linkcolor=black,citecolor=black,urlcolor=black}
Appendix~\ref{app:regularity} further gives $\min_{j\le r}\zeta_j\ge\zeta_0>0$ uniformly in $n_k$.
\endgroup
We next show that
\[
\bigl| \langle \hat{\mathbf{u}}_j^{(1)}, \hat{\mathbf{u}}_j^{(2)} \rangle^2 - \zeta_j^2 \bigr|
\prec_{\mathrm{p}} \phi_k,
\qquad j\in\mathcal S.
\]

Condition on the first half-sample. Then $\hat{\mathbf u}_j^{(1)}$ is deterministic and independent of the second half-sample.
Applying Lemma~\ref{lem:4} to the second half-sample with $\mathbf v=\hat{\mathbf u}_j^{(1)}$ yields
\begin{align}
\Bigl|
\langle \hat{\mathbf u}_j^{(1)}, \hat{\mathbf u}_j^{(2)} \rangle^2
-
\langle \hat{\mathbf u}_j^{(1)}, \mathcal Z_j \hat{\mathbf u}_j^{(1)} \rangle
\Bigr|
\prec_{\mathrm{p}}\;
& [v_j^{(1)}]^2(\phi_k+n_k^{-1/2}) + (\phi_k+n_k^{-1/2})^2  \nonumber\\
& + \langle \hat{\mathbf u}_j^{(1)}, \mathcal Z_j \hat{\mathbf u}_j^{(1)} \rangle^{1/2}
\Bigl(\sum_{\textcolor{black}{i\ne j}} [v_i^{(1)}]^2\Bigr)^{1/2}(\phi_k+n_k^{-1/2}) \nonumber\\
\prec_{\mathrm{p}}\; & \phi_k,\label{jS}
\end{align}
where $v_\ell^{(1)}=\langle \mathbf u_\ell,\hat{\mathbf u}_j^{(1)}\rangle$ and $\mathcal Z_j=\zeta_j\,\mathbf u_j\mathbf u_j^\top$.
Indeed, $|v_j^{(1)}|\le 1$, $\sum_{\textcolor{black}{i\ne j}} [v_i^{(1)}]^2 \le 1$, $n_k^{-1/2}\leq \phi_k$, and
\[
\langle \hat{\mathbf u}_j^{(1)}, \mathcal Z_j \hat{\mathbf u}_j^{(1)} \rangle
= \zeta_j [v_j^{(1)}]^2 \le \zeta_j,
\]
so each term on the right-hand side is $O_{\prec_{\mathrm{p}}}(\phi_k)$.

On the other hand, applying Lemma~\ref{lem:4} to the first half-sample with $\mathbf v=\mathbf u_j$ gives
\[
\bigl|[v_j^{(1)}]^2-\zeta_j\bigr|
=
\bigl|\langle \mathbf u_j,\hat{\mathbf u}_j^{(1)}\rangle^2-\zeta_j\bigr|
\prec_{\mathrm{p}}\phi_k,
\]
which implies
\(
\bigl|\zeta_j\,[v_j^{(1)}]^2-\zeta_j^2\bigr|
\prec_{\mathrm{p}}\phi_k.
\)
Collecting the above results yields
\[
\bigl| \langle \hat{\mathbf{u}}_j^{(1)}, \hat{\mathbf{u}}_j^{(2)} \rangle^2 - \zeta_j^2 \bigr|
\prec_{\mathrm{p}} \phi_k.
\]

\noindent\textbf{Bulk indices.}
Fix an integer $\omega>r$ and consider any bulk index $r<j\le \omega$.
We show that the split-sample inner product is asymptotically negligible:
\[
\bigl|\langle \hat{\mathbf u}_j^{(1)},\hat{\mathbf u}_j^{(2)}\rangle\bigr|^2
\prec_{\mathrm{p}} n_k^{-1}+\phi_k^3.
\]
Condition on the first half-sample. Then $\hat{\mathbf u}_j^{(1)}$ is deterministic and independent of the second half-sample.
Applying the bulk part of Lemma~\ref{lem:4} to the second half-sample with $\mathbf v=\hat{\mathbf u}_j^{(1)}$ yields
\begin{align}\label{jB}
\bigl|\langle \hat{\mathbf u}_j^{(1)},\hat{\mathbf u}_j^{(2)}\rangle\bigr|^2
=
\bigl|\langle \mathbf v,\hat{\mathbf u}_j^{(2)}\rangle\bigr|^2
\prec_{\mathrm{p}} n_k^{-1}+\phi_k^3,
\end{align}
conditionally on the first half-sample. Since the bound does not depend on the particular realization of $\hat{\mathbf u}_j^{(1)}$,
the same stochastic domination holds unconditionally.

\noindent\textbf{Conclusion for $\hat r^{\mathrm{{ESC}}}$.}  
Consider the averaged absolute inner products from $m$ \textcolor{black}{fixed} replications,
\[
\bar I_j
\triangleq
\frac{1}{m}\sum_{t=1}^m
\bigl| \langle \hat{\mathbf{u}}_j^{(1,t)}, \hat{\mathbf{u}}_j^{(2,t)} \rangle \bigr|.
\]

From \textcolor{black}{the preceding spike estimate} and \eqref{jB}, we have, for any fixed \textcolor{black}{$\omega\ge r+2$},
\[
\bar I_j
=
\begin{cases}
\zeta_j+O_{\prec_{\mathrm{p}}}(\phi_k),& j\le r,\\[3pt]
O_{\prec_{\mathrm{p}}}(n_k^{-1/2}+\phi_k^{3/2}),& r<j\le \omega.
\end{cases}
\]
Here, we used $\zeta_j\ge \zeta_0>0$ and the identity
$|\sqrt{x}-\sqrt{y}|= |x-y|/(\sqrt{x}+\sqrt{y})$ to pass from \textcolor{black}{the same spike estimate} to the bound for $|\langle\hat{\mathbf u}_j^{(1)},\hat{\mathbf u}_j^{(2)}\rangle|$.
Recall that the estimator is defined as
\[
\hat r^{\mathrm{{ESC}}}
=
\min \Bigl\{ l :
\bar I_{l+1}<\varepsilon(n_k)
\text{ and }
\bar I_{l+2}<\varepsilon(n_k)
\Bigr\},
\]
with the threshold $\varepsilon(n_k) = n_k^{\varepsilon_0}\bigl(n_k^{-1/2}+\phi_k^{3/2}\bigr)$, where $\varepsilon_0>0$ is chosen sufficiently small so that $\varepsilon(n_k)\to0$ and $\varepsilon(n_k) \gg n_k^{-1/2}+\phi_k^{3/2}$.
With probability tending to $1$, we have
\[
\bar I_j > \varepsilon(n_k) \ \text{ for } j \le r,
\qquad
\bar I_j < \varepsilon(n_k) \ \text{ for } r< j \le \omega.
\]

Therefore, the first index $l$ such that both $\bar I_{l+1}$ and $\bar I_{l+2}$ fall below $\varepsilon(n_k)$ is exactly $l=r$ \textcolor{black}{with probability tending to one}, and hence
\[
\mathbb{P}\!\left( \hat r^{\mathrm{{ESC}}} = r \right) \to 1.
\]

\subsection{{Proof of Theorem~\ref{th3}}}
\begingroup

{
\noindent\textbf{Step 1: Estimation of the correction factors.}
By Theorem~\ref{th1}, it suffices to work on the event $\{\hat r_k=r_k\text{ for all }k\}$, whose probability tends to one. Let $\hat\zeta_{k,j}=\theta_{kj}^2$ be computed from \eqref{eq:zeta-estimator} and recall the full-sample factors $\zeta_{k,j}$ defined in Lemma~\ref{lem:4}. The uniform separation in Assumption~\ref{ass:3} gives $\min_{k,j}\zeta_{k,j}\ge\zeta_0>0$.

Suppress the source index $k$. Set $q_n=\phi_n+n^{-1/2}$, $\rho_j=g_{c_{n,r}}(-\lambda_j^{-1})$, and
\[
\widehat{\underline m}(z)=-\frac{1-c_{n,r}}{z}
+\frac1n\sum_{i=r+1}^{p}\frac1{\hat\lambda_i-z}.
\]
Apply the averaged local law (Theorem~S.3.9 in \href{https://arxiv.org/abs/1905.13060v3}{Ding and Yang (2020)}), with the truncation in Corollary~3.19 of \citet{Ding2021}, to the noise model with population spikes set to zero. Finite-rank interlacing and Cauchy's formula then give, uniformly on fixed complex neighborhoods of the $\rho_j$ separated from the bulk,
\begin{equation}\label{eq:bulk-transform-rate}
\sup_z\left\{
|\widehat{\underline m}(z)-\underline m_0(z)|+
|\widehat{\underline m}'(z)-\underline m_0'(z)|
\right\}=O_{\prec_{\mathrm p}}(q_n).
\end{equation}
By Assumption~\ref{ass:3}, the deterministic denominator $\rho_j\underline m_0'(\rho_j)$ is uniformly bounded away from zero. Lemmas~\ref{lem:3}--\ref{lem:4} and Taylor expansion of $\hat\zeta_j=-\widehat{\underline m}(\hat\lambda_j)/[\hat\lambda_j\widehat{\underline m}'(\hat\lambda_j)]$ about $\rho_j$ therefore give
\[
\max_{k}\max_{j\le r_k}|\hat\zeta_{k,j}-\zeta_{k,j}|
=O_{\prec_{\mathrm p}}(\phi_*+n_*^{-1/2}).
\]
Here $n_*=\min_k n_k$ and $\phi_*=n_*^{2/a-1/2}$. Hence the estimators $\hat\zeta_{k,j}$ are well defined and positive with probability tending to one. Since $\gamma_*=n_*^{\varepsilon_0}\phi_*$ with $\varepsilon_0>0$,
\begin{equation}\label{eq:zeta-rate}
\max_{k,j}|\hat\zeta_{k,j}-\zeta_{k,j}|=o_{\mathrm p}(\gamma_*).
\end{equation}
}

\smallskip
\noindent\textbf{Step {2}: Approximation of the cross inner product.}
Condition on the $k$th sub-dataset, so that $\hat{\mathbf u}_{k,j}$ is deterministic and independent of the $(k+1)$th sub-dataset.
Applying Lemma~\ref{lem:4} to the $(k+1)$th sub-dataset with $\mathbf v=\hat{\mathbf u}_{k,j}$ and the spike index $m$ yields
\[
\Bigl|
\langle \hat{\mathbf u}_{k,j},\hat{\mathbf u}_{k+1,m}\rangle^2
-
\langle \hat{\mathbf u}_{k,j},\mathcal Z_{k+1,m}\hat{\mathbf u}_{k,j}\rangle
\Bigr|
=O_{\prec_{\mathrm{p}}}({\phi_*}),
\]
where $\mathcal Z_{k+1,m}=\zeta_{k+1,m}\mathbf u_{k+1,m}\mathbf u_{k+1,m}^\top$.
Thus,
\begin{equation}\label{eq:cross1}
|\langle \hat{\mathbf u}_{k,j},\hat{\mathbf u}_{k+1,m}\rangle|^2
=
\zeta_{k+1,m}\,|\langle \hat{\mathbf u}_{k,j},\mathbf u_{k+1,m}\rangle|^2
+O_{\prec_{\mathrm{p}}}({\phi_*}).
\end{equation}
Next, applying Lemma~\ref{lem:4} to the $k$th sub-dataset with $\mathbf v=\mathbf u_{k+1,m}$ and the spike index $j$ gives
\begin{equation}\label{eq:cross2}
|\langle \mathbf u_{k+1,m},\hat{\mathbf u}_{k,j}\rangle|^2
=
\zeta_{k,j}\,|\langle \mathbf u_{k+1,m},\mathbf u_{k,j}\rangle|^2
+O_{\prec_{\mathrm{p}}}({\phi_*}).
\end{equation}
Plugging \eqref{eq:cross2} into \eqref{eq:cross1} yields
\begin{equation}\label{eq:cross3}
|\langle \hat{\mathbf u}_{k,j},\hat{\mathbf u}_{k+1,m}\rangle|^2
=
\zeta_{k,j}\zeta_{k+1,m}\,|\langle \mathbf u_{k,j},\mathbf u_{k+1,m}\rangle|^2
+O_{\prec_{\mathrm{p}}}({\phi_*}).
\end{equation}

\begingroup\hypersetup{linkcolor=black,citecolor=black,urlcolor=black}
\smallskip
\noindent\textbf{Step 3: Recovery of adjacent matches.} Let $\widehat{\mathcal T}$ be the set of index tuples retained by the sequential aggregation in Algorithm~\ref{alg:identification}:
\[
\widehat{\mathcal T}
=\bigl\{(j_1,\ldots,j_K):
(j_k,j_{k+1})\in\widehat{\mathcal E}_k
\text{ for every }k=1,\ldots,K-1\bigr\}.
\]
Define $\mathcal T^\star$ analogously using the oracle sets $\mathcal E_k^\star$. The algorithm sets $\hat r_{\mathrm s}=|\widehat{\mathcal T}|$ and takes $\widehat{\mathcal A}_k^{\mathrm s}$ to be the $k$th-coordinate projection of $\widehat{\mathcal T}$. Let $\mathcal S_k=\{1,\ldots,r_k\}$ be the index set of spikes of the $k$th dataset and define the \emph{oracle matching set}
\[
{\mathcal E}_{k}^{\star}
\;\triangleq\;
\bigl\{(j,m)\in\mathcal S_k\times\mathcal S_{k+1}:\ \mathbf u_{k,j}=\pm \mathbf u_{k+1,m}\bigr\},\quad k=1,\ldots,K-1.
\]
For each source $k\in\{1,\ldots,K\}$, let $\{\hat{\mathbf u}_{k,j}\}_{j\in\mathcal S_k}$ denote the sample eigenvectors
corresponding to the spike eigenvalues, and let $\{\zeta_{k,j}\}_{j\in\mathcal S_k}$ be the deterministic constants in
Lemma~\ref{lem:4}, with estimators $\hat\zeta_{k,j}$.
For $k=1,\ldots,K-1$, define the normalized similarity
\[
T_{k}(j,m)
\triangleq
\frac{\bigl|\langle \hat{\mathbf u}_{k,j},\hat{\mathbf u}_{k+1,m}\rangle\bigr|}
{\sqrt{\hat\zeta_{k,j}\hat\zeta_{k+1,m}}},
\qquad \textcolor{black}{j\in[\hat r_k],\ m\in[\hat r_{k+1}]},
\]
and the thresholded edge set
\[
\widehat{\mathcal E}_{k}
\triangleq
\bigl\{(j,m)\in\textcolor{black}{[\hat r_k]\times[\hat r_{k+1}]}:\ T_{k}(j,m)\ge 1-\textcolor{black}{\textcolor{black}{\gamma_n}}\bigr\},
\]
where $\{\textcolor{black}{\gamma_n}\}$ is a deterministic sequence satisfying $\textcolor{black}{\gamma_n}\downarrow 0$ sufficiently slowly.
Equations~\eqref{eq:cross3} and~\eqref{eq:zeta-rate}, with $O_{\prec_{\mathrm p}}(\phi_*)=o_{\mathrm p}(\gamma_*)$, give
\begin{equation}\label{eq:T-approx2}
T_k(j,m)^2=|\langle\mathbf u_{k,j},\mathbf u_{k+1,m}\rangle|^2+o_{\mathrm p}(\gamma_*).
\end{equation}
For common directions, the leading term is $1$, and $1-(1-\gamma_*)^2\sim2\gamma_*$. For noncommon directions, it is at most $(1-\eta)^2<1$ by Assumption~\ref{ass:A4}.
\endgroup
Therefore, for each fixed pair $(j,m)$,
\[
\mathbb P\!\bigl(\mathbf 1\{(j,m)\in\widehat{\mathcal E}_k\}\neq \mathbf 1\{(j,m)\in{\mathcal E}_k^\star\}\bigr)\to 0.
\]
Finally, since $|\mathcal S_k|$ and $|\mathcal S_{k+1}|$ are finite, a union bound over
$(j,m)\in\mathcal S_k\times\mathcal S_{k+1}$ yields
$\mathbb P(\widehat{\mathcal E}_k={\mathcal E}_k^\star)\to 1$.
A further union bound over the fixed set $k\in\{1,\ldots,K-1\}$ shows that
\[
\mathbb P\!\left(
\widehat{\mathcal E}_k=\mathcal E_k^\star
\text{ for all }k=1,\ldots,K-1
\right)\to 1.
\]

\smallskip
\noindent\textbf{Step 4: Recovery of the common rank and locations.}
On the event that every adjacent matching set is recovered, the path construction gives $\widehat{\mathcal T}=\mathcal T^\star$. A tuple in $\mathcal T^\star$ links population spike eigenvectors that coincide up to sign across all $K$ sources, and every common feature yields such a tuple. Distinct tuples represent distinct common features because the spike eigenvectors within each source are orthogonal. Therefore $|\mathcal T^\star|=r_{\mathrm s}$, and its $k$th-coordinate projection is $\mathcal A_k^{\mathrm s}$ for every $k$. Since $\hat r_{\mathrm s}=|\widehat{\mathcal T}|$, this yields
\[
\mathbb P\!\left(\hat r_{\mathrm s}=r_{\mathrm s}\right)
\ge \mathbb P\!\left(\widehat{\mathcal T}=\mathcal T^\star\right)\to 1.
\]
Likewise, because $\widehat{\mathcal A}_k^{\mathrm s}$ is the $k$th-coordinate projection of $\widehat{\mathcal T}$,
\[
\mathbb P\!\left(\widehat{\mathcal A}_k^{\mathrm s}=\mathcal A_k^{\mathrm s}
\text{ for all }k=1,\ldots,K\right)
\ge \mathbb P\!\left(\widehat{\mathcal T}=\mathcal T^\star\right)\to 1.
\]
\endgroup

\clearpage

\bibliography{bibliography.bib}

\end{document}